\documentclass[aps, prb, twocolumn, showpacs, preprintnumbers,
amssymb, floatfix]{revtex4}

\usepackage{amsmath}
\usepackage{amsfonts}
\usepackage{amsbsy}
\usepackage{xcolor}
\usepackage{soul}
\usepackage{graphicx}

\newcommand{\spvec}[1]{\ensuremath{\mathbf{#1}}}
\newcommand{\unitvec}[1]{\ensuremath{\mathbf{\hat{#1}}}}
\newcommand{\sptensor}[1]{\ensuremath{\boldsymbol{\mathbf{#1}}}}

\newcommand{\colvec}[1]{\ensuremath{\mathrm{#1}}}

\renewcommand{\(}{\left(}
\renewcommand{\)}{\right)}

\newcommand{\commentout}[1]{{}}

\newcommand{\eq}[1]{Eq.~\eqref{#1}}
\newcommand{\rv}{\spvec{r}}
\newcommand{\Ev}{\spvec{E}}
\newcommand{\Dv}{\spvec{D}}
\newcommand{\Bv}{\spvec{B}}
\newcommand{\Hv}{\spvec{H}}
\newcommand{\Pv}{\spvec{P}}
\newcommand{\Mv}{\spvec{M}}
\newcommand{\pol}{\unitvec{e}}
\newcommand{\qv}{\spvec{q}}
\newcommand{\kv}{\spvec{k}}
\newcommand{\eo}{\epsilon_0}
\newcommand{\beq}{\begin{equation}}
\newcommand{\eeq}{\end{equation}}

\begin{document}
\title{Theoretical formalism for collective electromagnetic response
  of discrete metamaterial systems}
\author{Stewart D. Jenkins}
\author{Janne Ruostekoski}
\affiliation{School of Mathematics and Centre for Photonic
  Metamaterials, University of Southampton,
  Southampton SO17 1BJ, United Kingdom}

\begin{abstract}
  We develop a general formalism to describe the propagation of a
  near-resonant electromagnetic field in a  medium composed of
  magnetodielectric resonators.
  As the size and the spatial separation of nanofabricated resonators
  in a metamaterial array is
  frequently less than the wavelength, we describe them as discrete
  scatterers, supporting a single mode of current oscillation
  represented by a single dynamic variable.
  We derive a Lagrangian and Hamiltonian formalism for the coupled
  electromagnetic fields and oscillating currents in the \emph{length}
  gauge, obtained by the Power-Zienau-Woolley
  transformation.
  The response of each resonator to electromagnetic field is then
  described by polarization and magnetization densities that, to the
  lowest order in a multipole expansion, generate electric and
  magnetic dipole excitations.
  We derive a closed set of equations for the coherently scattered
  field and normal mode amplitudes of current oscillations of each
  resonator both within the rotating wave approximation, in which case
  the radiative decay rate is much smaller than the resonance
  frequency, and without such an assumption.
  The set of equations
  includes
  the radiative couplings between a
  discrete set of resonators mediated by the electromagnetic field,
  fully incorporating recurrent scattering processes to all orders.
  By considering an example of a two-dimensional split ring resonator
  metamaterial array,
  we show that the system
  responds cooperatively to near-resonant field, exhibiting collective
  eigenmodes, resonance frequencies, and radiative linewidths that
  result from strong radiative interactions between closely-spaced
  resonators.
\end{abstract}
\date{\today}
%42.25.Bs  -- Wave propagation, transmission and absorption.... not so sure
%42.50.Ct --Quantum Description of interaction of light with matter .. this is way down the list
%45.20.Jj -- Lagrangian and Hamiltonian Mechanics
\maketitle

\section{Introduction}
\label{sec:introduction}

Recent advances in nanofabrication provide a
 variety of tools for engineering the electromagnetic (EM) response
 of metamaterials in the radiofrequency, microwave,
 and  optical domains.
Metamaterials consist of arrays of artificially constructed
magnetodielectric resonators
which typically interact strongly with the incident and scattered EM
fields.
These resonator structures frequently extend over length scales
smaller than the wavelength of the EM field with which they interact.
For example, a metamaterial might comprise isolated circuit elements,
or meta-atoms, embedded in a dielectric host medium.
Whereas the EM properties of natural atoms are fixed, modifying the
design of artificially constructed meta-atoms can endow them with a
wide range of electric and/or magnetic responses.
Such control allows one to produce materials with EM properties such
as negative index of refraction \cite{ShelbySci2000, SmithEtAlPRL2000,
  SmithEtAlSCI2004} or negative group velocities.\cite{DollingSci2006}
These materials could conceivably be employed to create  perfect
lenses\cite{PendryPRL2000, VeselagoSPU1968} and electromagnetic
cloaks.\cite{PendryEtAlSCI2006, LeonhardtSCI2006, SchurigSci2006}

The exciting EM phenomena of nanofabricated metamaterials can often
depend on the effective bulk properties of the sample.
Homogenization theories have met with substantial success in
describing these properties.\cite{SmithEtAlPRB2002,
  KoschnyEtAlPRL2004, PendryEtAlIEEE1999, SmithEtAlPRL2000,
  BelovSimovskiPRE2005, LiuEtAlPRE2007, SimovskiMetaMat2007,
  LiEtAlPRE2009,  SmithPRE2010, SmithPendryJOSAB2006,
  FietzShvetsPRB2010}
Homogenization leads to
effective continuum models
that
strive to
treat excitations using
averaged polarization and magnetization densities formed by current
oscillations within the unit-cell resonators.
Analyzing an EM response using uniform medium descriptions, however,
can be complicated by the fact that recurrent scattering events, in
which a photon scatters more than once off the same resonator, produce
interactions which can strongly influence a system's EM
response.\cite{vantiggelen90,MoriceEtAlPRA1995, Ishimaru1978,
  RuostekoskiJavanainenPRA1997L, RuostekoskiJavanainenPRA1997,
  JavanainenEtAlPRA1999,devries98,fermiline,muller01,pinheiro04,optlattice}
In certain circumstances, the bulk permittivity and permeability can
be inferred by analyzing the transmission and reflection properties of
a metamaterial with finite thickness,\cite{SmithEtAlPRB2002,
  KoschnyEtAlPRL2004} or from the scattering properties of a
metamaterial's constituent
slabs.\cite{SimovskiMetaMat2007,LiEtAlPRE2009, SmithPRE2010}
But, accurately accounting for strong interactions between a
metamaterial's unit cells often requires simplifying assumptions such
as the elements being arranged in an infinite
lattice.\cite{KoschnyEtAlPRL2004, PendryEtAlIEEE1999,
  SmithEtAlPRL2000, BelovSimovskiPRE2005, LiuEtAlPRE2007}
The discrete translational symmetry of the infinite lattice can be
exploited, e.g., to approximate the local field corrections in a
medium of discrete magneto-electric scatterers.\cite{Kastel07}

The discrete nature of metamaterials becomes apparent when the
infinite lattice symmetry is broken.
The strongly interacting nature of these structures renders them very
sensitive to finite size effects
\cite{FedotovEtAlPRL2010,SzaboIEEE2010} and to disorder in the
lattice.\cite{papasimakis2009,SavoEtAlPRB2012} 
In systems of discrete resonators, interference of different
scattering paths between the elements can result, e.g., in light
localization.\cite{WiersmaNat1997, vanTiggelen99}
This effect is analogous to Anderson localization of electrons in
solids.
Even in regular arrays, strong interactions between resonators can
find important applications in metamaterial systems, providing precise
control and manipulation of EM fields on a subwavelength scale, e.g.,
by localizing sub-diffraction field
hot-spots.\cite{SentenacPRL2008,KAO10}
As another example, a system of interacting resonant wires was used to
produce a meta-lens able to transfer subwavelength features of an
evanescent field to propagating waves.\cite{LemoultPRL10}
In essence, recurrent scattering events produce strong interactions between meta-atoms that contribute to these effects.
As a result, ensembles of interacting resonators exhibit collective
mode of oscillation with discrete resonance frequencies and radiative
emission rates.
In principle, one can calculate the scattered field profile in a
metamaterial by having knowledge of how the material comprising the
meta-atoms reacts to the EM field.
One could then numerically integrate Maxwell's equations with a
numerical mesh small enough to resolve the features of each
meta-atom.
This, however, becomes computationally intractable when the system
contains more than a few resonator elements.

In this article, we develop a simplified, computationally efficient
formalism that captures the fundamental physical properties of a
finite metamaterial.
In particular, we show how EM mediated interactions can form a
cooperative response of the metamaterial's constituent resonators.
In this model, each unit cell element, or metamolecule, of the
metamaterial array is formed by combinations of circuit elements
acting as resonators that interact with the incident and scattered EM
fields.
In several metamaterial realizations, a metamolecule may further be
divided into separate sub-elements, e.g., isolated circuit elements
that can naturally be considered as the elementary building blocks of
the metamaterial sample.
We refer to such elementary building blocks as meta-atoms.
We assume each meta-atom supports a single mode of current
oscillation represented by a single dynamic variable.

The theoretical formalism we introduce describes the collective
  response of a metamaterial array to an incident EM field.
  To develop this formalism, we begin with the
  Lagrangian and Hamiltonian representations for charge and current
  distributions interacting with EM fields.
  Analyzing the system in the \emph{length} gauge, obtained by the
  Power-Zienau-Woolley transformation,\cite{PowerZienauPTRS1959,
  PowerBook, PZW} we derive coupled equations for the
  EM fields and resonators.
  A single resonator interacting with its self-generated fields
  behaves as an LC circuit in which emission of EM radiation damps the
  current oscillation.
  An incident EM field drives each resonator.
  However, each meta-atom is also driven by
  fields scattered from all other meta-atoms in the metamaterial array.
  By integrating out the EM fields, we derive a set of equations for
  the meta-atom current oscillations which describes the collective
  response of the array to the incident field.
  Each eigenmode of this system of equations represents a collective
  oscillation distributed over the resonators with a particular
  resonance frequency and radiative decay rate.
  Some modes are superradiant, with emission rates enhanced by
  collective interactions.
  In other modes, EM mediated interactions result in subradiant
  emission in which
  radiation repeatedly scattered between resonators remains trapped,
  slowly leaking away from the metamaterial.
  As an example, we analyze a 2D array of split ring resonators and examine
  several of its characteristic collective modes.
  We find that EM mediated interactions can produce a broad distribution of
  collective emission rates, and that the width of this
  distribution is sensitive to  the inter-resonator spacing.
  For example, in a $33\times 33$ array in which the resonators are
  separated by half a wavelength of the resonant light, the radiative
  emission rate can be   suppressed by five orders of magnitude.
  On the other hand, when the
  spacing is increased to 1.4  wavelengths, the emission rate is only
  suppressed by a factor of five.

In previous Lagrangian treatments, the interaction between elements of
a single metamolecule was accounted for by a phenomenological
coupling between meta-atom dynamic variables.\cite{LiuEtAlPRB2007}
Similar phenomenological coupling between nearest neighbor resonators can also
describe the propagation dynamics of excitations in a one-dimensional
chain of metamolecules.\cite{LiuEtAlPRL2006}
Radiative losses were accounted for by additional dissipative terms.
However, important effects such as superradiance or subradiance of
collective modes
cannot be modeled in this way.
By contrast, in our treatment the interactions between meta-atoms are
mediated entirely by the scattered EM fields; the radiation lost
through decay of one meta-atom can drive another and vice-versa.
The resulting collective modes of the system can, therefore, exhibit
either subradiant or superradiant decay.

This article is organized as follows.
We highlight the
main results of the developed formalism for the collective response of a metamaterial sample to EM fields
in Sec.~\ref{sec:dyn_summary}.
In Sec. \ref{sec:system-description}, we set up our description of
the metamaterial.
We provide a theoretical description of the system dynamics in
Sec.~\ref{sec:system-dynamics} where we  introduce the Lagrangian
and derive the Hamiltonian for the system and the equations of motion for the
meta-atoms.
We also arrive at expressions for the scattered EM fields that
drive the meta-atom dynamics.
A derivation of our Lagrangian from that describing arbitrary charged
particles interacting with the EM field in the Coulomb gauge is
provided in Appendix~\ref{sec:lagrangian},
and we elaborate on the
derivation of the Hamiltonian in Appendix~\ref{sec:elim-inst-non}.
We combine the field and meta-atom dynamics in
Sec.~\ref{sec:analysis-model} to arrive at coupled equations of
motion between meta-atoms in the rotating wave approximation, in which
the meta-atom decay rates are much less than their resonance
frequencies.
A more general model for collective interactions is
provided in Appendix~\ref{sec:ensemble-meta-atoms}.
In Sec.~\ref{sec:an-ensemble-asymm}, we apply the theoretical
formalism to describe collective modes of oscillation in
an array of
symmetric split ring
resonators.
Collective modes of these resonators are connected to the linewidth
narrowing \cite{JenkinsLineWidthArxiv} of a transmission resonance
observed in Ref.~\onlinecite{FedotovEtAlPRL2010}.
In Sec. \ref{sec:quant-metam-dynam}, we quantize this formalism in
the special case that the resonators do not suffer from thermal or
ohmic losses.
Conclusions follow in Sec.~\ref{sec:conclusions}.

\section{Key results: collective dynamics arising from recurrent EM scattering}
\label{sec:dyn_summary}

In this section, we summarize key results presented in this article.
Ultimately, we describe the collective dynamics arising from
an ensemble of magnetodielectric resonators interacting via
  a
  near-resonant EM field.
When such resonators are placed close to each other, the system can
respond to EM fields cooperatively.
In order to provide a computationally efficient description,
we consider a metamaterial
array
composed of
a set of
$N$
discrete
meta-atoms.
We assume each meta-atom $j$ ($j=1\ldots N$) supports a single mode
of current oscillation whose behavior is described by a single
dynamic variable $Q_j$, with units of charge, and its rate of change
$I_j=\dot{Q}_j$, with units of current.
As described in Sec.~\ref{sec:system-description}, the current
oscillation produces a polarization density $\spvec{P}_j(\rv,t)$
proportional to $Q_j$ and a magnetization density $\spvec{M}_j(\rv,t)$
proportional to $I_j$. An incident wave with electric field
$\spvec{E}_{\mathrm{in}}(\rv,t)$ and magnetic
induction field $\spvec{B}_{\mathrm{in}}(\rv,t)$ impinges on the
system.

For the coupled set of circuit elements and EM fields we derive
a Lagrangian and Hamiltonian formalism in Sec.~\ref{sec:lagr-hamilt-meta}.
The Lagrangian is expressed in the \emph{length} gauge, obtained by the
Power-Zienau-Woolley transformation.\cite{PowerZienauPTRS1959,
  PowerBook, PZW}
For the dynamical variables $Q_j$ and the EM vector potential
$\spvec{A}(\spvec{r})$ we obtain the corresponding conjugate momenta
$\phi_j$ [Eq.~\eqref{eq:conjMomofQ_j}] and $-\spvec{D} (\spvec{r})$
[Eq.~\eqref{eq:conMom}], respectively.
Here $\spvec{D} \equiv \epsilon_0\spvec{E}+\spvec{P}$ denotes the electric displacement field.
The joint dynamics of the meta-atom and EM fields are then
governed by the Hamiltonian
\begin{eqnarray}
  \mathcal{H} &=& \mathcal{H}_{\mathrm{EM}} + \frac{1}{2\epsilon_0} \int
  d^3r \left|\spvec{P}\right|^2 + \sum_j
  \Bigg[\frac{1}{2l_j}(\phi_j-\Phi_j)^2 \nonumber \\
  & & - \frac{1}{\epsilon_0} \int d^3r\,
  \spvec{D}(\spvec{r},t) \cdot \spvec{P}_j(\spvec{r})
  \Bigg] \, \textrm{,}
  \label{eq:Ham_intro}
\end{eqnarray}
where $\mathcal{H}_{\mathrm{EM}}$ [Eq.~\eqref{eq:HEM}] is the
Hamiltonian for the free EM field and $\Phi_j$ [Eq.~\eqref{eq:PhiDef}]
is an effective magnetic flux through the meta-atom.
The final term of the Hamiltonian accounts for interactions between
electric dipoles distributed in the current oscillations and the
electric field, while magnetic interactions are contained in
$(\phi_j-\Phi_j)^2$ and arise in the relationships between $\phi_j$,
$\Phi_j$ and $I_j$.

From the Hamiltonian we derive a coupled set of equations for the EM fields
and the meta-atoms.
The incident EM fields drive current oscillations within the
meta-atoms, thereby inducing polarization and magnetization
densities.
In Sec.~\ref{sec:scattered-em-fields} we derive and integrate the
equations for the total EM fields that are expressed in terms of the
incident fields and the fields scattered from the polarization and
magnetization densities of the meta-atoms.
Specifically, currents in meta-atom $j$, when oscillating at a
frequency $\Omega$, produce the monochromatic scattered electric field
$\spvec{E}_{\mathrm{S},j}$ and magnetic field
$\spvec{H}_{\mathrm{S},j}$ given by
\begin{subequations}
  \label{eq:scatteredFields_pre}
  \begin{align}
    \spvec{E}_{\mathrm{S},j}(\spvec{r},\Omega) &=
    \frac{k^3}{4\pi\epsilon_0} \int d^3 r' \,
    \Big[\sptensor{G}(\spvec{r} - \spvec{r}',\Omega) \cdot
    \spvec{P}_{j}(\spvec{r}',\Omega)  \nonumber\\
    & \qquad + \frac{1}{c} \sptensor{G}_\times
    (\spvec{r}-\spvec{r}',\Omega) \cdot
    \spvec{M}_{j}(\spvec{r}',\Omega) \Big] , \label{eq:E_Sj_pre}\\
    \spvec{H}_{\mathrm{S},j}(\spvec{r},\Omega) & =
    \frac{k^3}{4\pi} \int d^3 r' \, \Big[\sptensor{G}(\spvec{r} -
    \spvec{r}',\Omega)\cdot
    \spvec{M}_{j}(\spvec{r}',\Omega) \nonumber\\
    & \qquad - c \sptensor{G}_\times(\spvec{r}-\spvec{r}',\Omega)\cdot
    \spvec{P}_{j}(\spvec{r}',\Omega) \Big]\,,
    \label{eq:H_Sj_pre}
  \end{align}
\end{subequations}
where $\sptensor{G}(\rv-\rv',\Omega)$ is the radiation kernel
connecting an oscillating electric (magnetic) dipole at position
$\rv'$ to the electric (magnetic) field at position $\rv$, while
$\sptensor{G}_{\times}(\rv-\rv',\Omega)$ connects an electric
(magnetic) dipole at $\rv'$ to its radiated magnetic (electric) field
at $\rv$.\cite{Jackson}
Expressions for these radiation kernels are given in
Eqs.~\eqref{eq:RadKerPmFreq} and \eqref{eq:CrossKerPmFreq}.

The total electric and magnetic fields are obtained as a sum of
  the incident fields and the fields scattered by all the meta-atoms
  in the system
  \begin{align}
    \spvec{D}(\rv,t) &=\spvec{D}_{\mathrm{in}}(\rv,t)+\sum_j
    \spvec{D}_{{\mathrm{S}},j}(\rv,t)
    \label{eq:TotalDFromInScatt_pre} \,\textrm{,}\\
    \spvec{B}(\rv,t) &=\spvec{B}_{\mathrm{in}}(\rv,t)+\sum_j \spvec{B}_{\mathrm{S},j}(\rv,t) \, \textrm{,}
\end{align}
where
we have the scattered magnetic induction
$\spvec{B}_{\mathrm{S},j} \equiv
\mu_0(\spvec{H}_{\mathrm{S},j} + \spvec{M}_j)$
and the scattered electric displacement $\spvec{D}_{\mathrm{S},j} \equiv
\epsilon_0\spvec{E}_{\mathrm{S},j} + \spvec{P}_j$  from the meta-atom $j$.

Although, according to Eqs.~\eqref{eq:scatteredFields_pre}, the
  polarization and magnetization densities of all the meta-atoms
  act as source terms that determine the scattered EM fields, there
  is, in general, no simple way of solving for the
  polarization $\spvec{P}_{j}(\spvec{r},\Omega)$ and magnetization
  $\spvec{M}_{j}(\spvec{r},\Omega)$ densities themselves.
The equations for
near-resonant EM fields and closely-spaced
  resonators are strongly coupled, and the meta-atoms are
  driven by both the incident fields and fields scattered by all
other meta-atoms in the system.
  This is illustrated by
Hamilton's equations of motion
  for the resonators
\begin{subequations}
  \label{eq:HamEqs1_intro}
  \begin{eqnarray}
    \label{eq:Qmotion_intro}
    \dot{Q}_j(t) &=& I_j(t), \\
    \dot{\phi}_j(t) &=& \mathcal{E}_j(t) \text{,}
  \end{eqnarray}
\end{subequations}
where the total electric field induces an effective electromotive force
(EMF) $\mathcal{E}_j$ [Eq.~\eqref{eq:PhiDef}], driving the meta-atoms.

Solving the coupled dynamical equations for the resonators
\eqref{eq:HamEqs1_intro} and the EM fields
\eqref{eq:scatteredFields_pre} constitute the central results of the
paper. We begin in  Sec.~\ref{sec:single-meta-atom} by considering a
single meta-atom.
A meta-atom not only feels the influence of the incident EM field and
the fields scattered from other meta-atoms, but also its
self-generated field.
We show that interactions between a meta-atom $j$ and its
self-generated field produces an effective damped LC circuit for the
current oscillations with capacitance $C_j$ [Eq.~\eqref{eq:C_j_xxx2}],
self-inductance $L_j$ [Eq.~\eqref{eq:LjDef}] and
resonance frequency $\omega_j = 1/\sqrt{C_jL_j}$.
The oscillating electric and magnetic dipoles of the meta-atom scatter
EM fields and therefore induce a radiative decay at rates
$\Gamma_{\mathrm{E},j}$ and $\Gamma_{\mathrm{M},j}$, respectively.

In a metamaterial array of several meta-atoms we then solve the
coupled set of equations \eqref{eq:HamEqs1_intro} and
\eqref{eq:scatteredFields_pre} when each meta-atom is also driven by
the scattered fields from all the other meta-atoms.
This results in multiple scattering events and yields EM field
mediated interactions between the meta-atoms.
In particular, when the multiple scattering between closely-spaced
resonators becomes dominant, so that the EM wave is scattered more
than once by the same scatterer (this is called recurrent scattering),
the system responds to EM fields cooperatively. In order to analyze
the eigenmodes of such a system, it is beneficial to introduce the
excitation amplitudes of each meta-atom LC circuit in terms of its
dynamical coordinates and the canonical momenta.
In particular, when
the decay rates are much less than the resonance frequencies, the
meta-atom dynamics are well described by the slowly varying normal
variables
\begin{equation}
  b_j(t)  \equiv  \frac{e^{i\Omega_{0}t}}{\sqrt{2\omega_j}}
  \left( \frac{Q_j(t)}{\sqrt{C_j}} + \frac{\phi_j(t)}
    {\sqrt{L_j}}\right) \text{.}
  \label{eq:bSingAtDef_pre}
\end{equation}
In terms of the normal variables we then derive a linear set of equations for the meta-atoms whose interactions are mediated by the EM fields by explicitly integrating out the scattered fields
\begin{equation}
  \label{eq:EqsOfM_pre}
  \dot{b}_j = \sum_{j'} {\mathcal{C}}_{j,j'} b_{j'} +
  f_{\mathrm{in},j} \textrm{.}
\end{equation}
The expressions for the coupling matrix $\mathcal{C}$ between the
meta-atoms and the driving terms by the incident fields
$f_{\mathrm{in},j}$ are derived in Sec.~\ref{sec:coll-inter-rotat}.
The diagonal elements of $\mathcal{C}$ reflect the resonance
frequencies and decay rates of the single meta-atoms in isolation,
while the off-diagonal elements arise from scattered electric and
magnetic fields interacting with the meta-atom electric and magnetic
dipoles.
The coupled equations \eqref{eq:EqsOfM_pre} include the recurrent
scattering events between the meta-atoms to all orders.
We generalize our treatment to account for stronger interactions
between meta-atoms, i.e., where interactions mediated by scattered
fields are comparable to the effects of the self-generated field, in
Appendix~\ref{sec:ensemble-meta-atoms}.

A system of $N$ meta-atoms supports $N$ collective modes of current
oscillation, each matched to an eigenvector of the matrix
$\mathcal{C}$.
Each mode $i$ has its own collective resonance frequency and decay
rate given in terms of its eigenvalue $\lambda_i$ as
\begin{subequations}
  \begin{eqnarray}
    \label{eq:EigenFreq}
    \Omega_i &=& -\operatorname{Im}(\lambda_i) + \Omega_0, \\
    \gamma_i &=& -2 \operatorname{Re}(\lambda_i)\,,
    \label{eq:EigenGamma}
  \end{eqnarray}
  \label{eq:EigenValRels}
\end{subequations}
respectively.
As a result of the interactions, the collective emission rates
can be either much less than (subradiant) or much greater than
(superradiant) the constituent single meta-atom decay rates.
We demonstrate this in Sec.~\ref{sec:an-ensemble-asymm} where we
consider the collective effects on a 2D metamaterial array of
symmetric split ring resonators (SRRs), metamolecules possessing
reflection symmetry which consist of two concentric circular arcs of
equal length.
Even in a relative small metamaterial sample of $33\times 33$
unit-cell resonators for the lattice spacing of a half-wavelength we
find that the spectrum of resonance frequencies exhibits a long tail
of strongly subradiant eigenmodes.
The most subradiant mode of the system possesses a radiative decay
rate of about five orders of magnitude less than that of an isolated
meta-atom.
This eigenmode exhibits a checkerboard phase-pattern of dominantly
electric dipole excitations.
We also find that the strong response of the metamaterial sample is
very sensitive to the spacing between the resonators.
We analyze the spectrum for the lattice spacing of 1.4 wavelength in
which case the distribution of the decay rates is considerably
narrowed.
The most subradiant mode now has a resonance linewidth that is five
times narrower than the one of the isolated unit-cell resonator.
Finally, we also provide an example how the propagation dynamics of excitations in a metamaterial array 
can be analyzed using the collective eigenmodes. We find that the lattice spacing, and hence the interactions 
between the resonators, strongly influence the rate at which excitations spread over the array. 

\section{Discrete resonator model of a metamaterial}
\label{sec:system-description}

To develop the formalism characterizing
interactions of magnetodielectric resonators in EM fields, we first
provide a detailed description of the metamaterial and the model we
use to represent it.
We consider an ensemble of metamolecules, unit-cell elements
that comprise the metamaterial, driven by an incident EM field.
Each metamolecule can be decomposed into some number of meta-atoms,
which may correspond, for example, to individual circuit elements.
We model our metamaterial as an ensemble of $N$ meta-atoms.
The position of the meta-atom $j$ is denoted by $\spvec{R}_j$
($j=1,\ldots,N$).
An external beam with electric field
$\spvec{E}_{\mathrm{in}}(\spvec{r},t)$ and associated magnetic
induction
$\spvec{B}_{\mathrm{in}}(\rv,t)$ impinges on the ensemble, driving the
meta-atoms.
We assume the incident field is bandwidth limited with a spectrum
centered at angular frequency $\Omega_0$, and that the spatial extent
of each meta-atom lies well within a carrier wavelength $\lambda =
2\pi c/\Omega_0$.

The meta-atoms may be composed of, e.g., metallic circuit elements
supporting plasmonic oscillations, allowing charges and currents to
flow internally.
The current and charge distributions produce EM fields,
which in turn, influence the dynamics of these distributions.
As such, each element supports various eigenmodes of current
oscillation.\cite{ProdanEtAlSCI2003, WangEtAlACR2006}
For simplicity, we identify meta-atom $j$ with a single eigenmode of
current oscillation whose state can be described by a single dynamic
variable $Q_j(t)$ with units of charge and whose spatial profile is
described by time-independent functions $\spvec{p}_j(\spvec{r})$ and
$\spvec{w}_j(\spvec{r})$.
These mode functions are defined such that the polarization
$\spvec{P}_j(\rv,t)$ and magnetization $\spvec{M}_j(\rv,t)$ densities
associated with atom $j$ are
\begin{subequations}
  \begin{align}
    \spvec{P}_j(\rv,t) &= Q_j(t) \spvec{p}_j(\spvec{r}),\\
    \spvec{M}_j(\rv,t) &= I_j(t) \spvec{w}_j(\spvec{r})\,,
  \end{align}
  \label{eq:PandMDensDefs}
\end{subequations}
where $I_j(t) \equiv dQ_j/dt$ is the current.
The definitions of the polarization and magnetization lead to the expressions
of the charge and current densities within each meta-atom,
\begin{subequations}
  \begin{align}
    \rho_j(\spvec{r},t) & = - Q_j(t) \nabla \cdot
    \spvec{p}_j(\spvec{r})
    \label{eq:chargeDensity}\\
    \spvec{j}_j(\spvec{r},t) & = \left(\spvec{p}_j(\spvec{r}) + \nabla
      \times \spvec{w}_j(\spvec{r})\right) I_j(t) \text{.}
    \label{eq:currentDensity}
  \end{align}
  \label{eq:chargeAndCurrentDensities}
\end{subequations}
The total polarization and magnetization densities of the system are
obtained from a sum over the polarization and magnetization densities
of every meta-atom $j$, respectively,
\begin{align}
\spvec{P}(\spvec{r},t) &= \sum_j \spvec{P}_j(\spvec{r},t),\\
\spvec{M}(\spvec{r},t) &= \sum_j \spvec{M}_j(\spvec{r},t)\,.
\end{align}
We choose the mode functions
so that they are zero outside the neighborhood of the meta-atom.

We note that, in general, the various parts of a circuit element
contain charge and current densities that could behave independently
of one and other; they could therefore be represented by separate
dynamic variables.
These extra degrees of freedom could be described by assigning
multiple modes of current oscillation to the element, each with its
own dynamic variable and mode functions to describe the corresponding
polarization and magnetization densities.
The resulting set of mode function dynamic variables could then
interact with one and other via the EM fields.
In essence, one could view a circuit element as an ensemble of
meta-atoms that touch or overlap with one and other.
In this work, however, we have assumed that the mode functions have
been chosen so that they are eigenmodes of elements, i.e., there
is a zero net interaction between the modes in a given circuit element
within a metamolecule.
We therefore identify a meta-atom with an eigenmode of current
oscillation within a circuit element and treat each meta-atom as
possessing only a single mode of current oscillation.
This is analogous to approximating an atom interacting with the
EM field as a two-level atom. In the present work, we will not address
how the eigenmodes of current oscillations are calculated.
For isolated
circuit elements they could be computed numerically solving Maxwell's equations
using actual material parameters.
Alternatively, one could obtain the meta-atom resonance properties
directly from experimental measurements, or estimate them using
geometrical arguments.

\section{System dynamics}
\label{sec:system-dynamics}

In this section we introduce the Lagrangian and Hamiltonian formalism
for a magnetodielectric medium interacting with EM fields,
specifically derived for a system consisting of circuit elements whose
dynamic variables represent
eigenmodes of current
oscillations.
The Lagrangian is expressed in the \emph{length} gauge, obtained by
the Power-Zienau-Woolley transformation.\cite{PowerZienauPTRS1959,
  PowerBook, PZW}
We find that
this
particular representation of electromagnetism turns
out to be especially useful for describing localized, collectively
interacting circuit elements.
The specific details of the Power-Zienau-Woolley transformation are
covered in Appendix \ref{sec:lagrangian}.

From the Lagrangian we derive the conjugate momenta for the dynamic
variables of the meta-atoms and the EM fields, and the Hamiltonian for
the system.
The dynamics of the model describe charge and current densities of
the system interacting with the EM fields.
We derive a coupled set of equations for the EM fields and the
resonators in which
both the electric and magnetic fields drive the meta-atom dynamics.
The expressions for the electric and magnetic fields are obtained in
terms of the incident fields illuminating the sample and the fields
scattered from the polarization and magnetization densities that
represent the meta-atoms in the medium.

\subsection{The Lagrangian and Hamiltonian formalism for meta-atoms interacting with EM fields}
\label{sec:lagr-hamilt-meta}

We treat the dynamics of the system in the Coulomb gauge beginning
with the Lagrangian formalism.
It is particularly advantageous to study the EM response in a gauge
representation obtained by the Power-Zienau-Woolley transformation.\cite{PowerZienauPTRS1959,  PowerBook, PZW}
In Appendix \ref{sec:lagrangian}, we show that the Lagrangian in the
Power-Zienau-Woolley picture \cite{CohenT} can be written in terms of
meta-atom dynamic variables as
\begin{equation}
  \label{eq:Lagrange2}
  \mathcal{L} = \mathcal{K} +\mathcal{L}_{\mathrm{EM}} + V_{\mathrm{Coul}} +
  \sum_j \left[ Q_j(t) \mathcal{E}_j +   I_j(t) \Phi_j \right]\,,
\end{equation}
where
$\mathcal{K}$ is an effective kinetic energy given by
\begin{equation}
  \mathcal{K} = \sum_j \frac{1}{2} l_j I_j^2 \,\textrm{.}
  \label{eq:KinEnergy}
\end{equation}
The phenomenological kinetic inductance $l_j$ of meta-atom $j$
provides, within the effective single-particle
description of the system, an inertia to the current oscillation
that would be present in the absence of EM field interactions.
This inertia could result, for example, from the effective mass of
charge carriers or surface plasmons within the meta-atom.
Excitation of a current oscillation displaces charge carriers from
their equilibrium configuration producing a charge density
$\rho_j(\rv)$ [Eq.~\eqref{eq:chargeDensity}] within meta-atom $j$.
The meta-atom charge densities interact via the instantaneous Coulomb
interaction
\begin{equation}
  V_{\mathrm{Coul}} =
  \frac{1}{8\pi\epsilon_0}\sum_{j,j'}
  \int d^3r d^3r'\,
  \frac{\rho_j(\rv)\rho_{j'}(\rv')}{\left|\rv-\rv' \right|} \,\text{.}
  \label{eq:V_Coul}
\end{equation}
The current oscillation in meta-atom $j$ interacts with external EM fields via
an effective electromotive force (EMF) $\mathcal{E}_j$
and an effective magnetic flux $\Phi_j(t)$ through that
meta-atom:
\begin{align}
  \mathcal{E}_j(t) &\equiv \int d^3r\, \spvec{E}(\spvec{r},t) \cdot
  \spvec{p}_j(\spvec{r}), \label{eq:EMFDef}\\
  \Phi_j(t) &\equiv \int d^3r\, \spvec{B}(\spvec{r},t) \cdot
  \spvec{w}_j(\spvec{r})\,. \label{eq:PhiDef}
\end{align}
The EMF interacts with the charge $Q_j$ on the circuit, while the
current $I_j$ interacts with the magnetic flux.
The Lagrangian for the free EM field
$\mathcal{L}_{\mathrm{EM}}$ is given in terms of the Coulomb gauge
vector potential $\spvec{A}$ as
\begin{equation}
  \label{eq:L_EM}
  \mathcal{L}_{\mathrm{EM}} = \frac{\epsilon_0}{2} \int d^3r\, \left[
    \left|\frac{d\spvec{A}}{dt}\right|^2 - c^2 \left|\nabla \times
      \spvec{A}\right|^2\right] \,\text{.}
\end{equation}
The Lagrangian for the free EM field represents the radiative fields that are responsible for the excitations of the meta-atoms.

We now wish to determine the Hamiltonian for the system.
We proceed by identifying the conjugate  momenta of the dynamic
variables.
Those for charges are given by
\begin{equation}
  \phi_j \equiv \frac{\partial \mathcal{L}} {\partial I_j}
  = l_j I_j +\Phi_j \,. \label{eq:conjMomofQ_j}
\end{equation}
Note that in the limit where $l_j$ is vanishingly small, the conjugate
momentum of the charge is dominated by the flux through the circuit.
This is often the case in microwave metamaterials, where the EM
interactions dwarf the effects of charge carrier inertia.
The vector potential represents a continuous field of dynamic
variables which possess a corresponding continuum of conjugate momenta
defined as
\begin{equation}
  \label{eq:conMomDef}
  \spvec{\Pi}(\spvec{r},t) \equiv
  \frac{\partial \mathcal{L}}{\partial \dot{\spvec{A}}(\rv,t)} \text{.}
\end{equation}
This conjugate momentum will have a contribution from
$\mathcal{L}_{\mathrm{EM}}$ and pick up a contribution from the
interaction term $\sum_j\mathcal{E}_jQ_j = \int d^3r\, \spvec{P} \cdot
\spvec{E}$.
For a system of neutral meta-atoms,
the conjugate momentum of the vector potential
is given by \cite{CohenT}
\begin{equation}
  \label{eq:conMom}
  \spvec{\Pi}(\spvec{r},t) = -\spvec{D}(\rv,t) \,\text{,}
\end{equation}
where
\begin{equation}
  \spvec{D} = \epsilon_0 \spvec{E} + \spvec{P}
  \label{eq:DDef}
\end{equation}
is the electric displacement field.

In treating the field dynamics, it is often convenient to express
these fields in terms of the normal variables $a_{\qv,\lambda}(t)$
describing a plane wave with wavevector $\spvec{q}$ and transverse
polarization $\unitvec{e}_{\unitvec{q},\lambda}$.
These normal variables are defined such that the
electric displacement and magnetic fields are given by
\begin{align}
  \spvec{D} (\spvec{r},t) &= i\int d^{3}q\,
  \xi_q \sum_{\lambda} \unitvec{e}_{\unitvec{q},\lambda}
  {a}_{\spvec{q},\lambda}(t) e^{i\spvec{q}\cdot\spvec{r}} +
  \mathrm{C.c.}, \label{eq:elecDisp} \\
 \spvec{B}(\spvec{r},t) &= i \sqrt{\frac{\mu_0}{\epsilon_0}}  \int
  d^{3}q\, \xi_q
  \sum_{\lambda}
  \left( \unitvec{q} \times \unitvec{e}_{\unitvec{q},\lambda} \right)
  a_{\qv,\lambda}(t)  e^{i \spvec{q}\cdot\spvec {r}} \nonumber\\
  &+ \mathrm{C. c.}, \label{eq:magField}
\end{align}
respectively, where
\begin{equation}
  \xi_q \equiv \sqrt{\frac{cq\epsilon_0}{2(2\pi)^3}}\text{ .}
\end{equation}
The normal variables for the EM field satisfy the following relations
in terms of the Poisson brackets \cite{Goldstein}
\begin{equation}
  \left\{ a_{\spvec{q},\lambda},
    a_{\spvec{q'},\lambda'}^\ast\right\}  = -i
  \delta_{\lambda,\lambda'} \delta(\spvec{q} - \spvec{q}'),
\end{equation}
and $\left\{ a_{\spvec{q},\lambda},  a_{\spvec{q}',\lambda'}\right\} = \left\{
  a_{\spvec{q},\lambda}, Q_j \right\} =
\left\{ a_{\spvec{q},\lambda}, \phi_j \right\} = 0$.

Having obtained the conjugate momenta and normal field variables, one
may write the Hamiltonian for the system by applying the Legendre
transform
\begin{equation}
  \label{eq:H_Legendre}
  \mathcal{H} = \sum_j \dot{Q}_j\phi_j + \int d^3r\, \dot{\spvec{A}}
  \cdot \spvec{\Pi} - \mathcal{L} \text{.}
\end{equation}
It is beneficial to decompose the Hamiltonian $\mathcal{H} =
\mathcal{H}_{\mathrm{mm}} +  \mathcal{H}_{\mathrm{E}}$ into a
component containing contributions from the meta-atom conjugate
momenta, $\mathcal{H}_{\mathrm{mm}}$, and a component accounting for
electric field interactions and the free EM field,
$\mathcal{H}_\mathrm{E}$.
The former contribution is given explicitly by
\begin{equation}
  \label{eq:H_mam}
  \mathcal{H}_{\mathrm{mm}} \equiv \sum_j \left(\phi_j I_j - \Phi_j
    I_j\right)- \mathcal{K} \text{.}
\end{equation}
But, because $\phi_j-\Phi_j = l I_j$ [Eq.~\eqref{eq:conjMomofQ_j}],
\begin{equation}
  \mathcal{H}_{\mathrm{mm}} = \mathcal{K} = \sum_j \frac{\left(\phi_j
      - \Phi_j\right)^2}{2l_j}
\end{equation}
reduces to the kinetic energy of the current oscillations.
The terms involving the electric field contribution, on the other
hand, are given explicitly by
\begin{equation}
  \label{eq:H_EDef}
  \mathcal{H}_{\mathrm{E}} = - \int d^3r\, \left(\dot{\spvec{A}}\cdot
    \spvec{D} + \spvec{E}\cdot \spvec{P} \right) - V_{\mathrm{Coul}}
  -\mathcal{L}_{\mathrm{EM}}
\end{equation}
It is beneficial to simplify the contribution of Eq.~\eqref{eq:H_EDef}.
We carry out this simplification in Appendix~\ref{sec:elim-inst-non}.
The total system Hamiltonian may thus be written in the the
Power-Zienau-Woolley picture as\cite{CohenT}
\begin{eqnarray}
 \mathcal{H} &=& \mathcal{H}_{\mathrm{EM}} + \frac{1}{2\epsilon_0} \int
  d^3r \left|\spvec{P}\right|^2 + \sum_j
  \Bigg[\frac{1}{2l_j}(\phi_j-\Phi_j)^2 \nonumber \\
  & & - \frac{1}{\epsilon_0} \int d^3r\,
  \spvec{D}(\spvec{r},t) \cdot \spvec{P}_j(\spvec{r})
  \Bigg] \, \textrm{,}
  \label{eq:Ham}
\end{eqnarray}
where the Hamiltonian for the free EM field is
\begin{align}
  \label{eq:HEM}
  \mathcal{H}_{\mathrm{EM}} = & \frac{\epsilon_0}{2} \int d^3r\, \left[
    \left|\frac{\spvec{\Pi}}{\epsilon_0}\right|^2 + c^2 \left|\nabla \times
      \spvec{A}\right|^2\right]\nonumber\\ =& \int d^3 q\, \sum_\lambda cq\,
  a_{\qv,\lambda}^\ast a_{\qv,\lambda} \,.
\end{align}

To understand the dynamics that will arise from this Hamiltonian, we
examine the physical role of each term individually.
The interaction between the displacement field and polarization
density can be written in terms of the emitter dynamic variables as
\begin{equation}
  - \int d^3r\,
  \frac{\spvec{D}(\rv,t)}{\epsilon_0} \cdot \spvec{P}_j(\rv,t) = -
  \frac{Q_j}{\epsilon_0} \int d^3r\, \spvec{D}(\rv,t) \cdot
  \spvec{p}_j(\rv) \, \textrm{.}
  \label{eq:DispFieldInteraction}
\end{equation}
This represents an interaction energy between the electric
displacement and the spatial distribution of electric dipoles
contained in the polarization density.
On the other hand, the interaction with the magnetic field becomes
apparent when expanding $\mathcal{K}$, which yields
\begin{equation}
  \label{eq:magInter}
  \frac{(\phi_j-\Phi_j)^2}{2l_j} = \frac{\phi_j^2}{2l_j} -
  \frac{\phi_j}{l_j}\Phi_j + \frac{\Phi_j^2}{2l_j} \,\text{,}
\end{equation}
The interaction of meta-atom $j$ with the magnetic field
arises in the second term.
The physical significance of this interaction can be understood by
expressing that contribution in terms of the magnetization density
[Eq.~\eqref{eq:PandMDensDefs}] as
\begin{equation}
  \label{eq:magneticDipInt}
  -\frac{\phi_j}{l_j}\Phi_j = -\int d^3r\, \spvec{M}_j\cdot\spvec{B} -
  \frac{\Phi_j^2}{l_j} \text{.}
\end{equation}
Equation~\eqref{eq:magneticDipInt} effectively contains
the interaction between the magnetization density and the magnetic
field.
Additionally, Eq.~\eqref{eq:magneticDipInt} includes a term
proportional to the square of the magnetic flux.
This artifact appears because the magnetization density is a function
of the meta-atom current $I_j$ rather than its conjugate momentum.
When this portion of the interaction is written as entirely in terms
of the meta-atom conjugate momentum, the term proportional to the
square of the flux disappears.
The last term of Eq.~\eqref{eq:magInter} represents a diamagnetic
interaction proportional to the square of the magnetic field flux
through a meta-atom.
These interactions are analogous to the effective magnetization and
diamagnetic interactions found in the Hamiltonian for electrically
charged point particles in Ref.~\onlinecite{CohenT}.

Finally, we examine the local polarization self-energy term appearing
in the Hamiltonian [Eq.~\eqref{eq:HEM}].
This can be expressed in terms of the dynamic variables as
\begin{align}
  & \frac{1}{2\epsilon_0}
  \int d^3r\, \spvec{P}(\rv,t)\cdot\spvec{P}(\rv,t) \nonumber\\
  &= \sum_{j,j'}
  \frac{Q_jQ_{j'}}{2\epsilon_0} \int d^3r \, \spvec{p}_j(\rv) \cdot
  \spvec{p}_{j'}(\rv) \,\textrm{.}
  \label{eq:contactInteraction_full}
\end{align}
When the meta-atoms are spatially separated, however, their
polarization mode functions do not overlap, i.e., $\spvec{p}_j\cdot
\spvec{p}_{j'} \equiv 0$ for $j\ne j'$.
The presence of $\spvec{P}(\rv,t)\cdot\spvec{P}(\rv,t)$ results only
in an interaction of the meta-atom with itself, which manifests itself
as
\begin{equation}
  \frac{1}{2\epsilon_0}
  \int d^3r\, \spvec{P}(\rv,t)\cdot\spvec{P}(\rv,t) = \sum_{j}
  \frac{Q_j^2}{2\epsilon_0} \int d^3r \, \left|\spvec{p}_j(\rv)\right|^2 \,\textrm{.}
  \label{eq:contactInteraction}
\end{equation}
If the meta-atoms were to overlap, a contact potential proportional to
$Q_jQ_{j'}$ would appear between the overlapping elements $j$ and
$j'$.
In the initial Lagrangian [Eq.~\eqref{eq:Lagrange2}], direct
interactions between the meta-atoms appeared via the instantaneous
Coulomb interaction.
An advantage of the Hamiltonian treatment in the Power-Zienau-Woolley
picture is that such interactions do not appear explicitly; other
instantaneous, non-causal contributions to the dynamics cancel out
those of the Coulomb interaction.
This leaves the meta-atom dynamic variables to interact exclusively
with the vector potential and its conjugate momentum.
Any effective interactions between meta-atoms are thus mediated by
these field dynamic variables.\cite{CohenT}

\subsection{The meta-atom dynamics}
\label{sec:meta-atom-dynamics}

The meta-atoms' interaction with the EM fields are illustrated by
Hamilton's equations of motion describing the current oscillations
\begin{subequations}
  \label{eq:HamEqs1}
  \begin{eqnarray}
    \label{eq:Qmotion}
    \dot{Q}_j(t) &=& \left\{Q_j,\mathcal{H}\right\}
    =I_j(t) = \frac{\phi_j(t)-\Phi_j(t)}{l} \\
    \dot{\phi}_j(t) &=& \left\{\phi_j,\mathcal{H}\right\}
    = \mathcal{E}_j(t) \text{.}
\end{eqnarray}
\end{subequations}
The conjugate momentum $\phi_j$ is driven entirely by the EMF
$\mathcal{E}_j$, while Eq.~\eqref{eq:Qmotion} is nothing more than a
statement that the rate of change of the charge is the current.
At first glance, it may appear that the magnetic field does not drive
the meta-atoms.
However, its effects manifest themselves indirectly through a
relationship between the conjugate momentum $\phi_j$ and the current
$I_j$ that will be discussed in Sec.~\ref{sec:analysis-model}.
Effective interactions between the resonators come about through
multiple scattering of the EM field between resonators.

\subsection{The scattered EM fields}
\label{sec:scattered-em-fields}

In the previous subsection we derived the equations for the meta-atoms
driven by EM fields.
In order to find a coupled set of equations for the fields and the
resonators, we need find how the EM fields depend on the state of the
meta-atom charges and currents.
In this subsection we derive integral expressions for the scattered EM
fields where the metamaterial medium acts as a source with effective
polarization and magnetization densities.
The total electric and magnetic fields are then represented as sums of
the incident fields and the scattered fields from the medium.
The resulting equations for the resonators and the EM fields are
strongly coupled: the resonator dynamics are driven by the EM fields
and the fields themselves depend on the excited meta-atom current
oscillations.

We begin with the equation of motion for the normal field operators
\begin{eqnarray}
  \lefteqn{\frac{d a_{\qv,\lambda}}{dt}
    =  \left\{a_{\qv,\lambda},
      \mathcal{H}\right\} = -icq a_{\qv,\lambda}}  \nonumber\\
  &  & + e^{icq t}\frac{\xi_q} {\epsilon_{0}}
  \int d^{3}r^{\prime} \, e^{-i\spvec{q}
     \cdot\spvec{r}^{\prime}}
   \bigg[  \left(\pol_{\unitvec{q},\lambda} \cdot
     \spvec{P}(\spvec{r}^{\prime},t)  \right)  \nonumber \\
   & & \qquad\qquad\qquad + \frac{1}{c}  \left(\unitvec{q} \times
       \pol_{\unitvec{q},\lambda}\right) \cdot
     \spvec{M}(\spvec{r}^{\prime},t) \bigg] \text{,}
\label{eq:normVarEqsOfm}
\end{eqnarray}
where the first term results in the oscillation of the free EM field
and in the second term we find the polarization and magnetization
densities arising from meta-atom currents that act as sources for
radiation.
Upon integrating Eq.~(\ref{eq:normVarEqsOfm}), one obtains
\begin{eqnarray}
  a_{\qv,\lambda}(t) &=&
  e^{-icqt}a_{\qv,\lambda}^{(\mathrm{in})}
  \nonumber \\
  &+&  \frac{\xi_q}{\epsilon_{0}}
  \int d^3 r'
    \int_{-\infty}^t  dt'\, e^{-icq(t-t')}
    e^{-i \spvec{q} \cdot  \spvec{r}'} \nonumber\\
    & & \times \bigg[\pol_{\unitvec{q},\lambda} \cdot
     \spvec{P}(\spvec{r}^{\prime},t') \nonumber\\
     & & \qquad + \frac{1}{c}  \left(\unitvec{q} \times
       \unitvec{e}_{\unitvec{q},\lambda}\right)
     \cdot \spvec{M}(\spvec{r}^{\prime},t')\bigg] \text{,}
  \label{eq:NormalVarSol}
\end{eqnarray}
where $a_{\qv,\lambda}^{(\mathrm{in})} = \lim_{t_0\to-\infty}
e^{icq t_0} a_{\qv,\lambda}(t_0)$ is the initial state of the plane
wave normal variable before it interacts with the resonators.

The incident fields ($\spvec{D}_{\rm in}, \spvec{B}_{\rm in}$) and the
scattered fields ($\spvec{D}_{\rm S}, \spvec{B}_{\mathrm{S}}$)
radiated by the meta-atoms comprise the total electric displacement
and magnetic induction fields
\begin{align}
  \spvec{D}(\rv,t) &=\spvec{D}_{\rm in}(\rv,t)+\spvec{D}_{\rm S}(\rv,t)
  \label{eq:TotalDFromInScatt} \,\textrm{,}\\
  \spvec{B}(\rv,t) &=\spvec{B}_{\rm
    in}(\rv,t)+\spvec{B}_{\mathrm{S}}(\rv,t) \, \textrm{,}\\
  \spvec{D}_{\mathrm{S}}(\rv,t) &= \sum_j \spvec{D}_{{\rm S},j}(\rv,t)  \, \textrm{,}\\
  \spvec{B}_{\rm S}(\rv,t) &= \sum_j \spvec{B}_{{\rm S},j}(\rv,t)\, \textrm{,}
  \label{eq:TotalBFromScatt}
\end{align}
where $\spvec{D}_{{\rm S},j}(\rv,t)$ and
$\spvec{B}_{\mathrm{S},j}(\rv,t)$ denote the fields emitted by the
meta-atom $j$.
The incident fields $\spvec{D}_{\mathrm{in}}$ and
$\spvec{B}_{\mathrm{in}}$ are given in the plane-wave representation
by Eqs.~\eqref{eq:elecDisp} and \eqref{eq:magField} with
$e^{-icqt}a_{\qv,\lambda}^{(\mathrm{in})}$ substituted for
$a_{\qv,\lambda}$.
For the field observation point $\rv$ located outside the meta-atoms
we have $\spvec{D}_{\mathrm{in}}(\rv,t) = \epsilon_0
\spvec{E}_{\mathrm{in}}(\rv,t)$ from \eq{eq:DDef}.
A common situation in experiments corresponds to an
  illumination of a metamaterial sample by a non-focused,
  monochromatic beam that can be approximated by a single plane-wave
  component, with $|\kv_{\rm in}| =\Omega_{0}/c$,
\begin{align}
  \spvec{D}_{\mathrm{in}}(\rv,t) &= D_{\mathrm{in}}
  \,\pol_{\mathrm{in}} e^{i (\kv_{\mathrm{in}}\cdot \rv-  \Omega_{0}
    t)}  + \mathrm{C.c.},\\
  \spvec{B}_{\mathrm{in}}(\rv,t) &= \sqrt{{\mu_0\over \epsilon_0}} \,
  \unitvec{k}_{\mathrm{in}}\times \spvec{D}_{\mathrm{in}}(\rv,t)\,.
\end{align}

One obtains explicit expressions for $\spvec{D}$ and $\spvec{B}$, by
substituting Eq.~\eqref{eq:NormalVarSol} into Eqs.~\eqref{eq:elecDisp}
and \eqref{eq:magField}, summing over the two transverse polarizations
$\unitvec{e}_{\unitvec{q},\lambda}$, and integrating over $\qv$.
Following this procedure, one obtains the scattered fields
\begin{eqnarray}
  \spvec{D}_{\mathrm{S}}(\spvec{r},t) &=& \int_{-\infty}^{t}\int d^3r' \,
  \big[\sptensor{S}(\spvec{r}-\spvec{r}',t-t') \cdot
  \spvec{P}(\spvec{r}',t') \nonumber\\
  & & \qquad+ \frac{1}{c}
  \sptensor{S}_{\times}(\spvec{r}-\spvec{r}',t-t') \cdot
  \spvec{M}(\spvec{r}',t')\big]
  \label{eq:DFullSol}
\end{eqnarray}
and
\begin{eqnarray}
  \spvec{B}_{\mathrm{S}}(\spvec{r},t) &=& \mu_0 \int_{-\infty}^{t}\int d^3r' \,
  \big[\sptensor{S}(\spvec{r}-\spvec{r}',t-t') \cdot
  \spvec{M}(\spvec{r}',t') \nonumber\\
  & & \qquad - c
  \sptensor{S}_{\times}(\spvec{r}-\spvec{r}',t-t') \cdot
  \spvec{P}(\spvec{r}',t')\big] \, \text{,}
  \label{eq:BFullSol}
\end{eqnarray}
where $\sptensor{S}$ is the propagator that takes the electric
(magnetic) field from the electric (magnetic) dipole source $\rv'$ to
the observation point at $\rv$, and $\sptensor{S}_\times$ represents
the propagation of the radiated electric (magnetic) field from the
magnetic (electric) dipole sources to the observation points.
The propagator $\sptensor{S}$ is given by
\begin{align}
  \sptensor{S}(\spvec{r},t) & = \frac{ic}{16\pi^3} \int d^3 q\,
  k \left(\sptensor{1}  - \unitvec{q}\unitvec{q} \right)
    e^{i \spvec{q} \cdot \spvec{r}} \left(
      e^{-icqt} - e^{icqt} \right)
  \nonumber\\
  & = \frac{c}{4\pi} (\nabla\nabla -
  \mathbf{1} \nabla^2) \frac{\delta(r-ct) - \delta(r+ct)}{r}\,.
  \label{eq:propIntegrated}
\end{align}
The two delta-functions produce retarded and advanced time
contributions to the scattered fields, with the advanced time's
contribution arising only at $r=0$.
The derivatives acting on $1/r$ result in a contact interaction
proportional to $\delta(\spvec{r})$.
At first glance, the retarded and advanced time contributions may
appear to cancel out at $r=0$, thus nullifying such a contact
interaction.
However, by examining the frequency components of
$\sptensor{S}$, one can show that this contact interaction does survive.\cite{RuostekoskiJavanainenPRA1997L}
The corresponding expressions for $\sptensor{S}_{\times}$ are
\begin{align}
  \sptensor{S}_{\times}(\spvec{r},t)  &=
  -\frac{ic}{16\pi^3} \int d^3 q\,
  e^{i \spvec{q} \cdot \spvec{r}}
  \left( e^{-ic{q}t} + e^{ic{q}t} \right) \spvec{q} \times \sptensor{1}\nonumber\\
  & = -
  \nabla \times \frac{1}{4\pi r}\frac{\partial}{\partial t} \left[\delta(r-ct) -
      \delta(r+ct)\right]
  \label{eq:propCrossIntegrated}
\end{align}

From the oscillator equations of motion [Eqs.~(\ref{eq:HamEqs1})],
we find that the fields emitted from one resonator will drive all of
the others.
The driven resonators, in turn, re-scatter these fields to yet other
resonators in the metamaterial.
To more easily account for the cumulative effects of these multiple
scatterings and identify collective modes in the system, we
analyze the field and oscillator dynamics in the frequency domain.

We therefore decompose the source fields $\spvec{P}$ and $\spvec{M}$
into their frequency components and compute the scattered
monochromatic constituents of the EM fields.
Specifically, we write an arbitrary source field,
\beq
\spvec{f}(\rv,t) =
{1\over\sqrt{2\pi}} \int_{-\infty}^\infty d\Omega \, \spvec{f}(\spvec{r},\Omega)
e^{-i\Omega t}\,,
\eeq
in terms of the Fourier components
$\spvec{f}(\spvec{r},\Omega)$.
In evaluating the response of the field to each monochromatic source
component, one encounters integrals of the form
\begin{align}
  \int_{-\infty}^t dt'\,& \sptensor{S}(\spvec{r}-\spvec{r}',t-t')
  \spvec{f}(\rv',\Omega)   e^{-i\Omega t'} \nonumber\\
  & = e^{-i\Omega t}  \sptensor{\tilde
    S}(\spvec{r}-\spvec{r}',\Omega)\spvec{f}(\rv',\Omega)\label{simplify}
\end{align}
and
\begin{align}
  \int_{-\infty}^t dt'\,& \sptensor{S}_\times(\spvec{r}-\spvec{r}',t-t')
  \spvec{f}(\rv',\Omega)   e^{-i\Omega t'} \nonumber\\
  & = e^{-i\Omega t}  \sptensor{\tilde
    S}_\times(\spvec{r}-\spvec{r}',\Omega)\spvec{f}(\rv',\Omega)
  \label{simplifyCross} \,\text{,}
\end{align}
where $\sptensor{\tilde{S}}(\rv,\Omega)$ and
$\sptensor{\tilde{S}}_\times(\rv,\Omega)$ are the monochromatic
versions of the expressions given in Eqs.~\eqref{eq:propIntegrated}
and~\eqref{eq:propCrossIntegrated} that describe the propagation of
the radiated fields from the source to an observation point.
In evaluating these propagators, we find it convenient to treat
positive and negative frequencies $\Omega$ separately.
We therefore decompose the propagators as
\begin{align}
  \sptensor{\tilde{S}}(\spvec{r},\Omega) & =
  \sptensor{\tilde{S}}^{(+)}(\rv,\Omega) \Theta(\Omega) +
  \sptensor{\tilde{S}}^{(-)}(\rv,\Omega) \Theta(-\Omega) \\
  \sptensor{\tilde{S}}_\times(\spvec{r},\Omega) & =
  \sptensor{\tilde{S}}_\times^{(+)}(\rv,\Omega) \Theta(\Omega) +
  \sptensor{\tilde{S}}_\times^{(-)}(\rv,\Omega) \Theta(-\Omega)
\end{align}
where $\Theta$ is the Heaviside function, and the  propagators'
positive and negative frequency components are given by
\begin{equation}
  \sptensor{\tilde{S}}^{(\pm)}(\rv,\Omega)  = \frac{1}{4\pi}
  \left(\nabla\nabla - \sptensor{1}\nabla^2\right)
  \frac{e^{\pm i k r}}{r} \, \text{,}
  \label{eq:monoProp} \\
\end{equation}
and,
\begin{equation}
  \sptensor{\tilde{S}}_{\times}^{(\pm)}(\rv,\Omega)
  = \pm \frac{ik}{4\pi} \left(\nabla\times\frac{e^{\pm i k r}}{r}\right)
   \sptensor{1} \,\text{,}
  \label{eq:monoPropCross}
\end{equation}
where $k \equiv |\Omega|/c$ is the angular wavenumber of the radiation
emitted from a monochromatic source of frequency $\Omega$.

One of our goals is to provide radiated electric and magnetic fields
$\Ev$ and $\Hv$, respectively.
These are related to the electric displacement $\Dv$ and magnetic
induction $\Bv$ by the familiar expressions
\begin{subequations}
  \begin{align}
    \Ev(\rv,t) & = {1\over \eo} [\Dv(\rv,t) -\Pv(\rv,t)], \label{ev}\\
    \Hv (\rv,t) &= {1\over \mu_0}\Bv(\rv,t)
    -\Mv(\rv,t) \label{hv}\,.
  \end{align}
\end{subequations}
We thus define dimensionless radiation kernels
\begin{align}
  \label{eq:dimlessRadDef}
  \sptensor{G}(\spvec{r}, \Omega) &= \frac{4\pi}{k^3}\left[
  \sptensor{\tilde S}(\spvec{r},\Omega) -\delta(\rv)\right],
  \\
  \sptensor{G}_\times(\spvec{r}, \Omega) &= \frac{4\pi}{k^3}
  \sptensor{\tilde S}_\times(\spvec{r},\Omega)\,,
  \label{eq:dimlessRadDefCross}
\end{align}
where, by the relations of Eqs.~\eqref{ev} and \eqref{hv}, the delta
function in Eq.~\eqref{eq:dimlessRadDef} transforms the monochromatic
propagators of $\Dv$ and $\Bv$ to those of $\Ev$ and $\Hv$, respectively.
The Fourier components of the corresponding EM fields are thus given
by
\begin{align}
  \spvec{E}(\rv,\Omega) & = {1\over\eo}\spvec{D}_{\mathrm{in}}(\rv,\Omega)
  + \sum_{j} \spvec{E}_{\mathrm{S},j}(\rv,\Omega) \label{eq:ETot}\\
  \spvec{H}(\rv,\Omega) & = {1\over\mu_0}
  \spvec{B}_{\mathrm{in}}(\rv,\Omega)
  +\sum_j \spvec{H}_{\mathrm{S},j}(\rv,\Omega)   \label{eq:HTot}
\end{align}
where the fields scattered from meta-atom $j$ are
\begin{subequations}
  \label{eq:scatteredFields}
  \begin{align}
    \spvec{E}_{\mathrm{S},j}(\spvec{r},\Omega) &=
    \frac{k^3}{4\pi\epsilon_0} \int d^3 r' \,
    \Big[\sptensor{G}(\spvec{r} - \spvec{r}',\Omega) \cdot
    \spvec{P}_{j}(\spvec{r}',\Omega)  \nonumber\\
    & \qquad + \frac{1}{c} \sptensor{G}_\times
    (\spvec{r}-\spvec{r}',\Omega) \cdot
    \spvec{M}_{j}(\spvec{r}',\Omega) \Big] , \label{eq:E_Sj}\\
    \spvec{H}_{\mathrm{S},j}(\spvec{r},\Omega) & =
    \frac{k^3}{4\pi} \int d^3 r' \, \Big[\sptensor{G}(\spvec{r} -
    \spvec{r}',\Omega) \cdot
    \spvec{M}_{j}(\spvec{r}',\Omega) \nonumber\\
    & \qquad - c \sptensor{G}_\times(\spvec{r}-\spvec{r}',\Omega) \cdot
    \spvec{P}_{j}(\spvec{r}',\Omega) \Big]\,.
    \label{eq:H_Sj}
  \end{align}
\end{subequations}
As with the monochromatic propagators, we decompose the radiation
kernels into their positive and negative frequency components as
\begin{align}
  \label{eq:RadKerPmFreq}
  \sptensor{G}(\rv,\Omega) &= \sptensor{G}^{(+)}(\rv,\Omega)
  \Theta(\Omega)
  + \sptensor{G}^{(-)}(\rv,\Omega) \Theta(-\Omega) \,.\\
  \label{eq:CrossKerPmFreq}
  \sptensor{G}(\rv,\Omega) &= \sptensor{G}_\times^{(+)}(\rv,\Omega)
  \Theta(\Omega)
  + \sptensor{G}_\times^{(-)}(\rv,\Omega) \Theta(-\Omega) \,.
\end{align}
The radiation kernel
$\sptensor{G}^{(\pm)}(\spvec{r}-\spvec{r}',\Omega)$ corresponds to a
familiar expression for a radiated electric (magnetic) field at the
observation point $\rv$, originating from an oscillating electric
(magnetic) dipole residing at $\rv'$.\cite{Jackson}
Similarly, an oscillating electric (magnetic) dipole at $\rv'$
generates a magnetic (electric) field at $\rv$ that is represented by
the radiation kernel
$\sptensor{G}_{\times}^{(\pm)}(\spvec{r}-\spvec{r}',\Omega)$.
The explicit expressions for these read
\begin{align}
  \label{eq:Green'sfunc}
  \sptensor{G}^{(\pm)}(\spvec{r},\Omega) &= i\left[\frac{2}{3}
    \sptensor{1}
    h_0^{(\pm)}(kr) + \left(\frac{\spvec{r}\spvec{r}}{r^2} -
      \frac{1}{3}
      \sptensor{1} \right) h_2^{(\pm)}(kr) \right] \nonumber\\
  & - \frac{4\pi}{3}
  \sptensor{1} \delta(k\spvec{r})\, \text{,}\\
 \label{eq:CrossGreen}
 \sptensor{G}_{\times}^{(\pm)} (\spvec{r},\Omega) &=
 \pm \frac{i}{k} \nabla\times \frac{e^{\pm i k r}}{kr}\,\sptensor{1}\,\textrm{,}
\end{align}
where $h_n^{(+)}$  and $h_n^{(-)}$ are the spherical Hankel functions
with order $n$ of the first and second kinds, respectively, defined by
\begin{subequations}
  \begin{align}
    h_0^{(\pm)}(x) &= \mp i \frac{e^{\pm ix}}{x}\text{,} \label{eq:h0}\\
   h_2^{(\pm)}(x) & = \pm i \left(\frac{1}{x} \pm i \frac{3}{x^2} -
      \frac{3}{x^3} \right) e^{\pm i x} \text{.} \label{eq:h2}
  \end{align}
  \label{eq:hankelFuncs}
\end{subequations}

Equations \eqref{eq:ETot}-\eqref{eq:CrossKerPmFreq}, together with the
radiation kernels of Eqs.~\eqref{eq:Green'sfunc}
and~\eqref{eq:CrossGreen}, constitute the main results of this
subsection.
They provide the total electric and magnetic fields both inside and
outside the metamaterial sample as a function of polarization and
magnetization densities that are produced by oscillating currents in
the meta-atoms.
Although we have derived the integral expressions for the scattered EM
fields in terms of the resonator excitations in
\eq{eq:scatteredFields}, in general there is no simple way of solving
for $\spvec{P}(\spvec{r})$ and $\spvec{M}(\spvec{r})$.
Together with Hamilton's equations for the dynamic variables of the
electric charges of the meta-atoms [Eqs.~\eqref{eq:HamEqs1}], the
formulas for the radiated fields form a coupled set of equations for
the EM fields and the matter.
The scattered fields from each meta-atom drive the dynamics of the
other meta-atoms in the system, with the EM fields mediating
interactions between the resonators.
For the case of near-resonant field excitation and closely-spaced
circuit elements the coupling between the EM fields and the meta-atoms
can be strong due to multiple scattering processes leading to
collective behavior of the system.

In evaluating the scattered fields of \eq{eq:scatteredFields}, we note
that because $h_2^{(\pm)}(r)$ contains a $1/r^3$ divergence near
$r=0$, the spatial integral of
$\sptensor{G}^{(\pm)}(\spvec{r},\Omega)$ in \eq{eq:Green'sfunc} is not
absolutely convergent around the origin.
However, as Ref.~\onlinecite{RuostekoskiJavanainenPRA1997L} points out one
can handle such a singularity by carving an infinitesimal spherical
region around $\spvec{r}=0$ from the integral and treating this region
separately.
The integral over the radiation kernel \eqref{eq:Green'sfunc} is then
\emph{defined} using the convention that the term inside the brackets
vanishes over an infinitesimal volume enclosing the origin.
Mathematically, this is achieved by carrying out the integral in this
region in spherical coordinates, first integrating over the spherical
angles, so that only the $\delta$-function contributes to the
integral.\footnote{Physically, this indicates an isotropic
  high-momentum cut-off in the formulation of the non-relativistic
  electromagnetism.}
With this integration procedure, the $\delta$-function appearing in
Eq.~\eqref{eq:Green'sfunc} is required for the scattered fields to
satisfy Gauss' law, as well as to produce the correct Maxwell's
equations, $\nabla \cdot \spvec{D}=0$ for a neutral system and $\nabla
\cdot \spvec{B}=0$.
The requirement that these conditions are satisfied also confirms that
we have duly selected the correct field terms in the Hamiltonian
\eqref{eq:Ham} (e.g., electric displacement, instead of electric
field) and that the integration procedure of the contact terms
[\eq{eq:propIntegrated}] has been performed correctly.
While the $\delta$-function singularity in $\sptensor{G}$ does not
play a role in the interactions between non-overlapping meta-atoms, we
find in Sec.~\ref{sec:single-meta-atom} that it does contribute to
interactions of a meta-atom with its self-generated field.

The EM fields derived from the Hamiltonian are indeed consistent with
Maxwell's equations.
To verify this, we check that the positive and negative frequency
components of a monochromatic field with wavenumber $k$ satisfy the
wave equations with sources $\Pv^{(\pm)}$ and $\Mv^{(\pm)}$
\cite{Jackson}:
\begin{eqnarray}
  (\nabla^2+k^2) \Dv^{(\pm)} &=&
  -\nabla\times(\nabla\times\Pv^{(\pm)}) \nonumber \\
   & & \mp i\frac{k}{c}\nabla\times \spvec{M}^{(\pm)}\label{eq:Dwave} \\
  (\nabla^2+k^2) \Bv^{(\pm)} &=&
  -\mu_0\nabla\times(\nabla\times\Mv^{(\pm)}) \nonumber \\
  & &\pm i\mu_0 ck \nabla\times
  \Pv^{(\pm)} \label{eq:Bwave}
\end{eqnarray}
We confirm that the total fields produced by our system satisfy
Eqs.~\eqref{eq:Dwave} and \eqref{eq:Bwave} by applying the operator
$(\nabla^2+k^2)$ to the total electric and magnetic fields
[Eqs.~\eqref{eq:ETot} and \eqref{eq:HTot}].
Because the incident waves are composed of superpositions of plane
waves, the action of the operator $(\nabla^2 +k^2)$ on these fields
trivially reduces to
\begin{equation}
  (\nabla^2+k^2)\Dv_{\mathrm{in}} = (\nabla^2+k^2) \Bv_{\mathrm{in}}=0 \text{.}
  \label{eq:waveEq}
\end{equation}
Therefore, the only contributions to $(\nabla^2 +k^2)\Ev$ and
$(\nabla^2 + k^2)\Hv$ come from the scattered fields
$\spvec{E}_{\mathrm{S},j}$ and $\spvec{H}_{\mathrm{S},j}$
[Eqs.~\eqref{eq:E_Sj} and \eqref{eq:H_Sj}].
These contributions are most readily determined by expressing the
tensor components of the radiation kernels in the differential form
\begin{align}
  &\sptensor{G}_\times^{(\pm)\mu,\eta}= \pm {i\over
    k}\epsilon_{\mu\nu\eta} {\partial\over\partial r_\nu}\(
  {e^{\pm{}ikr}\over k r} \) \\
  \sptensor{G}^{(\pm)\mu,\eta}& = {1\over k^2}\({\partial\over\partial
    r_\mu}{\partial\over\partial r_\eta}-\delta_{\mu\eta}\nabla^2\)\(
  {e^{\pm{}ikr}\over kr}\) \nonumber \\
  & -4\pi \delta_{\mu\eta}\delta(k\rv)
\end{align}
Because the differential operators involved in the radiation kernels
readily commute with $(\nabla^2+k^2)$, the expressions for this
operator acting on the scattered fields involve contributions of the
form
\begin{equation}
  (\nabla^2+k^2)\( {e^{\pm ik \left|\rv - \rv'\right|}\over \left|\rv -
      \rv'\right| }\)=-4\pi \delta(\spvec{r} - \spvec{r}')\,.
\end{equation}
appearing under the integral.
Physically, the $\delta$-function represents a point source away from which a monochromatic spherical wave ($e^{\pm ikr}/r$) propagates.
The resulting expressions for
$(\nabla^2+k^2)\Ev_{\mathrm{S},j}^{(\pm)}$ and
$(\nabla^2+k^2)\spvec{H}_{\mathrm{S},j}^{(\pm)}$ thus contain
integrals over $\delta$-functions which are readily evaluated.
Explicitly, for the component $\mu$ of the scattered electric field,
we have
\begin{eqnarray}
  \label{eq:WaveEqComp}
  \left( \nabla^2 +k^2\right) D_{\mathrm{S},j}^{(\pm)\mu} &=&
    - \left(\frac{\partial}{\partial r_{\mu}}
      \frac{\partial}{\partial r_{\eta}} - \delta_{\mu,\eta}
      \nabla^2\right) P_j^{(\pm)\eta}  \nonumber \\
  & &\mp i\frac{k}{c}
  \epsilon_{\mu\nu\eta} \frac{\partial}{\partial r_\nu} M_j^{(\pm)
    \eta} \,\text{,}
  \label{eq:EcompWaveEq}
\end{eqnarray}
where
\begin{equation}
  D_{\mathrm{S},j}^{(\pm)\mu} \equiv \epsilon_0 E_{\mathrm{S},j}^{(\pm)\mu} + P_j^{(\pm)\mu}
\end{equation}
is the $\mu^{\textrm{th}}$ component of the scattered displacement field from meta-atom $j$.
Adding the contributions of Eq.~\eqref{eq:EcompWaveEq} for all
meta-atoms $j$, produces the equivalent of the wave equation [Eq.~\eqref{eq:Dwave}], which
is the desired result.
Similarly, one finds that by adding the contributions $\sum_j
(\nabla^2 +k^2)H^{(\pm)\mu}_{\mathrm{S},j}$ for all meta-atoms, one
recovers the wave equation for the magnetic field [Eq.~\eqref{eq:Bwave}].

\section{Meta-atom interactions mediated by the EM field}
\label{sec:analysis-model}

In the previous section we established how current oscillations in the
meta-atoms respond to the EM field [Eqs.~\eqref{eq:HamEqs1}].
Additionally, we arrived at expressions for the
electric and magnetic
fields scattered by the meta-atoms
[Eqs.~\eqref{eq:ETot}-\eqref{eq:CrossKerPmFreq}].
These fields were solved in terms of the magnetization and the
polarization densities, generated by the resonator excitations.
The current oscillations in each meta-atom thus depend on
the excitation of all other meta-atoms via the scattered radiation.
In this section, we combine
the response of the meta-atoms to EM field and the expressions of
  the EM fields scattered by the meta-atoms in order
to
investigate
how the
radiation
mediates
interactions between the meta-atoms.

We begin by examining
the dynamics of a
single driven meta-atom
in Sec.~\ref{sec:single-meta-atom}.
There we
show that when radiative losses are much
weaker
than the resonance frequency, a single meta-atom's dynamics reduce to
those of the familiar damped LC circuit in which the energy is lost to
the scattered EM field.
We then examine
interactions between different meta-atoms in a collection of
  closely-spaced resonators. Due to the strong coupling between the EM
  fields and the current oscillations,
the emitted radiation leads to
the collective dynamics of the ensemble.
In Sec.~\ref{sec:coll-inter-rotat}, we explore the
collective response of the system
in the rotating wave approximation, in which each meta-atom's
radiative emission rate is much less than its resonance frequency.
We present an analysis for a more strongly interacting system outside
the rotating wave approximation in
Appendix~\ref{sec:ensemble-meta-atoms}.

In these treatments, we assume the spatial extent of each meta-atom is
much smaller than the wavelength of EM field with which it interacts.
As such, the radiation scattered from each meta-atom can often be
approximated as  that of electric and magnetic dipoles oscillating in
sync with one and other.
For simplicity,
when evaluating the interactions between meta-atoms, we
assume that the
electric
quadrupole and higher order multipole contributions to the radiation of a single meta-atom are much weaker than the dipole radiation and that they can be neglected.
This is by no means a necessary approximation.  We could extend
  the general formalism to incorporate multipole-field radiation
  components in a multipole expansion. The dipole approximation,
  however, will provide an advantage in maintaining
the tractability of
the
derivation of the collective metamaterial response to EM fields.
Moreover, in several practical situations, a unit-cell resonator
  of a metamaterial array may consist of two or more meta-atoms.
Hence, in the dipole approximation to a single meta-atom, the
unit-cell resonator would still exhibit multipole radiation
contributions.
The multipole fields radiated by unit-cell resonators are also weak in many cases.
For instance, metamaterial samples consisting of
asymmetric split ring metamolecules
have been experimentally employed in the studies of collective
  resonator response.\cite{FedotovEtAlPRL2007,papasimakis2009,SavoEtAlPRB2012,FedotovEtAlPRL2010}
  In an
asymmetric split ring metamolecule
the generated quadrupole field is notably suppressed when compared
  to the corresponding dipolar  field.\cite{PapasimakisComm}

The electric and magnetic dipole moments produced by the current
oscillation in meta-atom $j$ are
\begin{equation}
  \label{eq:dipoles}
  \begin{array}{ccc}
    \spvec{d}_{j}=Q_{j}h_{j}\unitvec{d}_{j} & \textrm{and} &
    \spvec{m}_{j}=I_{j}A_{j}\unitvec{m}_{j} \,,
  \end{array}
\end{equation}
respectively.
These are given in terms of the charge $Q_j$ and the current $I_j$
  of the meta-atom. The geometry-dependent proportionality
  coefficients
$h_{j}$ and $A_{j}$ have units of length and area and are defined such
that
\begin{equation}
  \label{eq:hAndAdefs}
  \begin{array}{ccc}
    h_{j}\unitvec{d}_{j}=\int d^{3}r\,\ \spvec{p}_{j}(\spvec{r})  &
    \textrm{and} &  A_{j}\unitvec{m}_{j}=\int
    d^{3}r\,\spvec{w}_{j}(\spvec{r}) \textrm{.}
  \end{array}
\end{equation}
The unit vectors $\unitvec{d}_{j}$ indicate the orientation of the
electric dipole while the unit vectors $\unitvec{m}_{j}$ indicate the
orientations of the magnetic dipoles.
The distributions $\spvec{p}_{j}(\spvec{r})$ and
  $\spvec{w}_{j}(\spvec{r})$ [see Eq.~\eqref{eq:PandMDensDefs}] represent
  the spatial profile of the polarization and magnetization densities
  in terms of $Q_j$ and $I_j$.
While, generally, the current resulting from the polarization density
[the first term in Eq.~\eqref{eq:currentDensity}] contributes to the
magnetic dipole, the polarization and magnetization densities (and
hence the mode functions) that lead to a given charge and current
distribution are not unique. \cite{Jackson}
We have therefore chosen
for each meta-atom $j$,
$\spvec{p}_j$, $\spvec{w_j}$ and
the position vector
$\spvec{R}_j$ such that the contribution of the polarization current
to the magnetic dipole moment about $\spvec{R}_j$ is zero.

To facilitate an understanding of how the EM field influences the
meta-atom dynamics, we consider a meta-atom's
self-generated fields separately from the fields generated externally.
Consider the dynamics of a single meta-atom $j$
interacting with the EM field.
The meta-atom's equations of motion are given by
Eq.~\eqref{eq:HamEqs1}.
To isolate the dynamics arising from the self-generated field, we
decompose the electric and magnetic fields into those generated by
meta-atom $j$ -- $\spvec{E}_{\mathrm{S},j}$ and
$\spvec{B}_{\mathrm{S},j}$ -- and those generated externally to
meta-atom $j$, $\spvec{E}_{j,\mathrm{ext}} $ and $\spvec{B}_{j,\mathrm{ext}}$.
We then obtain the following relationship between the different contributions
\begin{subequations}
  \begin{eqnarray}
    \spvec{E}_{j,\mathrm{ext}} \equiv \spvec{E}_{\mathrm{in}} +
    \sum_{j'\ne j} \spvec{E}_{\mathrm{S},j'} \\
    \spvec{B}_{j,\mathrm{ext}} \equiv \spvec{B}_{\mathrm{in}} +
    \sum_{j'\ne j} \spvec{B}_{\mathrm{S},j'} \text{.}
    \label{eq:9}
  \end{eqnarray}
\end{subequations}
These external fields include contributions from the incident field and the fields
scattered by
all the
other meta-atoms in the system.

In the previous section we derived the expressions for the
  scattered fields in terms of the polarization and the magnetization
  densities of the source medium. It was advantageous to represent the
  scattered fields in the frequency domain. We
similarly analyze here the Fourier components of the dynamic
variable $Q_j$ oscillating
at frequency $\Omega$.
As we did with the emitted fields, we find it convenient to decompose the
meta-atom variables
  \begin{align}
    Q_j(t) &= Q_j^{(+)}(t) +Q_j^{(-)}(t),\\
    \phi_j(t) &= \phi_j^{(+)}(t) + \phi_j^{(-)}(t)\,,
  \end{align}
into their positive and negative frequency components, with
\begin{equation}
  Q_j^{(-)}(t) = \left[Q_j^{(+)}(t)\right]^\ast,
  \quad \phi_j^{(-)}(t) = \left[\phi_j^{(+)}(t)\right]^\ast\,.
  \label{eq:PMFreqCompTimeDomain}
\end{equation}
The positive and negative frequency components for these variables are
defined such that, for a given frequency $\Omega$
\begin{subequations}
  \label{eq:QandPhiFreqComp}
  \begin{eqnarray}
    \label{eq:14}
    Q_j^{(\pm)}(\Omega) &\equiv& Q_j(\Omega) \Theta(\pm\Omega) \\
   \phi_j^{(\pm)}(\Omega) &\equiv& \phi_j(\Omega) \Theta(\pm\Omega) \text{.}
  \end{eqnarray}
\end{subequations}
With frequency components of $Q_j$ and
$\phi_j$ defined in this way, the positive (negative) frequency
components of the dynamic variables
are driven exclusively
by
the
positive (negative) frequency components of the EM fields.
Since the metamaterial system we consider in this model is linear, the
equations of motion in Fourier space become  the algebraic relationships
between Fourier components of a common frequency $\Omega$,
\begin{subequations}
  \label{eq:singleMAEqsFourierSpace}
  \begin{align}
    \label{eq:singleMAEqQFourierSpace}
    -i\Omega Q_j(\Omega)   & =  \frac{\phi_j(\Omega) -
      \Phi_{j,\mathrm{self}}(\Omega) - \Phi_{j,\mathrm{ext}}(\Omega)} {l_j} \\
    -i\Omega \phi_j(\Omega) & = \mathcal{E}_{j,\mathrm{self}}(\Omega)
    + \mathcal{E}_{j,\mathrm{ext}}(\Omega) \text{,}
  \end{align}
\end{subequations}
where $\mathcal{E}_{j,\mathrm{self}}$ and
$\spvec{\Phi}_{j,\mathrm{self}}$ are the self-generated EMF and flux,
respectively, while $\mathcal{E}_{j,\mathrm{ext}}$ and
$\spvec{\Phi}_{j,\mathrm{ext}}$ are the EMF and flux generated
externally to meta-atom $j$.
The current relates to conjugate momentum and
magnetic flux through Eq.~\eqref{eq:conjMomofQ_j}, and
the equation of motion for $Q_j$, Eq.~\eqref{eq:singleMAEqQFourierSpace}, is
nothing more that the statement that the rate of change of $Q_j$ is
the current $I_j$.
This translates to the relationship between
frequency components $-i \Omega Q_j(\Omega) = I_j(\Omega)$.
The EMF  and magnetic flux contain the
external driving induced by the external EM fields as
well as driving induced by the field that the current oscillation
itself generates.

The externally applied EMF and magnetic flux are given explicitly in terms of the
externally generated fields as
\begin{align}
  \label{eq:inputEMF}
  \mathcal{E}_{j,\mathrm{ext}}(\Omega) & \equiv \int
  d^{3}r\,\spvec{p}_j(\spvec{r})  \cdot
  \spvec{E}_{j,\mathrm{ext}}(\spvec{r},\Omega) \text{,} \\
  \Phi_{j,\mathrm{ext}}(\Omega) &\equiv \int d^{3}r\, \spvec{w}_j(\spvec{r}) \cdot
  \spvec{B}_{j,\mathrm{ext}}(\spvec{r}, \Omega) \, .
  \label{eq:extFlux661}
\end{align}
When the external fields vary slowly over the volume of meta-atom
$j$, $\mathcal{E}_{j,\mathrm{ext}}$  and $\Phi_{j,\mathrm{ext}}$
reduce to a direct driving of the meta-atoms' electric and magnetic
dipoles, respectively
\begin{subequations}
  \label{eq:ExtDipole}
  \begin{align}
    \label{eq:EjExtDipole}
    \mathcal{E}_{j,\mathrm{ext}}(\Omega) &\approx h_j\unitvec{d}_j \cdot
    \spvec{E}_{j,\mathrm{ext}}(\spvec{R}_j,\Omega) \text{,} \\
    \Phi_{j,\mathrm{ext}}(\Omega) & \approx A_j\unitvec{m}_j \cdot
    \spvec{B}_{j,\mathrm{ext}}(\spvec{R}_j,\Omega)  \text{,}
  \end{align}
\end{subequations}
where $\spvec{R}_j$ is the position of the meta-atom.
The external EMF and flux mediate the interactions between distinct
meta-atoms which we will discuss in Subsection
\ref{sec:coll-inter-rotat} and Appendix \ref{sec:ensemble-meta-atoms}.

\subsection{A single meta-atom interacting with the EM field}
\label{sec:single-meta-atom}

Before investigating how scattered EM fields facilitate
interactions between meta-atoms, we first shed light on how the
meta-atom's field influences the evolution of the meta-atom itself.
This is done by studying a single, isolated externally driven
  meta-atom.
We will present expressions for the self-generated fields'
contribution to both the EMF and the flux.
When the spatial extent of the meta-atom is much less
than a wavelength, the self-induced EMF can be written in terms of an
effective self-capacitance, and the magnetic flux can be written in
terms of a magnetic self-inductance.
We thus show how each meta-atom can be treated as a radiatively damped
LC circuit which is driven by external fields.
This analogy allows us to define slowly varying normal variables and
derive their dynamics.

\subsubsection{Self-induced EMF and magnetic flux}
\label{sec:self-induced-emf-and-flux}

The EMF and the magnetic flux
represent reactions of a meta-atom to EM fields generated by the
meta-atom itself, as well as to
external fields.
Self-generated electric and magnetic fields provide a major
  contribution to the EMF and magnetic flux, respectively.
We define the self-generated EMF and flux as
\begin{align}
  \label{eq:EMF_Self_def1}
  \mathcal{E}_{j,\mathrm{self}}(\Omega) & \equiv \int
  d^{3}r_j \,\spvec{p}_j(\spvec{r}_j)  \cdot
  \spvec{E}_{\mathrm{S},j}(\spvec{r}_j,\Omega) \text{,} \\
  \Phi_{j,\mathrm{self}}(\Omega) &\equiv \int d^{3}r\, \spvec{w}_j(\spvec{r}) \cdot
  \spvec{B}_{\mathrm{S},j}(\spvec{r}, \Omega) \, .
  \label{eq:Flux_self_def1}
\end{align}
The self-generated fields of meta-atom $j$,
i.e.,
the fields
$\spvec{E}_{\mathrm{S},j}$ and $\spvec{H}_{\mathrm{S},j} \equiv
\spvec{B}_{\mathrm{S},j}/\mu_0 - \spvec{M}_{j}$
scattered from meta-atom $j$, at a frequency $\Omega$
are given in
[Eq.~\eqref{eq:scatteredFields}].
From the expression for $\spvec{E}_{\mathrm{S},j}$
  [Eq.~\eqref{eq:E_Sj}], we obtain the self-induced
  EMF [Eq.~\eqref{eq:EMF_Self_def1}] in terms of the radiation kernels
  [Eqs.~\eqref{eq:Green'sfunc} and \eqref{eq:CrossGreen}]
\begin{eqnarray}
  \lefteqn{\mathcal{E}_{j,\mathrm{self}}(\Omega)  =  \frac{k^{3}} {4\pi\epsilon_{0}} \int d^{3}r \int
    d^{3}r^{\prime} \, } \nonumber   \\
  && \times \bigg[ \spvec{p}_{j}(\spvec{r})  \cdot
  \sptensor{G}(\spvec{r}-\spvec{r}^{\prime}, \Omega) \cdot
  \spvec{p}_j(\spvec{r}')
  Q_j(\Omega) \nonumber\\
  &  &+ \frac{1}{c} \spvec{p}_j(\spvec{r}) \cdot
  \sptensor{G}_{\times}(\spvec{r}-\spvec{r}',\Omega)  \cdot
  \spvec{w}_{j}(\spvec{r}^{\prime})  I(\Omega) \bigg] \mathrm{.}
  \label{eq:EMF_self75}
\end{eqnarray}
Similarly, the self-generated flux is obtained from the expression for
$\spvec{H}_{\mathrm{S},j}$ [Eq.~\eqref{eq:H_Sj}], and is given in terms
of the radiation kernels as

\begin{eqnarray}
  \lefteqn{\Phi_{j,\mathrm{self}}(\Omega)
    =  \frac{\mu_{0}k^{3}} {4\pi} \int d^{3}r\Bigg\{ |\spvec{w}_j(\rv)|^2}
  \nonumber\\
  & & + \int d^{3}r^{\prime}\, \Big[
    \spvec{w}_j(\spvec{r}) \cdot
    \sptensor{G}(\spvec{r}-\spvec{r}^{\prime},\Omega) \cdot
    \spvec{w}_j(\spvec{r}^{\prime})  I_j(\Omega)  \nonumber \\
  & &  - c \spvec{w}_j(\spvec{r})  \cdot
  \sptensor{G}_{\times}(\spvec{r}-\spvec{r}',\Omega) \cdot
  \spvec{p}_j(\spvec{r}')  Q_j(\Omega) \Big] \Bigg\}
  \label{eq:Phi76}
\end{eqnarray}
The first term of Eq.~\eqref{eq:Phi76} arises because the flux is
defined in terms of $\spvec{B}=\mu_0 \left(\spvec{H}+\spvec{M}\right)$
rather than $\spvec{H}$, whose scattered field components are
determined by the radiation kernels.
This results in different contact terms in Eqs.~\eqref{eq:EMF_self75}
and \eqref{eq:Phi76}.

Because we have assumed that the meta-atoms are much smaller than the
wavelength, we may expand the radiation kernels to lowest order in $k
r$ and thus approximate the self-interactions in the near field limit.
Since, in this limit, $\sptensor{G}_{\times}(\spvec{r} ,\Omega) /
\sptensor{G}(\spvec{r},\Omega)  \sim kr \ll 1$, we neglect the
contribution of $\sptensor{G}_\times$ to the self-interaction.
To leading order in $k r$, we have the positive and negative frequency
components of the radiation kernels
\begin{align}
  \label{eq:ReRadApprox}
  \operatorname{Re}\sptensor{G}^{(\pm)}(\spvec{r},\Omega)   &  \approx
  \frac{3\unitvec{r} \unitvec{r} - \sptensor{1}} {k^{3}r^{3}} - \frac{4\pi}{3k^{3}}
  \delta(\spvec{r})  ,\\
  \label{eq:ImRadApprox}
  \operatorname{Im}\sptensor{G}^{(\pm)}(\spvec{r},\Omega) &  \approx
  \pm \frac{2}{3}\text{.}
\end{align}

\subsubsection{Self-capacitance and self-inductance}
\label{sec:self-capac-self}

The long wavelength approximation allows us to simplify the
  expressions for the self-generated EMF and flux
  [Eq.~\eqref{eq:EMF_self75} and \eqref{eq:Phi76}] by neglecting the
contributions of $\sptensor{G}_\times$.
This approximation
implies that the
self-induced EMF is directly proportional to the charge $Q_j$, and
that
the
self-induced magnetic flux is directly proportional to the current
$I_j$.
We can thus draw an analogy between a typical meta-atom and a standard
LC circuit where the charge $Q_j$ and current $I_j$ are related to
$\mathcal{E}_{j,\mathrm{self}}$ and $\Phi_{j,\mathrm{self}}$ through
an effective capacitance $C_j$ and magnetic self-inductance
$L_j^{(\mathrm{M})}$.
From Eqs.~\eqref{eq:ReRadApprox} and \eqref{eq:ImRadApprox}, the
positive and negative frequency components of the EMF and flux arising
from the meta-atom's self-generated field become
\begin{align}
  \mathcal{E}^{(\pm)}_{j,\mathrm{self}}(\Omega) & = -\left(  \frac{1}{C_j}
    \mp i \frac{h_j^{2}k^{3}}{6\pi\epsilon_{0}}\right)
  Q_j^{(\pm)}
  (\Omega)  ,\label{eq:emfSingleMA}\\
  \Phi^{(\pm)}_{j,\mathrm{self}}(\Omega)  & = \left(  L_j^{(\mathrm{M})} \pm i
    \frac{\mu_{0}A_j^{2}k^{3}}{6\pi}\right)
  I_j^{(\pm)}(\Omega) \label{eq:fluxSingleMA} \text{.}
\end{align}
In addition to the capacitance and inductance
the EMF and flux have respective imaginary contributions that, as we
shall see later, represent dissipation of the current
oscillation due to radiation being emitted away from the meta-atom.
The self-capacitance $C_j$ is given by
\begin{widetext}
  \begin{eqnarray}
    \frac{1}{C_j}=\int d^{3}r\,  \frac{\left\vert
        \spvec{p}_j(\spvec{r})\right\vert^{2}}{3\epsilon_0}
    -
    \frac{1}{4\pi\epsilon_{0}} \int d^{3}r \int d^{3}r^{\prime}\,
    \frac{3\left(  \spvec{p}_j(\spvec{r}) \cdot \unitvec{n}\right)
      \left(  \unitvec{n} \cdot \spvec{p}_j(\spvec{r}^{\prime})
      \right)
      - \spvec{p}_j(\spvec{r}^{\prime}) \cdot \spvec{p}_j(\spvec{r})  }
    {\left\vert \spvec{r}-\spvec{r}^{\prime}\right\vert ^{3}}
    \text{,}
    \label{eq:C_j_xxx2}
  \end{eqnarray}
with $\unitvec{n} \equiv \left(  \spvec{r}-\spvec{r}^{\prime} \right)
/ \left\vert \spvec{r}-\spvec{r}^{\prime}\right\vert $, and the
magnetic self-inductance is
\begin{equation}
  L_j^{(\mathrm{M})} = \frac{2\mu_{0}}{3} \int d^{3}r\,\left|
    \spvec{w}_j(\spvec{r})\right|^{2} +\frac{\mu_{0}}{4\pi} \int d^{3}r
  \int d^{3}r^{\prime} \frac{3\left(  \spvec{w}_j(\spvec{r}) \cdot
      \unitvec{n} \right)
    \left(\unitvec{n} \cdot \spvec{w}_j(\spvec{r}^{\prime})\right)
    - \spvec{w}_j(\spvec{r}^{\prime})  \cdot \spvec{w}_j(\spvec{r})}
  {\left| \spvec{r}-\spvec{r}^{\prime}\right|^{3}} \text{,}
\end{equation}
\end{widetext}
In essence, excitation of the dynamic variable $Q_j$ produces a
distribution of electric dipoles (polarization density) proportional
to the mode function $\spvec{p}_j(\rv)$.
In the long-wavelength approximation, this distribution of dipoles
produces a quasi-static electric field in the vicinity of the
meta-atom generated by the real part of the radiation kernel
$\operatorname{Re}\sptensor{G}$  [Eq.~\eqref{eq:ReRadApprox}].
The current oscillation interacts with itself via the near field
electric dipole-dipole interactions, resulting in the effective
capacitance $C_j$ appearing in the self-induced EMF
[Eq.~\eqref{eq:emfSingleMA}].
Similarly, a nonzero current $I_j$ produces a distribution of magnetic
dipoles (magnetization density) proportional to the mode function
$\spvec{w}_j$.
The current oscillation then interacts with itself via the near field
magnetic dipole-dipole interactions, resulting in the magnetic
self-inductance $L_j^{(\mathrm{M})}$ appearing in the self-induced flux
[Eq.~\eqref{eq:fluxSingleMA}].

Because
the
self-induced
flux [Eq.~\eqref{eq:fluxSingleMA}] is
proportional to
$I_j$, we find it convenient to express the conjugate
momentum $\phi_j = l_jI_j + \Phi_j$
[Eq.~\eqref{eq:conjMomofQ_j}]  in terms of a total self-inductance
\begin{equation}
  \label{eq:LjDef}
  L_j \equiv l_j + L_j^{(\mathrm{M})} \, .
\end{equation}
This self-inductance includes contributions from both the magnetic and
the kinetic inductances.
When we include contributions from both the self-generated flux
[Eq.~\eqref{eq:fluxSingleMA}]
and the external flux
[Eq.~\eqref{eq:extFlux661}], the conjugate momentum for
meta-atom $j$ is given in terms of the total self-inductance by
\begin{equation}
  \phi^{(\pm)}_j(\Omega) = \left( L_j \pm
    i\frac{\mu_{0}A_j^{2}k^{3}}{6\pi} \right)
  I^{(\pm)}_j(\Omega) + \Phi^{(\pm)}_{j,\mathrm{ext}}(\Omega)  \text{.}
  \label{eq:phiSingleMA}
\end{equation}
This relation will be useful in determining the meta-atom equations of
motion.

\subsubsection{Equations of motion for a meta-atom interacting with
  its self-generated fields}
\label{sec:equat-moti-meta}

Having determined how the self-scattered fields affect the EMF and
flux, we now determine a closed set of equations of motion for the
meta-atom's dynamic variable and conjugate momentum.
The rate of change of the dynamic variable $Q_j$ is given by the
  current $I_j$.
Solving Eq.~\eqref{eq:phiSingleMA} for $I_j(\Omega)$ thus allows us to
determine an equation of motion for $Q_j$ in terms of its conjugate
momentum and magnetic flux
generated
by the external field.
Further, substituting the EMF from Eq.~\eqref{eq:emfSingleMA} into
Eq.~\eqref{eq:singleMAEqsFourierSpace} provides the corresponding
equation of motion for $\phi_j$.
Explicitly these equations of motion are given in the frequency domain
as
\begin{equation}
  -i\Omega Q_j^{(\pm)} = \left[ 1 \mp i\left(\frac{ck}
      {\omega_j} \right)^{3} \frac{\Gamma_{\mathrm{M},j}}
    {\omega_j D^{(\pm)}_j  } \right]  \frac{\phi^{(\pm)}_j}{L_j}
  - \frac{\Phi^{(\pm)}_{j,\mathrm{ext}}} {L_j D^{(\pm)}_j}
  \label{eq:saEqmQ}
\end{equation}
\begin{equation}
  -i\Omega \phi_j^{(\pm)} = -\frac{1}{C_j}\left[1 \mp i \left(
      \frac{ck} {\omega_j} \right)^{3}
    \frac{\Gamma_{\mathrm{E},j}}{\omega_j}\right]  Q^{(\pm)}_j +
  \mathcal{E}^{(\pm)}_{j,\mathrm{ext}}
  \label{eq:saEqmphi}
\end{equation}
where, as we demonstrate later,
\begin{equation}
  \omega_j \equiv \frac{1}{\sqrt{L_jC_j}}
  \label{eq:omegajDef}
\end{equation}
is the single meta-atom resonance frequency, $k \equiv |\Omega|/c$ is
the wavenumber of the field frequency component,
\begin{equation}
  \Gamma_{\mathrm{E},j} \equiv
  \frac{h_j^{2}C_j\omega_j^{4}}{6\pi\epsilon_{0}c^3}
  \label{eq:Gamma_EDef}
\end{equation}
is the emission rate due to electric dipole radiation,
\begin{equation}
  \Gamma_{\mathrm{M},j} \equiv \frac{\mu_{0}A_j^{2}\omega_j^{4}}{6\pi c^3
    L_j} \label{eq:Gamma_MDef}
\end{equation}
is the emission rate due to magnetic dipole radiation, and
\begin{equation}
  D^{(\pm)}_j(\Omega) = 1 \pm i \left(  \frac{ck}{\omega_j}\right)^{3}
  \frac{\Gamma_{j,\mathrm{M}}}{\omega_j}
  \label{eq:DenomDef}
\end{equation}
arises from the inversion of Eq.~\eqref{eq:phiSingleMA}.
The interaction of the meta-atom with its external fields are
parameterized by $h_j$ and $A_j$
[Eq.~\eqref{eq:hAndAdefs}] and hence by the radiative emission rates
$\Gamma_{\mathrm{E},j}$ and $\Gamma_{\mathrm{M},j}$.
This is made clear in the point dipole approximation where we have the
external EMF and magnetic flux which drive the meta-atom
[Eq.~\eqref{eq:ExtDipole}].
From Eqs.~\eqref{eq:Gamma_EDef} and \eqref{eq:Gamma_MDef}, one can
infer that, when the meta-atom geometry is altered such that the self-capacitance and self-inductance remain constant, an increased
interaction strength of the meta-atom with the external field
corresponds to increased radiative emission rates.

\subsubsection{A meta-atom as an LC circuit}
\label{sec:meta-atom-as}

If we neglect the radiative damping and consider a meta-atom interacting
exclusively with
its self-generated field, its dynamics are nothing more than
those of an LC circuit with resonance frequency  $\omega_j$, which
in the time domain satisfies the equations of motion
\begin{eqnarray}
  \label{eq:losslessEqM}
  \frac{d}{dt} \left(
    \begin{array}{c}
      Q_j(t)\\
      \phi_j(t)
    \end{array}
  \right) =
  \left(
    \begin{array}{cc}
      0 & L_j^{-1} \\
      - C_j^{-1} & 0
    \end{array}
  \right)
  \left(
    \begin{array}{c}
    Q_j(t) \\
    \phi_j(t)
    \end{array}
  \right) \text{.}
\end{eqnarray}
The meta-atom normal mode variables
\begin{equation}
  \beta_j \equiv \frac{1}{\sqrt{2\omega_j}}\left(\frac{Q_j}{\sqrt{C_j}} + i \frac{\phi_j}{\sqrt{L_j}}\right)
  \label{eq:10}
\end{equation}
 and
$\beta_j^\ast$ evolve with eigenfrequencies $\omega_j$ and $-\omega_j$,
respectively
\begin{equation}
  \beta_j(t) = \exp(-i\omega_j t) \beta_j(0).
\end{equation}

The collective dynamics within the metamaterial, of course, arise
from the interaction of each meta-atom with its external field,
necessitating the inclusion of radiative losses
$\Gamma_{\mathrm{E},j},\Gamma_{\mathrm{M},j}$.
But, as we will see later in this section, the presence of radiative
interactions not only results in energy being carried away from the
meta-atom by the radiated field, but also allows the meta-atom to be
driven by fields scattered from other meta-atoms.

\subsubsection{The meta-atom normal oscillator variables}
\label{sec:meta-atom-normal}

The variables $\beta_j$ represent eigenmodes of a single meta-atom in
the absence
of
 interactions with the external fields.
The presence of these interactions perturbs the single meta-atom
dynamics.
Since the incident EM field driving the metamaterial
oscillates at a central frequency $\Omega_0$, it is convenient to
analyze the effects of these perturbations using the slowly varying
normal oscillator variables
\begin{equation}
  b_j(t)  = e^{i\Omega_0 t} \beta_j(t) = \frac{e^{i\Omega_{0}t}}{\sqrt{2\omega_j}}
  \left( \frac{Q_j(t)}{\sqrt{C_j}} + \frac{\phi_j(t)}
    {\sqrt{L_j}}\right) \text{.}
  \label{eq:bSingAtDef}
\end{equation}
The oscillator variables satisfy the Poisson brackets
\begin{subequations}
  \label{eq:bjPoissonbrackets}
  \begin{eqnarray}
    \{b(t), b(t)\} &=& \{ b_j^{\ast}(t), b_{j'}^{\ast}(t)\}  = 0 \\
    \{b_j(t), b_{j'}^{\ast}(t)\} &=& -i\delta_{j,j'} \text{.}
  \end{eqnarray}
\end{subequations}
One can recover $Q_j$
and $\phi_j$
by solving the system of equations formed by
Eq.~\eqref{eq:bSingAtDef} and its complex conjugate.
This yields
\begin{align}
  \frac{Q_j(t)}{\sqrt{\omega_j C_j}} & = \frac{1} {\sqrt{2}} \left( e^{-i\Omega_0 t} b_j(t)
    + e^{i\Omega_0 t} b_j^{\ast}(t) \right), \label{eq:QasbSingle}\\
  \frac{\phi_j(t)}{\sqrt{\omega_jL_j}} & = -i \frac{1}{\sqrt{2}} \left(  e^{-i\Omega_0 t} b_j(t)
    - e^{i\Omega_0 t} b_j^{\ast}(t) \right)  \text{.} \label{eq:phiasbSingle}
\end{align}
As the incident electric field may consist of a range of frequencies
around $\Omega_0$ reflecting its variation in time, it is necessary,
in general,  to examine the
frequency components of the oscillator variables and how they are
related to those of $Q_j$ and $\phi_j$.
The Fourier components, for $\Omega > 0$, of $Q_j$ and $\phi_j$ are
given in
terms of the normal variables as
\begin{align}
  \frac{Q_j^{(+)}(\Omega)}{\sqrt{\omega_j C_j}} & = \frac{1} {\sqrt{2}} \left(  b_j(\delta)
    +b_j^{\ast}(-\delta-2
    \Omega_{0}
    ) \right), \label{eq:QasbSingle_freq}\\
  \frac{\phi_j^{(+)}(\Omega)}{\sqrt{\omega_j L_j}} & = -i\frac{1}{\sqrt{2}} \left(  b_j(\delta)
    - b_j^{\ast}(-\delta-2
    \Omega_{0}
    ) \right)  \text{,}
\end{align}
where
\begin{equation}
  \delta \equiv \Omega - \Omega_0.
  \label{eq:deltaDef}
\end{equation}
The negative frequency components of $Q_j$ and $\phi_j$
[given in terms of their positive frequency components in the
  time domain in Eq.~\eqref{eq:PMFreqCompTimeDomain}], when
$\Omega<0$, can be obtained from the relations, $Q_j^{(-)}(\Omega) =
\left[Q_j^{(+)}(-\Omega)\right]^\ast$ and $\phi_j^{(-)}(\Omega) =
\left[\phi_j^{(+)}(-\Omega)\right]^\ast$.

\subsubsection{Dynamics in the rotating wave approximation}
\label{sec:dynam-rotat-wave}

Radiative damping and driving of the meta-atom
by external fields
alter
the current
oscillation
represented by the normal variable $b_j$.
The interactions leading to these effects
are
often
sufficiently weak that we can
regard
their influence  as a small
perturbation.
We consider this weak interaction limit
here and in
Sec.~\ref{sec:coll-inter-rotat} where we examine the
\emph{collective} behavior of the meta-atoms comprising a metamaterial.
We thus assume
that
$b_j$ varies slowly with respect to the
dominant frequency $\Omega_0$ and
neglect the fast oscillating
components, i.e., we set $b_j(-\delta - 2\Omega_0) = 0$ for $|\delta| \ll
\Omega_0$.
The mode variables $b_j$ are
then
proportional to the slowly varying
envelope of the positive frequency components of the dynamic variables
$Q_j^{(+)}$ and their conjugate momenta $\phi_j^{(+)}$.
Neglecting
fast oscillating components of $b_j$ is
known as the rotating wave approximation (RWA), and is valid in the
limit $\Gamma_{\mathrm{E},j},\Gamma_{\mathrm{M},j}, |\Omega_0 -
\omega_j|, \delta\Omega \ll \Omega_0$, where $\delta\Omega \ll \Omega_0$ indicates a narrow bandwidth of the incident field.

In the RWA,
the meta-atom driving forces, i.e., the EMF and flux, can
be expressed in terms of their slowly varying envelopes
$\tilde{\mathcal{E}}_j$ and $\tilde{\Phi}_j(t)$ defined such that
\begin{eqnarray}
  \frac{\mathcal{E}_j^{(+)}(t)}{\sqrt{\omega_jL_j}} &=& e^{-i\Omega_0 t}
  \tilde{\mathcal{E}}_j(t)
  \label{eq:svEMFDef} \\
  \frac{\Phi_j^{(+)}(t)}{\sqrt{\omega_j L_j}} & = & e^{-i\Omega_0 t}
  \tilde{\Phi}_j(t) \label{eq:svFluxDev}
  \textrm{, }
\end{eqnarray}
where the overall factor of $\sqrt{\omega_j L_j}$ was included for
convenience.

The RWA
essentially
assumes
that all the dynamics are
dominated by the frequency
$\Omega_0$.
Because the RWA implies $\delta\Omega, |\omega_j - \Omega_0|
\ll \Omega_0$, we can
approximate the quantities
$(\Omega/\omega_j)^3$ appearing in the equations of
motion [Eqs.~\eqref{eq:saEqmQ} and \eqref{eq:saEqmphi}] as
$(\Omega/\omega_j)^3 \approx 1$.
In these limits, the equations of motion for the frequency components of
$Q_j$ and $\phi_j$
[Eqs.~\eqref{eq:saEqmQ} and \eqref{eq:saEqmphi}] yield the
relationship for the normal variables
\begin{eqnarray}
  -i\delta b_j(\delta)  &=& \left[ -i\left(\omega_j-\Omega_0\right) -\frac{\Gamma_{\mathrm{E},j} +
      \Gamma_{\mathrm{M},j} }{2}
  \right]  b_j(\delta)  \nonumber\\
  & & +
  f_{j,\mathrm{ext}}(\delta) , \label{eq:bEqmFreq}
\end{eqnarray}
where
the detuning of the meta-atom resonance $\omega_j$ from the
  frequency of the driving field $\Omega_0$ manifests itself as an
  oscillation of the normal variable $b_j$ at frequency $\omega_j - \Omega_0$,
while electric and magnetic dipole radiation emanating from the
  meta-atom results in the damping of $b_j$ at a rate
  $\Gamma_{\mathrm{E},j} + \Gamma_{\mathrm{M},j}$.
The forcing function
combines driving of the current oscillation by the external
  electric field via the EMF and the external magnetic field via the
  flux, and is given by
\begin{equation}
  f_{j,\mathrm{ext}}(\delta) =  i \frac{1}{\sqrt{2}}
  \left(
    \tilde{\mathcal{E}}_{j,\mathrm{ext}}(\delta) + i
    \omega_j\tilde{\Phi}_{j,\mathrm{ext}}(\delta)  \right)
  \textrm{. }
\end{equation}

\subsubsection{The meta-atom as a driven, RLC circuit}
\label{sec:meta-atom-as-1}

Here, we show that in the RWA, a meta-atom behaves as a damped, driven
RLC circuit interacting with the external driving field.
A source of loss
that
is typically present in a meta-atom which we
have thus far neglected is the ohmic losses due to resistance
to current flow
within the meta-atom.
We include the effects of this resistance phenomenologically through
the addition of the ohmic loss rate $\Gamma_{\mathrm{O},j}$ to the
radiative damping rate.
The approximations leading to Eq.~\eqref{eq:singleMARWA} are
still valid provided that $\Gamma_{\mathrm{O}}\ll\Omega_{0}$.
The total meta-atom damping becomes
\begin{equation}
  \Gamma_j \equiv \Gamma_{\mathrm{E},j} + \Gamma_{\mathrm{M},j} +
  \Gamma_{\mathrm{O},j} \text{.}\label{eq:11}
\end{equation}
We obtain  equation of motion for $b_j$ in the time domain
from Eq.~\eqref{eq:bEqmFreq} by multiplying by
$e^{-i\delta t}$ and integrating over the bandwidth of the external
field $-\delta\Omega < \delta < \delta\Omega$.
\begin{equation}
  \frac{db_j}{dt} = \left[ -i \left(\omega_j - \Omega_0\right) -
    \frac{\Gamma_j}{2} \right] 
  b_j(t) + f_{j,\mathrm{ext}}(t)
  \label{eq:singleMARWA}
\end{equation}
When the incident field is of finite duration,
i.e., $\spvec{E}_{\mathrm{in}}(\rv, \pm \infty) =
\spvec{B}_{\mathrm{in}}(\rv,\pm \infty) = 0$, $b_j$  satisfies Eq.~\eqref{eq:singleMARWA}
with the initial condition $b_j(-\infty) = 0$.

The interaction of the meta-atom's current oscillation with its
self-generated EM fields cause the current mode to oscillate at the
resonance frequency $\omega_j$ [Eq.~\eqref{eq:omegajDef}] analogous to that of
an LC circuit.
When a meta-atom current oscillation produces net electric and
  magnetic dipole moments, this oscillation can be driven by external
  fields as manifested by the term $f_{j,\mathrm{ext}}(t)$ in
  Eq.~\eqref{eq:singleMARWA}.
  Radiative and ohmic losses act as a resistance within the meta-atom,
and the external EMF $\tilde{\mathcal{E}}_{j,\mathrm{ext}}$ and
$\tilde{\Phi}_{j,\mathrm{ext}}$ provide the driving.

The dynamics of this effective RLC circuit can be derived from the
effective Hamiltonian
\begin{equation}
  \label{eq:effHam}
  H_{\mathrm{eff},j} = H_{\mathrm{LC},j} + H_{\mathrm{damp},j} + V_{\mathrm{ext},j}
\end{equation}
where $H_{\mathrm{LC},j}$ is the effective Hamiltonian for an undamped LC circuit,
\begin{equation}
  \label{eq:H_LC}
  H_{\mathrm{LC},j} = \omega_j b^\ast_j b_j = \frac{\phi_j^2}{2L_j} +
  \frac{Q_j^2}{2C_j} \text{,}
\end{equation}
the damping is provided by adding an imaginary term to the
effective Hamiltonian
\begin{equation}
  \label{eq:12}
  H_{\mathrm{damp},j} = -i\Gamma b^\ast_j b_j \text{,}
\end{equation}
and the interaction with the external field is provided by
\begin{equation}
  \label{eq:Vext}
  V_{\mathrm{ext},j} = - \mathcal{E}_{j,\mathrm{ext}}^{(+)}Q_j^{(-)} -
  \Phi_{j,\mathrm{ext}}^{(+)} \frac{\phi_j^{(-)}}{L_j} + \mathrm{C.c.}\textrm{ ,}
\end{equation}
The physical significance of the interaction term becomes clearer in
the dipole approximation.
When we neglect the spatial extent of the meta-atoms, the interaction potential with the external fields beomes
\begin{eqnarray}
  \label{eq:V_ext_dip}
  V_{\mathrm{ext},j} &=& -
  \spvec{E}^{(+)}_{j,\mathrm{ext}}(\spvec{R}_j,t) \cdot
  \spvec{d}_j^{(-)}(t) \nonumber \\
  & &- \spvec{B}_{j,\mathrm{ext}}^{(+)}(\spvec{R}_j,t) \cdot \spvec{m}^{\prime
    (-)}_j(t) + \mathrm{C.c.}\textrm{ ,}
\end{eqnarray}
where $\spvec{d}_j^{(\pm)}(t) \equiv h_jQ_j^{(\pm)}(t)$ is the electric
dipole of the meta-atom, and
  \begin{equation}
\spvec{m}^{\prime (\pm)}_j \equiv {A_j\over L_j}
\phi_j^{(\pm)}(t) \,,
\end{equation}
is an effective magnetic dipole of the
meta-atom.
To understand why $\spvec{m}_j^{\prime(\pm)}$ can be
interpreted in this way, consider the conjugate momentum expressed in
terms of the self-inductance [Eq.~\eqref{eq:phiSingleMA}] in the
limits of the RWA (namely $\Omega/\omega_j \approx 1$)
\begin{equation}
  \label{eq:phiInI2}
  \phi_j^{(\pm)} = L_j \left(1 \pm
    \frac{\Gamma_{\mathrm{M},j}}{\omega_j}\right) I_j^{(\pm)}
  + \Phi_{j,\mathrm{ext}}^{(\pm)}
\end{equation}
Because $\Gamma_{\mathrm{M},j} \ll \omega_j$, when the self-induced
magnetic flux dominates that generated by external fields, the
conjugate momentum is related to the current by
\begin{equation}
  \label{eq:IapproxphionL}
  \phi_{j}^{(\pm)} \approx L_j I_j^{(\pm)} \text{\,,}
\end{equation}
and $\spvec{m}_j^{\prime(\pm)} \approx \spvec{m}_j^{(\pm)}$ is
approximately the magnetic dipole created by the current oscillation
in meta-atom $j$.
The effective interaction Hamiltonian
[Eq.~\eqref{eq:V_ext_dip}] accounts for the energy of the
  meta-atom electric dipole interacting with externally generated
  electric fields and the meta-atom's magnetic dipole interacting with
externally generated magnetic fields.

The energy lost due to radiative damping is carried off by the
scattered fields.
The external fields contributing to the interaction
$V_{\mathrm{ext},j}$ include fields scattered from other meta-atoms
in the system.
In the following subsection we will explore how these scattered
fields drive and influence the dynamics of the meta-atoms.

\subsection{Collective interactions in the rotating wave approximation}
\label{sec:coll-inter-rotat}

In this subsection, we examine in detail how the fields emitted
externally to meta-atom $j$ drive the excitation in that meta-atom.
In particular, we will see how the fields emitted or scattered from
the ensemble of meta-atoms mediate interactions between them.
The EM field generated externally to each meta-atom has two
components: the incident field, and the fields scattered from all
other meta-atoms in the system.
The incident field impinges on the metamaterial driving all of its
constituent meta-atoms.
Each excited meta-atom, in turn, radiates an EM field which can
drive other meta-atoms
while undergoing multiple scattering between different resonators.
In order to calculate the response of the metamaterial array to
incident EM fields, we need to consider these multiple scattering
processes, which produce a coupling between meta-atom current
oscillations.
For near-resonant fields, recurrent scattering events in which the
field scatters off the same meta-atom multiple times dramatically
affect the potentially strong coupling between closely-spaced
resonators.

Here we will derive a coupled set of equations for the meta-atoms
where all the   multiple scattering processes are fully incorporated
in the EM field induced interactions between the meta-atoms.
We will then examine how the coupling can lead to a cooperative
response of the metamaterial to the incident field via excitation of
collective modes of current oscillation.
Such modes can have either superradiant character, where the
interactions enhance the radiation emitted from metamaterial, or
a subradiant character, where the radiation remains trapped as it
repeatedly scatters between meta-atoms leading to a suppressed
collective radiative emission rate.

In order to derive a coupled set of equations for the meta-atoms where
the interactions are mediated by the EM fields we consider the
meta-atom mode variables $b_j$ and investigate their dynamics within
the RWA.
As stated in Sec.~\ref{sec:single-meta-atom}, in order
for the RWA to be valid, we assume that the emission rates satisfy
$\Gamma_{\mathrm{E},j}, \Gamma_{\mathrm{M},j}, \Gamma_{\mathrm{O},j}
\ll \Omega_0$ and that the driving field's bandwidth and its detuning
from  meta-atom resonance are small compared to the frequency of the
driving field, i.e., $\delta\Omega, |\omega_j-\Omega_0| \ll \Omega_0$
for all meta-atoms $j$.
In these limits, the external field interactions act as
a slow perturbation on the fast oscillations caused by the meta-atoms'
self-generated fields.

A meta-atom $j$ experiences driving from the external electric and
magnetic fields.
These fields induce EMFs and fluxes, which by
Eq.~\eqref{eq:singleMARWA}, impact the dynamics of the current
oscillation.
The driving originates from both the incident field, and from the
fields scattered from all other meta-atoms $j' \ne j$ in the system.
As such, we decompose the EMF and flux into those generated directly by the
incident driving, $\tilde{\mathcal{E}}_{j,\mathrm{in}}$ and
$\tilde{\Phi}_{j,\mathrm{in}}$,  and those
induced by fields arriving from meta-atom $j'$,
$\tilde{\mathcal{E}}_{j,j'}$ and $\tilde{\Phi}_{j,j'}$.
Explicitly,
\begin{eqnarray}
  \tilde{\mathcal{E}}_{j,\mathrm{ext}} &=& \tilde{\mathcal{E}}_{j,\mathrm{in}} +
  \sum_{j'\ne j} \tilde{\mathcal{E}}_{j,j'}  \\
  \tilde{\Phi}_{j,\mathrm{ext}} & = & \tilde{\Phi}_{j,\mathrm{in}} + \sum_{j' \ne j}
  \tilde{\Phi}_{j,j'}
  \label{eq:externalDecomp}
\end{eqnarray}
The incident field directly drives each meta-atom, inducing a forcing term
\begin{equation}
  \label{eq:15}
  f_{j,\mathrm{in}} \equiv
  \frac{1}{\sqrt{2}}\left(i\tilde{\mathcal{E}}_{j,\mathrm{in}} -
    \omega_j \tilde{\Phi}_{j,\mathrm{in}}\right) \text{,}
\end{equation}
while the scattered fields produce a coupling between the resonators.
Below, we will show that in the RWA, the scattered fields emanating
from meta-atom $j'$ are proportional to the amplitude $b_{j'}$ of the
oscillation in meta-atom $j'$, and therefore that
$\tilde{\mathcal{E}}_{j,j'}$ and $\tilde{\Phi}_{j,j'}$ are proportional
to $b_{j'}$.
We will find that, by virtue of the scattered fields, the dynamics of
the individual meta-atoms are coupled.
The ensemble will exhibit collective modes of oscillation, each with
its own frequency and radiative decay rate.

Because the incident field has a narrow bandwidth around a frequency
$\Omega_0$, we find it convenient to define slowly varying quantities
to describe the dynamics of the system.
For any vector field $\spvec{F}(\rv,t) = \spvec{F}^{(+)}(\rv,t) +
\spvec{F}^{(-)}(\rv,t)$ with positive and negative frequency
components $\spvec{F}^{(+)}$ and $\spvec{F}^{(-)}$, respectively,
unless otherwise specified, we define the slowly
varying envelope $\tilde{\spvec{F}}(\rv,t)$ of the field
such that the positive frequency component
\begin{equation}
  \label{eq:vFieldPosFreq}
  \spvec{F}^{(+)}(\rv,t) \equiv e^{-i\Omega_0 t} \tilde{\spvec{F}}(\rv,t)
  \textrm{ ,}
\end{equation}
or equivalently in frequency space
\begin{equation}
  \label{eq:vFieldPosFreq_frqspc}
  \spvec{F}^{(+)}(\rv,\Omega) = \tilde{\spvec{F}}(\rv,\delta)
  \textrm{ ,}
\end{equation}
where again $\delta \equiv \Omega - \Omega_0$
[Eq.~\eqref{eq:deltaDef}]].
For the charge and conjugate momentum on meta-atom $j$, we define the
scaled slowly varying quantities $\tilde{Q}_j$ and $\tilde{\phi}_j$ such that
\begin{eqnarray}
  \frac{Q_j^{(+)}(t)}{\sqrt{\omega_j C_j}} &\equiv& e^{-i\Omega_0 t}
  \tilde{Q}_j(t) \label{eq:QsvDef} \textrm{ ,}\\
  \frac{\phi_j^{(+)}(t)}{\sqrt{\omega_j L_j}} &\equiv& e^{-i\Omega_0 t}
  \tilde{\phi}_j(t) \label{eq:phiSVDef} \textrm{ .}
\end{eqnarray}
In the RWA $\tilde{Q}_j$ and $\tilde{\phi}_j$
are trivially related to the normal variables by
\begin{subequations}
  \begin{eqnarray}
    \tilde{Q}_j(t) &=& \frac{b_j(t)}{\sqrt{2}}  \text{ ,}\\
    \tilde{\phi_j}(t) &=& -i \frac{b_j(t)}{\sqrt{2}}
    \textrm{ .}
  \end{eqnarray}
  \label{eq:svVarsAndbs}
\end{subequations}
Outside the RWA, $b_j$ contains fast
oscillating components whose origins we discuss in Appendix
\ref{sec:ensemble-meta-atoms}.
In this subsection, however, we will assume that
Eq.~\eqref{eq:svVarsAndbs} holds.
We also define the scaled current such that
\begin{equation}
  \sqrt{\frac{L_j}{\omega_j}} I_j^{(+)}(t) \equiv e^{-i\Omega_t} \tilde{I}_j(t)
  \label{eq:scaledCurrent}
\end{equation}
The relative scale factor of the current was chosen so that, for a
frequency $\delta \ll \Omega_0$, the Fourier components of
$\tilde{\phi}_j$ and $\tilde{I}_j$ are related by
\begin{equation}
  \tilde{\phi}_j(\delta) = D_j(\Omega_0+\delta) \tilde{I}_j(\delta) +
  \tilde{\Phi}_{j,\mathrm{ext}}(\delta) \textrm{ .}
  \label{eq:relationBetweendimlessIandPhi}
\end{equation}
The quantity $D_j$
[Eq.~\eqref{eq:DenomDef}]
serves as the dimensionless complex self-inductance.
Because we have assumed $\Gamma_{\mathrm{M},j} \ll \Omega_0$ in the
RWA, the quantity $D_j\approx 1$.

Next we will
determine the contribution of the fields scattered from each meta-atom
$j'$ to the normalized EMF,  $\tilde{\mathcal{E}}_{j,j'}$, and flux,
$\tilde{\Phi}_{j,j'}$, of meta-atom $j$.
We express the scattered fields from the meta-atom $j'$
in terms of the normalized variables $\tilde{Q}_{j'}$ and
$\tilde{I}_{j'}$.
We assume the bandwidth of the incident field is sufficiently small
that the time scale over which the fields vary, $1/\delta\Omega$, is
much longer than the time it takes for light to propagate across the
metamaterial sample.
We then obtain the slowly varying scattered fields by substituting
$\Omega_0$ for $\Omega$ in the radiation kernels, $\sptensor{G}$ and
$\sptensor{G}_\times$ [Eqs.~\eqref{eq:E_Sj} and \eqref{eq:H_Sj}],
and exploit Eq.~\eqref{eq:vFieldPosFreq_frqspc} to obtain
\begin{align}
  &\tilde{\spvec{E}}_{\mathrm{S},j'} = \frac{3}{2}
  \sqrt{\frac{k_0^3}{6\pi\epsilon_0}}
  \left(\frac{\Omega_0}{\omega_{j'}}\right)^{3/2} \nonumber\\
  & \times\Bigg[
  \sqrt{\Gamma_{\mathrm{E},j'}}\tilde{Q}_{j'} \int d^3r_{j'} \,
  \sptensor{G}(\rv-\rv_{j'},\Omega_0) \cdot
  \frac{\spvec{p}_{j'}(\rv_{j'})}{h_{j'}} \nonumber\\
  &+
  \sqrt{\Gamma_{\mathrm{M},j'}} \tilde{I}_{j'} \int d^3 r_{j'}
  \sptensor{G}_\times(\rv - \rv_{j'},\Omega_0) \cdot
  \frac{\spvec{w}_{j'}(\rv_{j'})}{A_{j'}}\Bigg]
  \label{eq:E_Sj_rwa}
\end{align}
and
\begin{align}
  &\tilde{\spvec{H}}_{\mathrm{S},j'} = \frac{3}{2}
  \sqrt{\frac{k_0^3}{6\pi\mu_0}}
  \left(\frac{\Omega_0}{\omega_{j'}}\right)^{3/2} \nonumber\\
  & \times\Bigg[
  \sqrt{\Gamma_{\mathrm{M},j'}}\tilde{I}_{j'} \int d^3r_{j'} \,
  \sptensor{G}^{(+)}(\rv-\rv_{j'},\Omega_0) \cdot
  \frac{\spvec{w}_{j'}(\rv_{j'})}{A_{j'}} \nonumber\\
  &-
  \sqrt{\Gamma_{\mathrm{E},j'}} \tilde{Q}_{j'} \int d^3 r_{j'}
  \sptensor{G}_\times^{(+)}(\rv - \rv_{j'},\Omega_0) \cdot
  \frac{\spvec{p}_{j'}(\rv_{j'})}{h_{j'}}\Bigg] \textrm{.}
  \label{eq:H_Sj_rwa}
\end{align}
The amplitude of the electric and magnetic fields emitted by the
electric dipole of meta-atom $j'$, driven by $\tilde{Q}_{j'}$, scale with
$\sqrt{\Gamma_{\mathrm{E},j'}}$.
Similarly, the fields emitted by the magnetic dipole of meta-atom
$j'$, driven by $\tilde{I}_{j'}$, scale with
$\sqrt{\Gamma_{\mathrm{M},j'}}$.

These scattered fields provide a portion of the slowly varying EMF,
\begin{equation}
  \tilde{\mathcal{E}}_{j,j'} = \frac{1}{\sqrt{\omega_j L_j}} \int d^3
  r_j \spvec{p}_{j}(\rv_j) \cdot \tilde{\spvec{E}}_{\mathrm{S},j'}(\rv_j)
  \label{eq:svEMF_jjprime}
\end{equation}
and flux
\begin{equation}
  \tilde{\Phi}_{j,j'} = \frac{\mu_0}{\sqrt{\omega_j L_j}} \int d^3
  r_j\, \spvec{w}_{j}(\rv_{j}) \cdot \tilde{\spvec{H}}_{\mathrm{S},j'}(\rv_j)
  \label{eq:3}
\end{equation}
at meta-atom $j$.
Substituting Eqs.~\eqref{eq:E_Sj_rwa} and \eqref{eq:H_Sj_rwa} into the
expressions for EMF and flux gives
\begin{align}
  \tilde{\mathcal{E}}_{j,j'} = &
 \left(\frac{\Omega_0}{\sqrt{\omega_j\omega_{j'}}}\right)^3 \Big[
  \sqrt{\Gamma_{\mathrm{E},j} \Gamma_{\mathrm{E},j'}}
  \left[\mathcal{G}_{\mathrm{E}}(\Omega_0)\right]_{j,j'}
  \tilde{Q}_{j'} \nonumber\\
  & \qquad \qquad+
  \sqrt{\Gamma_{\mathrm{E},j} \Gamma_{\mathrm{M},j'}}
  \left[\mathcal{G}_{\times}(\Omega_0)\right]_{j,j'}
  \tilde{I}_{j'} \Big]\text{ ,}
  \label{EMF_jjprime_rwa_final}
\end{align}
\begin{align}
  \tilde{\Phi}_{j,j'} = &\frac{1}{\omega_j}
  \left(\frac{\Omega_0}{\sqrt{\omega_j\omega_{j'}}}\right)^3  \Big[
    \sqrt{\Gamma_{\mathrm{M,j}} \Gamma_{\mathrm{M},j'}}
    \left[\mathcal{G}_{\mathrm{M}}(\Omega_0)\right]_{j,j'}
    \tilde{I}_{j'} \nonumber\\
    & \qquad \qquad- \sqrt{\Gamma_{\mathrm{M},j} \Gamma_{\mathrm{E},j'}}
    \left[\mathcal{G}_{\times}^T(\Omega_0)\right]_{j,j'} \tilde{Q}_{j'}
     \Big]  \text{\,,}
     \label{eq:Phi_jjprime_rwa_final}
\end{align}
where the matrices $\mathcal{G}_{\mathrm{E}}$,
$\mathcal{G}_{\mathrm{M}}$, and $\mathcal{G}_{\times}$ determine how
the meta-atoms' geometries and relative orientations influence the
respective contributions of the scattered electric fields to the EMFs, the scattered magnetic fields to the fluxes, and the scattered electric
(magnetic) fields the fluxes (EMFs).
These matrices have zero diagonal elements and off diagonal elements
given by
\begin{widetext}
  \begin{subequations}
    \begin{align}
      \left[ \mathcal{G}_{\mathrm{E}} (\Omega) \right]_{j,j'} & =
      \frac{3}{2} \int d^{3}r_{j} \int d^{3} r_{j'}
      \frac{\spvec{p}_{j}(\spvec{r}_{j})}{h_{j}} \cdot
      \sptensor{G}(\spvec{r}_{j}-\spvec{r}_{j'},\Omega) \cdot
      \frac{\spvec{p}_{j'}(\spvec{r}_{j'})}{h_{j'}}
      \\
      \left[\mathcal{G}_{\mathrm{M}} (\Omega)\right]_{j,j'} & =
      \frac{3}{2} \int d^{3}r_{j} \int d^{3}r_{j'}
      \frac{\spvec{w}_{j}(\spvec{r}_{j})}{A_{j}} \cdot
      \sptensor{G}(\spvec{r}_{j} - \spvec{r}_{j'},\Omega) \cdot
      \frac{\spvec{w}_{j'}(\spvec{r}_{j'})}{A_{j'}}
      \\
      \left[ \mathcal{G}_{\times}(\Omega)\right]_{j,j'} & =
      \frac{3}{2} \int d^{3}r_{j} \int d^{3}r_{j'}
      \frac{\spvec{p}_{j}(\spvec{r}_{j})}{h_{j}} \cdot
      \sptensor{G}_{\times}(\spvec{r}_{j}-\spvec{r}_{j'},\Omega) \cdot
      \frac{\spvec{w}_{j'}(\spvec{r}_{j'})}{A_{j'}} \textrm{.}
    \end{align}
    \label{eq:GCouplingMats}
  \end{subequations}
\end{widetext}
When the separation between two meta-atoms is much greater than the
spatial extent of the individual elements, these geometrical factors
depend exclusively on the relative positions and orientations of the
meta-atoms' electric and magnetic dipoles.
Explicitly, in that limit,
\begin{subequations}
  \begin{align}
    \left[ \mathcal{G}_{\mathrm{E}} (\Omega) \right]_{j,j'} & =
    \frac{3}{2} \unitvec{d}_j \cdot
    \sptensor{G}(\spvec{R}_{j}-\spvec{R}_{j'},\Omega) \cdot
    \unitvec{d}_{j'}
    \\
    \left[\mathcal{G}_{\mathrm{M}} (\Omega)\right]_{j,j'} & =
    \frac{3}{2} \unitvec{m}_j \cdot \sptensor{G}(\spvec{R}_{j} -
    \spvec{R}_{j'},\Omega) \cdot \unitvec{m}_{j'}
    \\
    \left[ \mathcal{G}_{\times}(\Omega)\right]_{j,j'} & = \frac{3}{2}
    \unitvec{d}_j \cdot
    \sptensor{G}_{\times}(\spvec{R}_{j}-\spvec{R}_{j'},\Omega) \cdot
    \unitvec{m}_{j'} \textrm{.}
  \end{align}
  \label{eq:GCouplingMatsDip}
\end{subequations}
The contribution of the electric field scattered by meta-atom $j'$ to
the EMF, $\tilde{\mathcal{E}}_{j,j'}$, scales with the geometric mean
of the electric dipole emission rates of the two meta-atoms,
$\sqrt{\Gamma_{\mathrm{E},j}\Gamma_{\mathrm{E},j'}}$.
Similarly, the magnetic field of element $j'$ contributes to the flux
$\tilde{\Phi}_{j,j'}$ with a strength proportional to
$\sqrt{\Gamma_{\mathrm{M},j}\Gamma_{\mathrm{M},j'}}$.
When the meta-atoms are sufficiently far away from one and other, the
electric field emitted by the magnetic dipoles and the magnetic field
emitted by the electric dipoles provide a significant contribution to
$\tilde{\mathcal{E}}_{j,j'}$ and $\tilde{\Phi}_{j,j'}$ that scale with
$\sqrt{\Gamma_{\mathrm{E},j} \Gamma_{\mathrm{M},j'}}$ and
$\sqrt{\Gamma_{\mathrm{M},j} \Gamma_{\mathrm{E},j'}}$, respectively.

We have set out to obtain coupled equations of motion for the
meta-atom normal variables $b_j$ mediated by the EM
field.
We have obtained contributions to the EMF and flux that are driven by
charges $\tilde{Q}_j$ and currents $\tilde{I}_j$.
However, only $\tilde{Q}_j$ and conjugate momenta $\tilde{\phi}_j$ are
trivially related to these normal variables
[Eq.~\eqref{eq:svVarsAndbs}].
The current, on the other hand obeys the more complex relationship
\begin{equation}
  \tilde{I}_{j'} = -i\frac{b_{j'}}{\sqrt{2}} -
  \sum_{j^{\prime\prime}\ne j'} \tilde{\Phi}_{j',j''} - \tilde{\Phi}_{j',\mathrm{in}}
  \label{eq:currentRelation}
\end{equation}
One can thus use Eqs.~\eqref{eq:svVarsAndbs} and
\eqref{eq:currentRelation} to express $\tilde{\mathcal{E}}_{j,j'}$ and
$\tilde{\Phi}_{j,j'}$ in terms of the normal variables $b_{j'}$.
We note, however, from Eq.~\eqref{eq:Phi_jjprime_rwa_final}, that
$\tilde{\Phi}_{j',j''}$ contains contributions that scale as
 $\sqrt{\Gamma_{\mathrm{M},j'}
  \Gamma_{\mathrm{M},j''}}/\omega_j$ and $\sqrt{\Gamma_{\mathrm{M},j'}
  \Gamma_{\mathrm{E},j''}}/\omega_j$, which under the conditions of
the RWA, are much less than 1.
Furthermore, $\tilde{I}_{j}$ contains a contribution from the incident
field flux $\tilde{\Phi}_{j,\mathrm{in}}$.
The contribution of the incident flux to
$\tilde{I}_j$
can also be ignored to lowest order
since it is about $\Gamma_j/\omega_j$ times the direct contribution of
the incident flux to the direct driving $f_{j,\mathrm{in}}$.
So, to determine $\tilde{\mathcal{E}}_{j,j'}$ and
$\tilde{\Phi}_{j,j'}$ to lowest order in $\Gamma_{j} /
\omega_j$,
we therefore exploit the approximate relationship $\tilde{I}_{j'}
\approx \tilde{\phi}_{j'} \approx -ib_{j'}/\sqrt{2}$.

Having computed the contributions of the scattered fields to the EMF
and flux of an individual meta-atom, we find that these scattered
fields produce a coupling between meta-atoms in the oscillator
equations of motion.
Substituting the EMF and flux into Eq. \eqref{eq:singleMARWA}, we find
the evolution of the column vector $\colvec{b}$ of normal variables is
governed by
\begin{equation}
  \dot{\colvec{b}} = \mathcal{C} \colvec{b} + \colvec{f}_{\mathrm{in}} \text{ ,}
  \label{eq:rwa_b_eqm}
\end{equation}
where we have introduced the following notation for  $\colvec{b}$ and
for the driving $\colvec{f}_{\mathrm{in}}$ caused by the incident
field:
\begin{equation} \label{eq:vectors}
  \colvec{b} \equiv
  \begin{pmatrix}
    b_1 \\
    b_2\\
    \vdots\\
    b_N
  \end{pmatrix},\quad \colvec{f}_{\mathrm{in}} \equiv
  \begin{pmatrix}
    f_{1,\mathrm{in}} \\
    f_{2,\mathrm{in}}\\
    \vdots\\
    f_{N,\mathrm{in}}
  \end{pmatrix}
  \,.
\end{equation}
The coupling matrix $\mathcal{C}$ is
given
to lowest order in $\Gamma_{\mathrm{E},j}/\omega_{j'}$, and
$\Gamma_{\mathrm{M},j}/\omega_{j'}$ by
\begin{align}
  \mathcal{C} & =  -i\mathrm{\Delta} - \frac{1}{2} \mathrm{\Upsilon}
  \nonumber\\
   &+ \frac{1}{2}\Big( i\mathrm{\Upsilon}_{\mathrm{E}}^{1\over 2}
    \mathcal{G}_{\mathrm{E}}\mathrm{\Upsilon}_{\mathrm{E}}^{1\over 2} +
    i \mathrm{\Upsilon}_{\mathrm{M}}^{1\over2 } \mathcal{G}_{\mathrm{M}}
    \mathrm{\Upsilon}_{\mathrm{M}}^{1\over 2} \nonumber\\
    & \qquad+ \mathrm{\Upsilon}_{\mathrm{E}}^{1\over 2} \mathcal{G}_\times
    \mathrm{\Upsilon}_{\mathrm{M}}^{1\over 2}
    + \mathrm{\Upsilon}_{\mathrm{M}}^{1\over 2} \mathcal{G}_\times^{T}
    \mathrm{\Upsilon}_{\mathrm{E}}^{1\over 2} \Big) \text{.}
  \label{eq:C_rwa}
\end{align}
Here the detunings of the incident field from the meta-atom resonances
are contained in the diagonal matrix $\mathrm{\Delta}$ with elements
\begin{equation}
  \Delta_{j,j} \equiv \omega_j-\Omega_0\,\text{.}
  \label{eq:DeltaMatDef}
\end{equation}
Moreover, the meta-atom emission rates are
incorporated in the diagonal matrices $\mathrm{\Upsilon}_{\mathrm{E}}$,
$\mathrm{\Upsilon_{\mathrm{M}}}$ and $\mathrm{\Upsilon}_{\mathrm{O}}$
with elements
\begin{subequations}
  \label{eq:UpsilonMatDefs_RWA}
  \begin{align}
    [\mathrm{\Upsilon}_{\mathrm{E}}]_{j,j} \equiv &
    \Gamma_{\mathrm{E},j}  \text{ ,} \\
    [\mathrm{\Upsilon}_{\mathrm{M}}]_{j,j} \equiv &
    \Gamma_{\mathrm{M},j} \text{ ,} \\
    [\mathrm{\Upsilon}_{\mathrm{O}}]_{j,j} \equiv &
    \Gamma_{\mathrm{O},j}\,,
  \end{align}
\end{subequations}
respectively, and we have defined
$\mathrm{\Upsilon} \equiv \mathrm{\Upsilon}_{\mathrm{E}} + \Upsilon_{\mathrm{M}}
+ \mathrm{\Upsilon}_{\mathrm{O}}$.

The interaction matrix $\mathcal{C}$ accounts for electric
dipole-dipole interactions, magnetic dipole-dipole interactions, as
well as interactions between electric and magnetic dipoles that arise
from magnetic (electric) fields emitted by electric (magnetic)
dipoles.
The diagonal elements of $\mathcal{C}$ result from interactions with
the self-generated fields and give rise to the meta-atoms' resonance
frequencies and radiative emission rates.

In the RWA, the dynamic equation [Eq.~\eqref{eq:rwa_b_eqm}] encapsulates
all the multiple scattering processes between the different
meta-atoms.
These are described by the interaction terms in the matrix
$\mathcal{C}$, mediated by the scattered EM fields.
The coupled set of equations  implies a
system of $N$ meta-atoms possesses $N$ collective modes of excitation.
These modes correspond to the eigenvectors of the matrix
$\mathcal{C}$.
For each collective eigenmode we have collective radiative
resonance linewidths  and resonance frequencies that are represented
by the eigenvalues of $\mathcal{C}$.
A strong coupling between the resonators can lead to a cooperative response of the
metamaterial sample to the EM fields, resulting in collective decay
rates which are substantially different from those of a single,
isolated meta-atom.
The interactions can either enhance radiative
  emission, producing a superradiant mode, or suppress emission,
  yielding a subradiant decay rate.
We will illustrate the effect of a cooperative response of a 2D
metamaterial array in Sec.~\ref{sec:an-ensemble-asymm} by considering
an example of closely-spaced split ring resonators.
We find that even in a relatively small sample the strong coupling
leads to a dramatic resonance linewidth narrowing of five orders of
magnitude and to a broad distribution of radiative decay rates.

In order to illustrate the coupling of an incoming field to
collective modes, suppose the incident field is engineered so that
it only excites the $i^{\mathrm{th}}$ collective mode, and then is
suddenly turned off.
The collective excitation is then
distributed over the sample
according to the eigenvector $\colvec{v}_i$ of $\mathcal{C}$.
Due to the repeatedly scattered fields that couple the meta-atoms,
the excitation oscillates at its resonance
frequency given by the eigenvalue $\lambda_i$,
\begin{equation}
  \label{eq:delta_i_def_1}
  \Omega_i \equiv \Omega_0 - \operatorname{Im}(\lambda_i) \text{\,,}
\end{equation}
and the amplitude of oscillations decay at a rate
\begin{equation}
  \label{eq:gammaCollectiveDef}
  \gamma_i \equiv -2 \operatorname{Re}(\lambda_i) \text{\,.}
\end{equation}
as radiation leaks out of the collective excitation and energy
dissipates through ohmic losses.
The vector of normal variables then evolves as
\begin{equation}
  \label{eq:2}
  \colvec{b}(t) \propto
  \exp\left\{\left[-i\left(\Omega_i-\Omega_0\right)
      -\frac{\gamma_i}{2}\right] t \right\} \colvec{v}_i\,.
\end{equation}

The nature of collective modes could also allow one to engineer a
cooperative response of the metamaterial to the incident field, addressing linear
combinations of modes by shaping the incident field's profile, or
adjusting its frequency.
Engineering of the collective response may then be used, for example, to
excite isolated subwavelength hot spots in a
  metamaterial.\cite{KAO10}

\subsection{Concluding remarks}
\label{sec:concluding-remarks}

In this section, we saw how the interaction of individual meta-atoms
with the EM field governs the collective dynamics of an ensemble of
meta-atoms that make up a metamaterial.
Each meta-atom experiences the influence of its current
oscillation's self-generated field, the field incident on the
metamaterial, and the fields scattered from all other meta-atoms in
the system.
We explored the influence of the self-generated fields in
Sec.~\ref{sec:single-meta-atom}.
In the RWA, the self-generated field dominates meta-atom dynamics.
Each meta-atom can be seen as an effective RLC circuit which
experiences damping due to electric and magnetic dipole radiation
carrying energy away from the meta-atom.
On the other hand, fields generated externally to the meta-atom,
i.e., the incident field and the fields radiated from all other
meta-atoms in the metamaterial, drive the current oscillations in
each meta-atom.
In Sec.~\ref{sec:coll-inter-rotat}, we saw how the fields scattered
by each meta-atom mediate interactions between them.
Fields emitted by one meta-atom drive the current oscillations
  in all the others, producing the dynamic inter-meta-atom coupling in
  Eq.~\eqref{eq:rwa_b_eqm}.
While Appendix \ref{sec:ensemble-meta-atoms}, develops a
formalism to account for arbitrarily strong interactions,
in this section
we have gained a significant physical insight in the RWA in
which we assume the meta-atoms' interact much more strongly with their
self-generated fields than with the fields generated externally.

In the following section, we will apply this formalism to examine
collective modes in an example metamaterial: an array of
symmetric split ring resonators.
This system will illustrate the vital role cooperative
interactions can play in the dynamics of a metamaterial composed of
closely spaced plasmonic resonators.
A metamaterial of $N$ resonators will have $N$ collective modes of
current oscillation, each with its own resonance frequency and
radiative emission rate.
Both of these quantities strongly influence how a given mode can be
excited.
The cooperative interactions lead to a broad distribution of
collective decay rates indicating strongly superradiant or
  subradiant modes.

\section{An ensemble of symmetric split ring resonators}
\label{sec:an-ensemble-asymm}

In this section, we apply the formalism developed in this article
to a metamaterial composed of split ring resonators (SRRs).
As the name suggests, these resonators are composed of loops with
segments that have been removed.
Owing to the curvature of the elements, current oscillations within
SRRs
can exhibit both an electric and a magnetic
response.
Variations of these resonators have been  used to produce
metamaterials which exhibit, e.g., negative indices of refraction.
\cite{ShelbySci2000,SmithEtAlPRL2000}
Here, we consider a particular realization of the
SRR
in which a single ring is cut into two disconnected concentric circular arcs of equal length. We then study the SRR metamolecule by assuming that the halves each form a meta-atom that supports a single mode of current oscillation.
The two halves could either oscillate in phase, producing a net
electric dipole, or out of phase, producing a net magnetic dipole.

In addition to active studies of metamaterial arrays of SRRs, there
has also been an increasing interest in fabricating metamaterials
consisting of split ring resonators in which the symmetry between the
two disconnected halves has been broken, e.g., by making one of them
longer.
Sheets of asymmetric split ring resonators (ASRs) have been shown to
exhibit transmission resonances \cite{FedotovEtAlPRL2007}
corresponding to excitations in which all magnetic dipoles in the
sheet oscillated in phase.
The quality factor of this resonance, however, was shown to depend
strongly on the number of ASRs in the system.\cite{FedotovEtAlPRL2010}
Furthermore, artificially adjusted disorder in the positions of the
unit-cell resonators was observed to destroy the
resonance.\cite{papasimakis2009}
If interactions mediated by the EM fields were not important, and the
ASRs behaved independently, system size or positional disorder of the
system would have little effect on the metamaterial response to the EM
fields.
These experimental observations provide ample evidence for the vital
role collective interactions play in this particular metamaterial.

Here we employ the formalism describing collective
interactions  to an ensemble of
SRRs in the RWA. We describe a single
SRR in subsection \ref{sec:asymm-split-ring},
while we examine the properties of collective modes of
SRRs
in a
lattice in subsection \ref{sec:collective-modes-an}.

\subsection{The symmetric split ring resonator}
\label{sec:asymm-split-ring}

\begin{figure}
  \centering
  \includegraphics{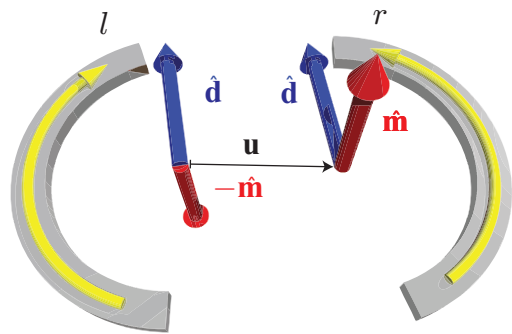}
  \caption{(color online) A schematic illustration of a split ring resonator.
    An excitation in the left meta-atom ($l$) produces an oscillating
    electric dipole (indicated by the blue arrow) in the direction
    $\unitvec{d}$ and a magnetic     dipole (indicated by the red
    arrow) in the direction $-\unitvec{m}$, while an excitation in the
    right meta-atom ($r$) produces an electric dipole in the direction
    $\unitvec{d}$ and a magnetic dipole in the direction
    $\unitvec{m}$.
    The meta-atoms, in the point source approximation, are separated
    by a vector $\spvec{u}$.
    When the meta-atoms are excited in phase, the electric dipoles
    reinforce each other and the magnetic dipoles cancel out.
  }
  \label{fig:ASRIllus}
\end{figure}
We begin by describing the interaction of a single SRR unit-cell
resonator with incident EM fields.
This particular realization of an SRR metamolecule
  consists of two meta-atoms formed by  two
concentric circular arcs labeled by $j \in\{l,r\}$ (for ``left'' and
``right''), as shown in Fig.~\ref{fig:ASRIllus}.
This metamolecule possesses reflection symmetry about a central
plane.

To illustrate this qualitative physical behavior of an
SRR,
we
approximate the meta-atoms as two point sources separated by
$\spvec{u} \equiv \spvec{R}_r - \spvec{R}_l$ (see Fig.~\ref{fig:ASRIllus}).
The current oscillations in meta-atoms produce electric dipoles
with orientation $\unitvec{d}_r = \unitvec{d}_l = \unitvec{d}$
associated with charge oscillating between the ends of the arcs.
Owing to the curvature of the meta-atoms, these currents also produce
magnetic dipoles with opposite orientations $\unitvec{m}_r =
-\unitvec{m}_l = \unitvec{m}$.
The generated electric dipoles lie in the plane of the
SRR
and are perpendicular to the displacement between the meta-atoms
($\unitvec{d}\perp\unitvec{u}$).
The generated magnetic dipoles, on the other hand, point out of the
plane in which the
SRR
resides ($\unitvec{m} \perp\unitvec{u},\unitvec{d}$).
Each meta-atom in isolation supports a single mode of oscillation with
resonance frequency $\omega_0$.
Here we consider a resonant driving with the frequency of the incident
field satisfying $\Omega_0 = \omega_0$.
For simplicity, we also assume each element
possesses identical radiative and
thermal decay rates $\Gamma_{\mathrm{E/M/O},l} =
\Gamma_{\mathrm{E/M/O},r} = \Gamma_{\mathrm{E/M/O}}$.

In the RWA, the normal variables $b_r$ and $b_l$
[Eq.~\eqref{eq:bSingAtDef} with $j\in \{r,l\}$] describe the
states of the right and left halves, respectively, of a single SRR metamolecule
in isolation.
We may now apply the previously developed theory for the EM
field mediated interactions between meta-atoms to a single SRR
unit-cell resonator consisting of these two meta-atoms.
According to
Eq.~\eqref{eq:rwa_b_eqm}, the normal variables $b_r$ and $b_l$ are
coupled by the EM fields as
\begin{equation}
  \label{eq:SSRDynEq}
  \left(
    \begin{array}{c}
      \dot{b}_r \\
      \dot{b}_l
    \end{array}
  \right) = \mathcal{C}_{\mathrm{SRR}}
  \left(
    \begin{array}{c}
      b_r \\
      b_l
    \end{array}
  \right)
  +
  \left(
    \begin{array}{c}
      f_{r,\mathrm{in}} \\
      f_{l,\mathrm{in}}
    \end{array}
  \right)\,.
\end{equation}
Here $\mathcal{C}_{\mathrm{SRR}}$ denotes the specific coupling matrix
in this case between the two meta-atoms, as described in detail
below.
The incident field impinging on the SRR produces the driving terms
$f_{j,\mathrm{in}}$ for each meta-atom $j=l,r$ [Eq.~\eqref{eq:15}].
Considering the meta-atoms as point emitters, the incident field
excites their electric and magnetic dipoles resulting in the
simplified driving terms
\begin{align}
  \label{eq:fSRRApprox}
  f_{r,\mathrm{in}} = &
  ih \frac{\unitvec{d} \cdot
    \tilde{\spvec{E}}(\spvec{R}_r,t)}{\sqrt{2\omega_0 L}}
  - \omega_0
  A  \frac{\unitvec{m} \cdot
    \tilde{\spvec{B}}(\spvec{R}_r,t)}{\sqrt{2\omega_0 L}} \text{\,,} \\
  f_{l,\mathrm{in}} = &
  i h \frac{\unitvec{d} \cdot
    \tilde{\spvec{E}}(\spvec{R}_l,t)}{\sqrt{2\omega_0 L}}
  + \omega_0
  A \frac{\unitvec{m} \cdot
    \tilde{\spvec{B}}(\spvec{R}_l,t)}{\sqrt{2\omega_0 L}}
  \text{\,.}
\end{align}
The quantity $h$ is an effective length along which charge flows to
form the meta-atoms' electric dipoles and is related to
$\Gamma_{\mathrm{E}}$ through Eq.~\eqref{eq:Gamma_EDef}.
Similarly, $A$ is an effective area that indicates the strength of
the magnetic dipole interaction and is related to the magnetic
  dipole emission rate   $\Gamma_{\mathrm{M}}$ through
  Eq.~\eqref{eq:Gamma_MDef}; $L$ is   the self-inductance of each
  meta-atom.
  Once excited, each half of the SRR scatters both electric and
  magnetic fields.
  These fields then impact the other meta-atom, driving its electric
  and magnetic dipoles.
  Repeated absorption and re-emission of scattered fields produces a
  dynamic interaction between the two halves of the SRR.
From Eq.~\eqref{eq:C_rwa}, the
coupling
matrix governing the
interaction
is given by
\begin{equation}
  \label{eq:SplitRing}
  \mathcal{C}_{\mathrm{SRR}} = \left(
    \begin{array}{cc}
      -\Gamma/2 &  i
      {\rm d}\Gamma  G -
      \bar{\Gamma} S
      \\
       i {\rm d}\Gamma  G
      - \bar{\Gamma} S &
      -\Gamma/2
    \end{array}
  \right)
\end{equation}
where a single meta-atom has a total decay rate
\begin{equation}
  \label{eq:TotalGammaDefsAgain}
  \Gamma \equiv \Gamma_{\mathrm{E}} + \Gamma_{\mathrm{M}} +
  \Gamma_{\mathrm{O} }\,
\end{equation}
appearing in the diagonal elements of $\mathcal{C}_{\mathrm{SRR}}$,
and we have defined
\begin{align}
  \label{eq:GammaGeoMean}
  \bar{\Gamma} & \equiv \sqrt{\Gamma_{\mathrm{E}}
    \Gamma_{\mathrm{M}}} \text{,}\\ \label{eq:GammaDiff}
  {\rm d}\Gamma & \equiv \Gamma_{\mathrm{E}} - \Gamma_{\mathrm{M}}\,.
\end{align}
Coupling between the two halves of the SRR, represented by
  the off diagonal elements of $\mathcal{C}_{\mathrm{SRR}}$, arises
  from interactions between the meta-atoms' electric dipoles, the
  meta-atoms' magnetic dipoles, as well as a cross interaction between
  the electric dipole of one meta-atom and the magnetic dipole of the
  other.
The strength of the electric
and magnetic
dipole-dipole interactions
is proportional to
the radiative decay rates
$\Gamma_\mathrm{E}$ and $\Gamma_{\mathrm{M}}$, respectively.
These dipole-dipole interactions also depend on
the spacing between the meta-atoms
and the relative orientations of the dipoles.
This geometrical dependence shows up in the factor
\begin{equation}
  G \equiv \frac{3}{4} \unitvec{d} \cdot \sptensor{G}(\spvec{u},\Omega_0)
  \cdot \unitvec{d}
  = \frac{3}{4} \unitvec{m}\cdot \sptensor{G}(\spvec{u},\Omega_0) \cdot
  \unitvec{m} \text{.}
\end{equation}
Notice that because identical meta-atom excitations (i.e., when
$b_l=b_r$)
produce parallel electric dipoles, but antiparallel
magnetic dipoles, the electric and magnetic dipole interactions work
against each other; the strength of interaction arising from the
geometrical factor $G$ is proportional to $\Gamma_{\mathrm{E}} -
\Gamma_{\mathrm{M}}$.
An additional interaction arises from the electric dipoles interacting
with fields scattered from the magnetic dipoles and vice versa.
The geometric mean of the radiative decay rates,
$\bar\Gamma$ [Eq.~\eqref{eq:GammaGeoMean}],
governs the strength of this interaction.
Relative orientations of the electric dipole of the left (right)
meta-atom and the magnetic dipole of the right (left) meta-atom appear
in the geometrical factor
\begin{equation}
  \label{eq:1}
  S  \equiv \frac{3}{4} \unitvec{d} \cdot \sptensor{G}_\times
  (\spvec{u},\Omega_0) \cdot  \unitvec{m}_r \,\textrm{,}
\end{equation}

To analyze the collective modes of the
SRR,
we consider the dynamics of symmetric $c_+$ and antisymmetric $c_-$
modes of oscillation defined by
\begin{equation}
  \label{eq:cpm_def}
  c_\pm \equiv \frac{1}{\sqrt{2}}\left(b_r \pm b_l\right) \textrm{.}
\end{equation}
These symmetric and antisymmetric  variables  represent
the eigenmodes of the
SRR.
From the dynamic equation [Eq.~\eqref{eq:rwa_b_eqm}] and the
SRR
coupling matrix [Eq.~\eqref{eq:SplitRing}], one finds
\begin{equation}
 \frac{d}{dt}
      c_\pm
   =
  \left(
      -\frac{\gamma_\pm}{2}  \mp i \Delta_{\mathrm{SRR}}
  \right)
    c_\pm
  +
  F_{\pm}
 \,\textrm{,}
  \label{eq:cpmEqs}
\end{equation}
where an incident field produces the driving terms
\begin{equation}
  \label{eq:ASRForce}
  F_\pm = \frac{1}{\sqrt{2}} \left(f_{r,\mathrm{in}} \pm f_{l,\mathrm{in}} \right) \text{ .}
\end{equation}
The interaction between the elements produces the decay rates
$\gamma_\pm$ and shifts the resonance frequencies of the symmetric and
antisymmetric modes by equal and opposite amounts,
$\Delta_{\mathrm{SRR}}$
\begin{align}
  \label{eq:3a}
  \gamma_\pm & = \Gamma_{\mathrm{E}} \left(1 \pm 2
    \mathrm{Im}(G)\right)  \nonumber\\
  & \quad+ \Gamma_{\mathrm{M}}\left(1 \mp
    2\mathrm{Im}(G)\right)  + 2\bar{\Gamma}
  \mathrm{Re} (S) +
  \Gamma_{\mathrm{O}}  \\
  \Delta_{\mathrm{SRR}} & =
  -2\mathrm{Re}(G)
  \left(\Gamma_{\mathrm{E}}-\Gamma_{\mathrm{M}}\right) - 2
  \bar{\Gamma}   \mathrm{Im} (S)  \text{ .}
\end{align}
An analogy can be drawn between these metamolecular current
oscillations and atomic or molecular energy
levels.\cite{ProdanEtAlSCI2003, WangEtAlACR2006}
The symmetric and antisymmetric modes have respective resonance
frequencies $\omega_0 +\Delta_{\mathrm{SRR}}$ and
$\omega_0-\Delta_{\mathrm{SRR}}$.
When excited, the symmetric mode decays at a rate $\gamma_+$, while an
excitation of the antisymmetric mode decays at rate $\gamma_-$.

Excitation of the symmetric mode ($c_+$) produces a net electric
dipole since the individual meta-atom electric dipoles oscillate in
phase while the meta-atom magnetic dipoles approximately cancel each other out.
Similarly, excitation of the antisymmetric mode ($c_-$) produces a net
magnetic dipole
and the net effect of the electric dipole approximately cancels out.
The symmetric and antisymmetric  excitations will thus be
referred to electric and magnetic dipole excitations, respectively.
When the spacing between the arcs $u \ll \lambda$, the decay rates
simplify to
\begin{subequations}
  \begin{eqnarray}
    \gamma_+ & \approx & 2\Gamma_{\mathrm{E}}+\Gamma_{\mathrm{O}} \text{ ,} \\
    \gamma_- & \approx & 2\Gamma_{\mathrm{M}} + \Gamma_ {\mathrm{O}} \text{ .}
  \end{eqnarray}
\end{subequations}
The electric mode loses energy via electric dipole radiation, while
the magnetic mode emits magnetic dipole radiation.
In the absence of magnetic dipole interactions, the symmetric and
antisymmetric modes are analogous to superradiant and
subradiant states in a pair of closely spaced two-level atoms: when
the two-level atoms are excited in phase, the radiative emission rate is
enhanced, and it is suppressed when the atoms are excited out of phase.
Furthermore, in the SRR metamolecule the electric and magnetic modes are driven
purely by the electric and magnetic fields, respectively, with $F_+
\propto \unitvec{d} \cdot
\tilde{\spvec{E}}_{\mathrm{in}}(\spvec{R},t)$ and $F_- \propto
\unitvec{m}
\cdot \tilde{\spvec{B}}_{\mathrm{in}}(\spvec{R},t)$, where $\spvec{R}$
denotes the center of mass coordinate of the
SRR.

When more than one
SRR
is present, radiation emitted from one
SRR
impacts and drives oscillations in another.
The resulting interactions produce collective modes of oscillation for the whole system.
We examine this collective behavior in the following subsection.

\subsection{Collective modes in an ensemble of symmetric split rings}
\label{sec:collective-modes-an}

Having discussed how EM field induced interactions arise between two
meta-atoms in a single SRR metamolecule,
we now explore how a collection of metamolecules can behave in concert
when brought together to form a metamaterial.
As an example we consider a 2D $N_x \times N_y$ array of
SRRs
arranged
in a square lattice with lattice vectors $\spvec{a}_1 = a
\unitvec{e}_x$ and $\spvec{a}_2 = a \unitvec{e}_y$.
This finite array resides in a region with free space (as
  opposed to e.g. periodic)
  boundary conditions.
A single
SRR
occupies each unit cell of the lattice.
They are oriented such that symmetric oscillations produce electric
dipoles along the direction $\unitvec{d} = \unitvec{e}_y$, and
antisymmetric oscillations produce magnetic dipoles pointing out of
the lattice in the direction $\unitvec{m} = \unitvec{e}_z$.
In this section, we quantify the collective interactions by examining
the collective eigenmodes of the system and showing how
the interactions can lead to strongly modified radiative emission
rates.
We also illustrate from this model how a subwavelength
inter-molecular spacing enhances the collective behavior of the
system. In particular, we find that
a subwavelength lattice spacing produces a much broader distribution of subradiant and superradiant collective decay rates.

\begin{figure}
  \centering
  \includegraphics{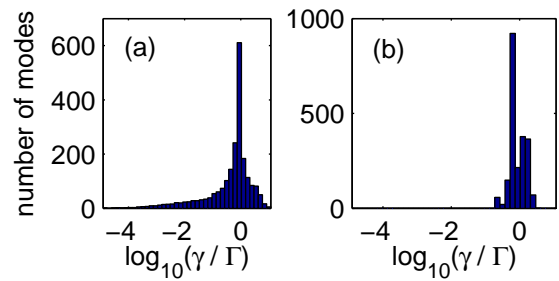}
  \caption{(color online) The distributions of collective radiative emission rates in
    a $33 \times 33$ square array of
    SRRs.
    The distribution is represented as a histogram of the $\log_{10}$
    of the collective emission rates.
    Panel (a) shows the distribution for an ensemble of
    SRRs
    with lattice spacing $a=0.5\lambda$.
    Panel (b) shows the distribution when the lattice spacing $a=
    1.4\lambda$.
    When the lattice spacing is larger, interactions become weaker,
    and cooperative effects are diminished, leading to a narrower
    distribution of collective mode line widths.
  }
  \label{fig:hist}
\end{figure}

While an
SRR
in isolation possesses two modes with two collective
resonance frequencies and two decay rates, the presence of
interactions in an ensemble can produce a broad distribution of
collective linewidths.
The lattice of $N_x \times N_y$ SRRs possesses $2N_x N_y$ collective
modes of oscillation, where the $i^{\textrm{th}}$ mode corresponds to
an eigenvector $\colvec{v}_i$ of the interaction matrix $\mathcal{C}$
[Eq.~\eqref{eq:C_rwa}].
The resonance frequency of this collective mode is shifted from
$\Omega_0$ by $\delta_i \equiv \Omega_i - \Omega_0$ and has a
collective decay rate $\gamma_i$.
These are given in terms of the mode's eigenvalue $\lambda_i$ as
\begin{subequations}
  \begin{eqnarray}
    \label{eq:detuningsAndRatesFromLambda}
    \delta_i &= & -\mathrm{Im}\, \lambda_i \text{ ,} \\
    \gamma_i &=& -2 \mathrm{Re}\, \lambda_i \,,
  \end{eqnarray}
\end{subequations}
respectively.
Here we consider an ensemble of SRRs whose elements have equal
single-meta-atom electric and magnetic decay rates
$\Gamma_E=\Gamma_M$, and we take the separation between constituent
meta-atoms of an SRR to be $u = 0.12\lambda$.
Because the thermal losses are equal in all meta-atoms, their presence would add to the decay rates of each collective mode equally.
Since here we are interested in how interactions
modify collective radiative decay rates, we
take the ohmic loss rate to be zero in this section.

We numerically calculate all the eigenmodes of the system that are
modified by the multiple scattering processes.
Figure \ref{fig:hist} illustrates how interactions mediated by the EM
field tend to broaden the distribution of collective linewidths in a
$33\times 33$ lattice of SRRs.
In Fig.~\ref{fig:hist}(a), where the lattice spacing is $a = 0.5\lambda$, the
radiative emission rates range from the very subradiant $1.2 \times
10^{-5} \Gamma$ to the superradiant $11\Gamma$, where $\Gamma$ is the decay rate of a single meta-atom in isolation.
Figure~\ref{fig:hist}(b), on the other hand, illustrates how the collective effects are
diminished when the lattice spacing $a=1.4\lambda$ exceeds a
wavelength.
The distribution of decay rates is considerably narrower with the
decreased inter-SRR interactions associated with lattice spacings
exceeding a wavelength.
Although the effects of collective interactions are significantly
reduced, they do not disappear entirely.
The radiative decay rates still range from $0.2\Gamma$ to $3\Gamma$.

The dramatically narrowed radiative resonance linewidth of some
of the collective modes and the sensitive dependence of the
narrowing on the spatial separation of the resonators indicates a
strong cooperative response of the system to EM fields.
For very closely-spaced resonators multiple scattering is
considerably influenced by recurrent scattering events in which the
field repeatedly scatters from the same meta-atoms.
In the example studied here,
this leads to the resonance linewidth narrowing of almost five
orders of magnitude.  Such narrowing
could not have been described by independent scatterer approach.

The recurrent scattering that is responsible for the dramatic linewidth narrowing
can be characterized by repeated scattering events between pairs of scatterers, triplets
of scatterers, etc.\cite{vantiggelen90,MoriceEtAlPRA1995,RuostekoskiJavanainenPRA1997L,
RuostekoskiJavanainenPRA1997,JavanainenEtAlPRA1999,fermiline} In the present work we have not analyzed the relative contribution of the different
processes to the distribution of linewidths. In the case of electric dipole scatterers the contribution,
for instance, of repeated exchanges of a photon between pairs of dipoles to the distribution of resonance
linewidths was studied in Ref.~\onlinecite{GoetschySkipetrovPRE2011}. A similar calculation could in principle be performed in our system, although
the interplay between the magnetic and electric dipoles may notably complicate the analysis. 

An alternative approach to quantify the contribution of different recurrent scattering processes was performed in Ref.~\onlinecite{JavanainenEtAlPRA1999}. Numerical simulation results were compared with the equations for correlation functions. One, in essence, constructs a hierarchy of equations in which the
$n$th level describes the  recurrent scattering between subsets of $n$ discrete resonators. Truncating the hierarchy after the $n$th level may therefore be used to quantify the contribution of the $n$th order recurrent scattering. In the case of randomly distributed, uncorrelated scatterers, the role of recurrent scattering between $n$ resonators scales with the $n$th power of density.\cite{JavanainenEtAlPRA1999,vantiggelen90,MoriceEtAlPRA1995,RuostekoskiJavanainenPRA1997L,
RuostekoskiJavanainenPRA1997,fermiline} Correlations in the positions of the scatterers modify this density dependence.\cite{JavanainenEtAlPRA1999,fermiline} It was found for the both correlated and uncorrelated samples\cite{JavanainenEtAlPRA1999} that changes in scattering resonance properties as a function of the density of scatterers corresponded to the increased role of recurrent scattering; at higher densities the higher order recurrent scattering processes become increasingly more important leading to the emergence of more strongly subradiant modes.\cite{RuostekoskiJavanainenPRA1997L,RuostekoskiJavanainenPRA1997,JavanainenEtAlPRA1999}

\begin{figure}
  \centering
  \includegraphics[width=8.6cm]{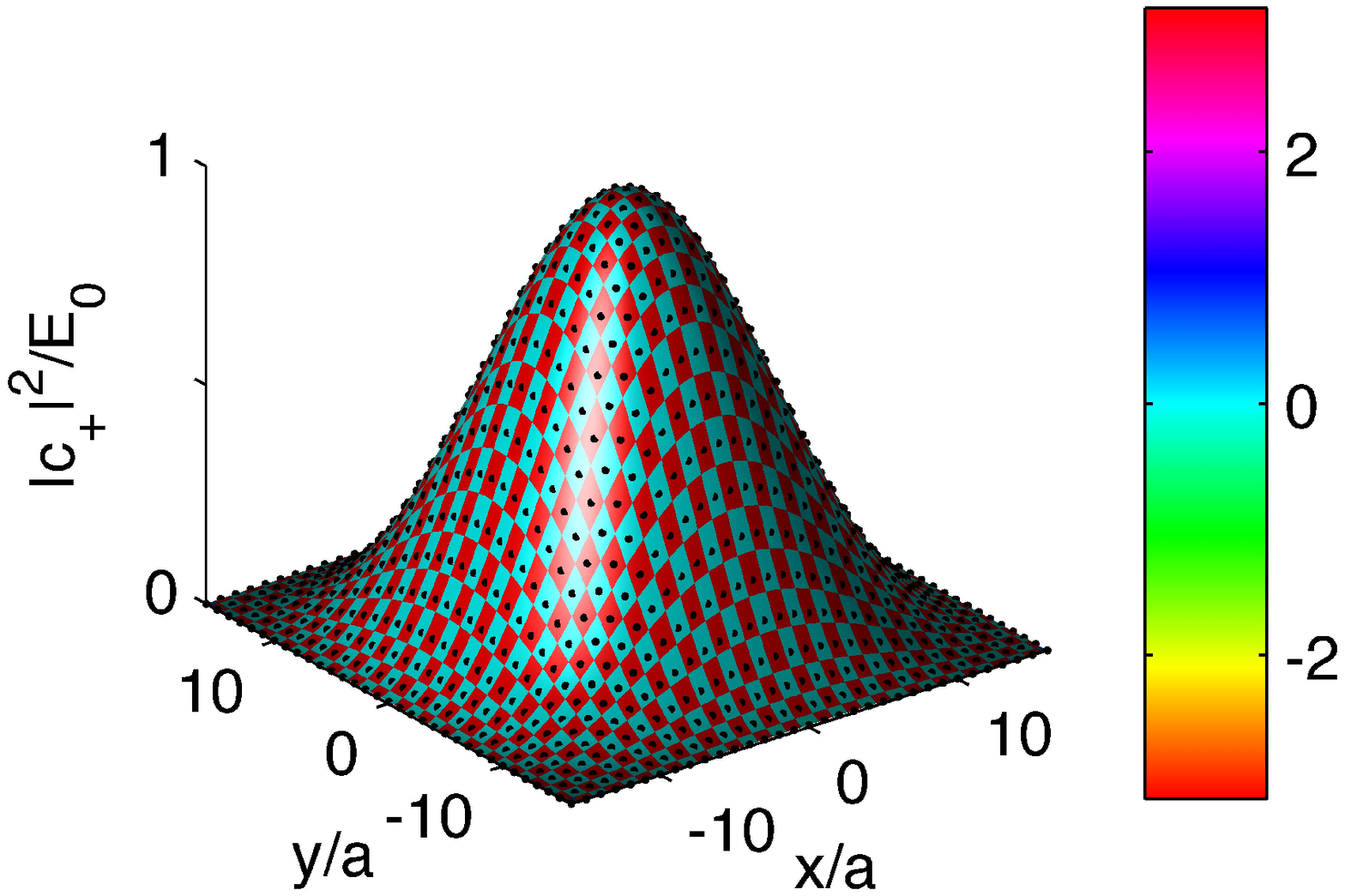}
  \caption{(color online) An illustration of the most subradiant of the collective modes in
    a $33 \times 33$ array of SRRs.
    The height of the surface represents the energy of the SRR
    symmetric oscillations $|c_+|^2$ normalized to the peak SRR energy
    $E_0 = \operatorname{max}_{\ell} (|c_{+,\ell}|^2 + c_{-,\ell}|^2)$.
    The colored patches indicate the phase of the electric dipole
    oscillations for each SRR.
    The black dots indicate the position of each SRR, while their
    height indicates the normalized total energy within each unit cell,
    so that the energy in the magnetic dipole oscillations is given by
    the difference between the black dots' height and the height of
    the surface.
    This subradiant mode is ferroelectric in nature, and has a
    radiative emission rate $1.2 \times 10^{-5} \Gamma$.
    The lattice spacing $a=0.5\lambda$, the
    meta-atom separation within the SRRs $u=0.12 \lambda$, and
    $\Gamma_{\mathrm{E}} = \Gamma_{\mathrm{M}}$.
  }
  \label{fig:subRadMode}
\end{figure}

We now examine the characteristics of some of the collective modes in
a $33\times 33$ lattice with an inter-SRR separation of $a = 0.5
\lambda$.
As with a single SRR, we can characterize the state of the system by
specifying a complex amplitude for both the symmetric (electric) and
antisymmetric (magnetic) oscillations.
Where the state of the system is fully specified by the vector of single meta-atom
amplitudes $(b_1,b_2,\ldots, b_{2N_xN_y})^T$,
we represent the electric and magnetic oscillations of a single SRR,
labelled by $\ell=1,\ldots,N_xN_y$, as $c_{+,\ell}$ and $c_{-,\ell}$,
respectively, where
\begin{equation}
  \label{eq:c_pm_l_def}
  c_{\pm,\ell} = \frac{1}{\sqrt{2}} \left(b_{2\ell-1} \pm b_{2\ell}\right) \text{.}
\end{equation}

As noted earlier, the subwavelength proximity of adjacent SRRs permits the creation of extremely subradiant collective modes.
We illustrate the most subradiant of these modes for a lattice spacing of $a=0.5\lambda$ in Fig.~\ref{fig:subRadMode}.
The energy of this mode resides almost exclusively in symmetric oscillations of the SRRs.
However, although the meta-atoms in each SRR oscillate symmetrically,
the electric dipole of each unit-cell resonator element points in the opposite direction to that of its nearest neighbor.
This mode is antiferroelectric in nature.
The phase of each electric dipole, indicated by the color of the unit cell,
forms a checkerboard pattern in the phase
profile.
This mode consists of more strongly excited electric dipole
oscillations in the center of the array with smaller contributions
from SRRs on the edges.
When this mode is excited, the fields emitted from the SRRs tend to remain trapped in the ensemble as they repeatedly scatter from one meta-atom to another.
The scattered fields will leak out if this mode very slowly as indicated by the collective emission rate of $1.2\times 10^{-5}\Gamma$.

\begin{figure}
  \centering
  \includegraphics[width=8.6cm]{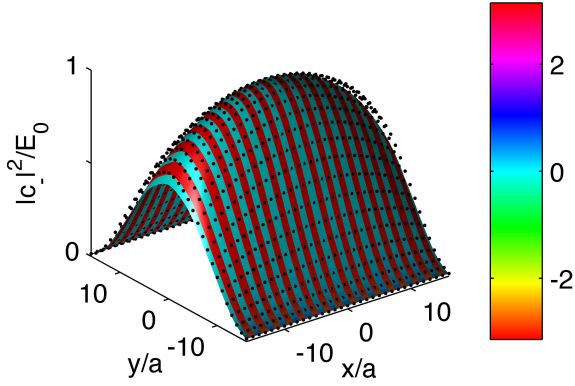}
  \caption{(color online) An illustration of the most superradiant of
    the collective modes in a $33 \times 33$ array of SRRs.
    The height of the surface represents the energy of the SRR
    antisymmetric oscillations $|c_-|^2$ normalized to the peak
    SRR energy
    $E_0 = \operatorname{max}_{\ell} (|c_{+,\ell}|^2 + c_{-,\ell}|^2$.
    The black dots indicate the position of each SRR, while their
    height indicates the normalized total energy within each unit cell.
    The total excitation, $|c_{+,\ell}|^2$, of the electric dipole oscillations is given by the difference between the black dots' height and the height of
    the surface.
    All parameters of the ensemble are as in
    Fig.~\ref{fig:subRadMode}.
    This superradiant mode has a radiative emission rate of about
    $11\Gamma$ and consists largely of magnetic dipole
    oscillations whose phase variation is matched with EM waves
    propagating in the $\pm x$-directions.
   }
  \label{fig:superRadMode}
\end{figure}

The most superradiant of the collective modes, shown in
Fig.~\ref{fig:superRadMode}, by contrast couples very strongly to
radiation propagating away from the ensemble.
This mode is almost entirely magnetic in nature with the SRRs
oscillating antisymmetrically.
These magnetic dipole oscillations consist of stripes of constant
phase in the $y$-direction, while the phase variation in the
$x$-direction is phase matched with radiation propagating along $\pm
\unitvec{e}_x$.
An EM plane wave propagating in the $\pm x$-direction whose magnetic field is polarized in the $z$-direction would have an electric field polarized along $\pm \unitvec{e}_y$.
Since the electric dipoles in this most superradiant of modes are largely unexcited,  this mode radiates into an equal superposition of EM fields propagating in the positive and negative $x$-directions.
The collective excitation coupling to these propagating fields results
in a spontaneous emission rate of $11\Gamma$, more than ten times
the single meta-atom emission rate.

In many experimental situations, however, a plane wave incident field,
with nearly uniform phase and intensity in the metamaterial plane,
drives the ensemble.
The incident field propagates perpendicular to the plane of the
metamaterial along the $z$-direction so that it drives the SRRs in
phase.
It is therefore worthwhile to examine modes whose oscillations are
phase matched with the incident field since they can be addressed
directly.
The two modes of interest are the uniform electric mode, with all electric dipoles oscillating in phase,  and
the uniform magnetic mode, where all magnetic dipoles
oscillate in phase.

\begin{figure}
  \centering
  \includegraphics[width=8.6cm]{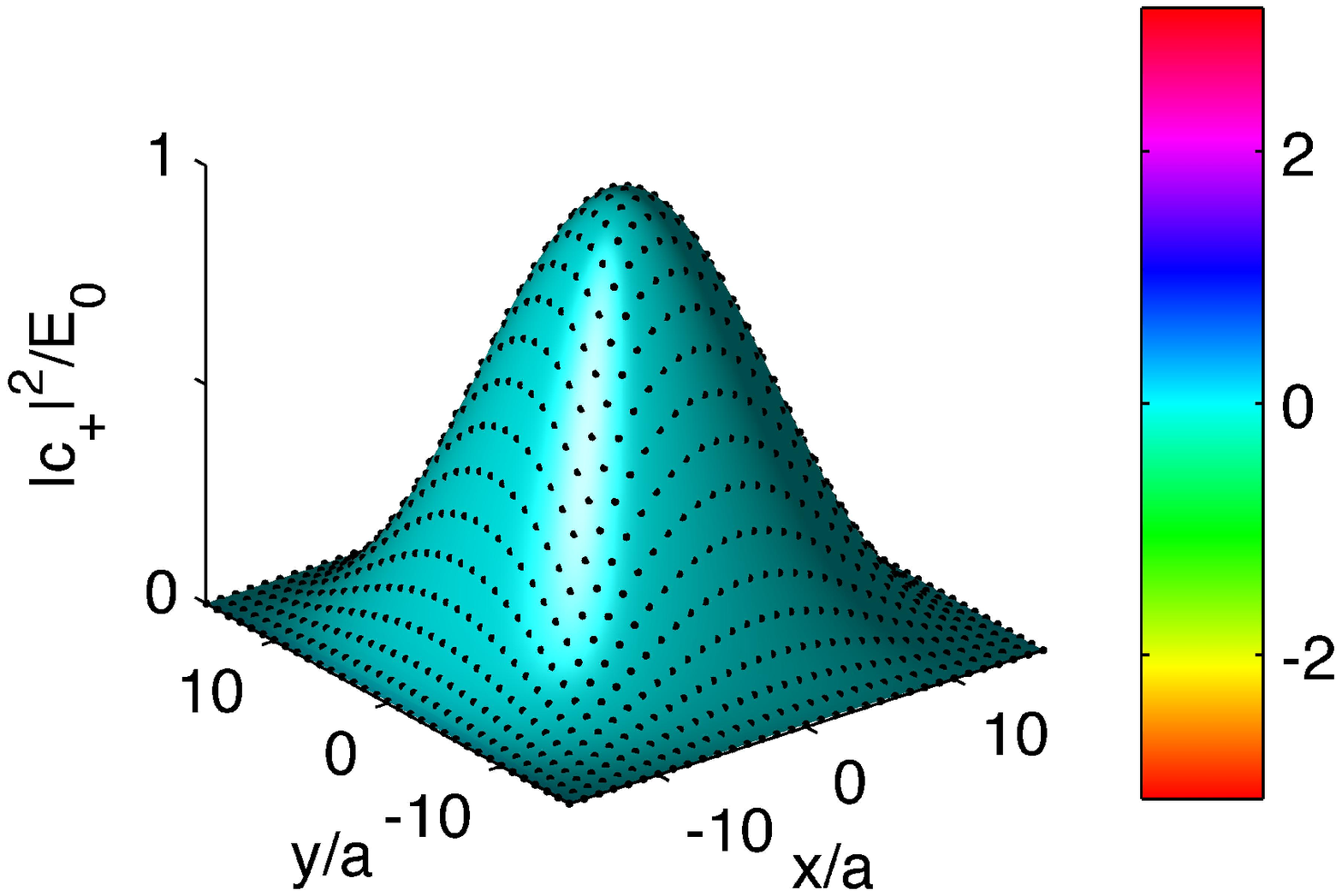}
  \caption{(color online) An illustration of the uniform electric collective mode in
    a $33 \times 33$ array of SRRs.
    The height of the surface represents the excitation energy of the SRR
    symmetric oscillations $|c_+|^2$ normalized to the peak SRR energy
    $E_0 = \operatorname{max}_{\ell} (|c_{+,\ell}|^2 + c_{-,\ell}|^2$.
    The colored patches indicate the phase of the electric dipole
    oscillations for each SRR.
    The black dots indicate the position of each SRR, while their
    height indicates the normalized total energy within each unit cel.
    All parameters of the ensemble are as in
    Fig.~\ref{fig:subRadMode}.
    This mode consists of the split ring electric dipoles oscillating
    in phase and has a radiative emission rate of approximately the
    single meta-atom emission rate $\Gamma$.
  }
  \label{fig:uniformElecMode}
\end{figure}

Figure \ref{fig:uniformElecMode} shows the structure of the uniform
electric mode.
As desired, an excitation in this mode has its energy almost purely in
electric dipole oscillations of the split rings.
Furthermore, because all electric dipoles oscillate in phase, this
mode efficiently couples to EM fields propagating out of the plane
along $\pm \unitvec{e}_z$ whose electric field polarization is along
the electric dipoles $\unitvec{d} = \unitvec{e}_y$.
Because the fields scattered by this mode propagate out of the plane,
excitation of the mode by an incident plane wave results in reflection
of the incident field from the metamaterial.
In the geometry considered here, the uniform electric mode has a
radiative decay rate of $\gamma_e \approx \Gamma$, about as strong as
the single meta-atom decay rate.

\begin{figure}
  \centering
  \includegraphics[width=8.6cm]{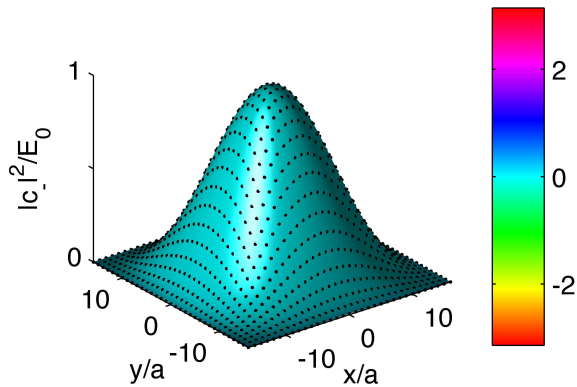}
  \caption{An illustration of the uniform magnetic collective mode in a
    $33 \times 33$ array of SRRs.
    The height of the surface represents the energy of the SRR
    antisymmetric oscillations $|c_-|^2$ normalized to the peak SRR energy
    $E_0 = \operatorname{max}_{\ell} (|c_{+,\ell}|^2 + c_{-,\ell}|^2$.
    The black dots indicate the position of each SRR, while their
    height indicates the normalized total energy within each unit cell.
    All parameters of the ensemble are as in
    Fig.~\ref{fig:subRadMode}.
    This mode is composed of all SRRs oscillating antisymmetrically,
    producing magnetic dipoles in phase.
    The radiative emission rate of this mode is $0.02\Gamma$.
  }
  \label{fig:uniformMagMode}
\end{figure}

The second phase matched mode, the uniform magnetic mode, is
illustrated in Fig.~\ref{fig:uniformMagMode}.
This uniform mode is almost purely magnetic in nature, with all of the
metamolecule magnetic moments oscillating in phase, producing a sheet
of magnetization pointing out of the metamaterial.
In contrast to the uniform electric mode, however, this mode
cannot strongly couple to fields propagating out of the plane.
In fact, we have found that for lattice spacings sufficiently less than a
wavelength, scattered radiation remains trapped in the ensemble and
this mode is subradiant.
Here, with a lattice spacing of $a = 0.5\lambda$, the radiative
emission rate is suppressed by about a factor of 50 below the single
meta-atom decay rate.
The form of the magnetic mode does not differ substantially
  from that in Fig.~\ref{fig:uniformMagMode} for larger lattice
  spacings; however, inter-resonator spacing affects cooperative
  interactions and strongly influence the mode's decay
  rate\cite{JenkinsLineWidthArxiv}. 
In Ref.~\onlinecite{JenkinsLineWidthArxiv}, it was shown how a subradiant mode
analogous to the uniform magnetic mode we discussed here is
responsible for the transmission resonance observed in an array of
ASRs.\cite{FedotovEtAlPRL2010}

The calculated collective modes of the system also determine the propagation
dynamics of localized excitations. The propagation of excitations are influenced
by strong interactions between the resonators. Specifically, in disordered systems,
where the locations of scatterers vary randomly, the transition to localization
can be characterized from transport properties.\cite{vanTiggelen99} In the studied system, the positions
of the resonators are fixed, so the propagation dynamics is determined by the particular
excitation. An initial excitation of SRR dipoles will be comprised of some linear
combination of collective modes. The more radiant components will quickly decay, leaving behind only
the contributions from subradiant modes which oscillate at differing
frequencies. This behavior manifests itself as a decaying propagation and
spreading of current oscillations through the metamaterial as EM fields scatter in the array.
The lifetime of the residual excitation strongly depends on the presence of recurrent scattering and subradiant modes.

In order to demonstrate the time dynamics of excitations we have
  studied 
  the specific example
of an excitation of the left-most strip of magnetic dipoles along the $y$ axis in the square
array. Such a pattern will lose $90\%$ of its energy, and propagate a single
lattice site in a time $t=10/\Gamma$ for a lattice spacing of $a=0.5\lambda$.
As the excitation propagates, it begins to broaden so that at time $t = 500 /\Gamma$, the remaining excitation, containing $2\times
10^{-4}$ of the initial energy, has 
spread through the sample. 
When the lattice spacing is larger, $a=1.5\lambda$, the
  excitation 
  spreads more quickly
through the sample (at time $20/\Gamma$), indicating weaker EM-mediated interactions between the resonators.
In this case only $5\times 10^{-8}$ of the initial energy has not been
 radiated away.

\section{Quantizing the metamaterial dynamics}
\label{sec:quant-metam-dynam}

In this article, we have developed a general formalism to describe
collective oscillations in ensembles of meta-atoms which comprise a
metamaterial.
In systems where thermal losses are suppressed and can be neglected,
however, this formalism can easily be quantized.
In the quantized system, the meta-atom dynamic variables $Q_j$ and
their conjugate momenta $\phi_j$, whose Poisson brackets are
$\{Q_j,\phi_{j'}\} = \delta_{j,j'}$, become quantum mechanical
operators $\hat{Q}_j$ and $\hat{\phi}_j$ which obey the commutation
relations
\begin{subequations}
  \label{eq:QuantCommRels}
  \begin{eqnarray}
    \left[\hat{Q}_j, \hat{Q}_{j'} \right] &=& \left[\hat{\phi}_j,
      \hat{\phi}_{j'} \right]  = 0 \\
    \left[\hat{Q}_j, \hat{\phi}_{j'} \right] & = & i\hbar \delta_{j,j'}
  \end{eqnarray}
\end{subequations}
When quantizing the system, the classical normal variables undergo the
transformations $b_j \rightarrow \sqrt{\hbar}\hat{b}_j$ and $b_j^\ast
\rightarrow \sqrt{\hbar}\hat{b}_j^\dag$.
The normal variables thus become harmonic oscillator creation and
annihilation operators which obey the commutation relations
\begin{subequations}
  \begin{eqnarray}
    \label{eq:bComm}
    \left[\hat{b}_j, \hat{b}_{j'}\right] & = & \left[\hat{b}_j^\dag,
      \hat{b}_{h'}^\dag\right] = 0 \\
    \left[\hat{b}_j, \hat{b}_{j'}^\dag \right] & = & \delta_{j,j'}
  \end{eqnarray}
\end{subequations}
 Similarly, the normal variables for the EM field
[Eqs.~\eqref{eq:elecDisp} and \eqref{eq:magField}] transform as
$a_{\qv,\lambda} \rightarrow \sqrt{\hbar}\hat{a}_{\qv,\lambda}$.
  The EM field normal variables then commute with those of the
  meta-atoms and satisfy the commutation relations
  \begin{subequations}
    \begin{eqnarray}
      \label{eq:aComm}
      \left[\hat{a}_{\spvec{q},\lambda},
        \hat{a}_{\spvec{q}',\lambda'}\right] & = &
      \left[\hat{a}_{\spvec{q},\lambda}^\dag,
        \hat{a}_{\spvec{q}',\lambda'}^\dag\right] = 0 \\
      \left[\hat{a}_{\spvec{q},\lambda},
        \hat{a}_{\spvec{q}',\lambda'}^\dag \right] & = &
      \delta_{\lambda,\lambda'}  \delta(\spvec{q} - \spvec{q}') \text{\,.}
    \end{eqnarray}
  \end{subequations}
The ability to easily quantize this formalism may be useful in
describing the interactions of low loss metamaterials with
nonclassical fields.
Furthermore, generalizations  of the formalism to nonlinear
metamaterials, e.g., involving superconductors, may in and of
itself produce nonclassical cooperative effects.

\section{Conclusions}
\label{sec:conclusions}

In conclusion, we
developed a theoretical formalism to describe cooperative
interactions of a magnetodielectic metamaterial sample with an EM field.
We modeled the metamaterial as an ensemble of discrete
EM resonators, or meta-atoms, that each support a single mode of current
oscillation. The meta-atoms
could, for example, be subwavelength circuit elements which
support plasmonic oscillations.
From a Lagrangian describing dynamics of the EM field and its
interactions with systems of charged particles,
we derived the conjugate momenta for the EM field and meta-atom
  dynamic variables, as well as Hamiltonian for the metamaterial system.
Hamilton's equations of motion then describe a coupled
dynamics between the meta-atoms and the EM field.

We showed how the EM fields are emitted from excited current
oscillations within each meta-atom, and in turn, how the EM fields
drive the meta-atom dynamics.
A single meta-atom interacting with its own self-generated field
behaves as a radiatively damped LC circuit. In an ensemble of resonators, the meta-atoms also interact with each other.
Initially excited by an external field, a meta-atom emits EM
radiation which then impinges on other meta-atoms.
The other meta-atoms then re-scatter the field. Multiple scattering
events mediate an interaction
between the meta-atoms' current oscillations.
The interactions culminate in a discrete, coupled set of
equations for the meta-atoms which describe the collective
metamaterial dynamics.
The coupled dynamics constituted the main results of this article.

In Sec.~\ref{sec:analysis-model}, we examined the collective
dynamics in a regime where the influence of a meta-atom's
self-generated fields dominates over that of the incident field or
the fields scattered by all other meta-atoms in the metamaterial.
This assumption allowed us to employ the rotating wave approximation
to simplify the description of the dynamics.
Appendix~\ref{sec:ensemble-meta-atoms}, on the other hand,
generalized the formalism to provide for a dynamical description
outside the limits of the RWA.

A metamaterial possesses as many collective modes as there are
  meta-atoms in the sample, each with its own
resonance frequency and decay rate.
These collective modes can behave very differently
from oscillations in a single, isolated meta-atom.
The cooperative interactions could result in superradiant modes in
which energy is radiated away more quickly than an ensemble of
meta-atoms acting independently.
Other modes, by contrast, are subradiant,
for which the mode's radiative emission rate is
suppressed.
As an example, we examined the dynamics of a planar metamaterial
  formed from a $33 \times 33$ square lattice of SRRs.
  When the resonators are closely spaced the collective modes have a
  broad distribution of radiative decay rates.
  For a lattice spacing of $0.5\lambda$, cooperative interactions
  suppress the most subradiant mode's emission rate by about five
  orders of magnitude, while the most super-radiant
  mode radiates eleven times faster than a single meta-molecule.
  Finally, we also provided an example how the propagation dynamics of excitations in a metamaterial array 
can be analyzed using the collective eigenmodes. We found that the lattice spacing, and hence the interactions 
between the resonators, strongly influence the rate at which excitations spread over the array. 
  In addition to SRRs, the formalism we developed could be used to
  describe interactions between emitters with other geometries,
  e.g. dielectric spheres \cite{BergmanStroudPRB1980}.

The collective dynamics derived from the discrete resonator
model can be successfully employed to explain
experimentally observed phenomena.
For example, in Ref.~\onlinecite{JenkinsLineWidthArxiv}, we used
this model to calculate the resonance
linewidth narrowing as a function of the system size, associated with
the experimental observations of the transmission resonance by Fedotov
\textit{et al}.\cite{FedotovEtAlPRL2010}.
The theoretical model provided an excellent agreement with experimental
  findings.
  This example illustrates how the formalism developed here lays the
  ground work allowing one to model collective dynamics in large
  metamaterial systems in which finite-size effects or irregularities
  may play a role.

\begin{acknowledgements}
  We would like to thank the EPSRC and the Leverhulme trust for
  financial support. We also thank N.\ Papasimakis, V.\ Fedotov, M.\ O.\ Borgh, and
  N.\ Zheludev for discussions.
\end{acknowledgements}
\appendix

\section{The Lagrangian and the Power-Zienau-Woolley transformation}
\label{sec:lagrangian}

In this Appendix, we derive the Lagrangian describing the dynamics of
meta-atoms interacting with the EM field given in
Eq.~\eqref{eq:Lagrange2}.
We start from the standard Lagrangian for the EM field in
the Coulomb gauge interacting with arbitrary charge and current
distributions.
Then, using the Power-Zienau-Woolley transformation,\cite{PowerZienauPTRS1959,
  PowerBook, PZW} we express
the equivalent Lagrangian in terms of polarization and magnetization
densities.
Given the expressions for the polarization and magnetization
densities in Eq.~\eqref{eq:chargeAndCurrentDensities}, we express the
Lagrangian in terms of effective magnetic fluxes and EMFs as in
Eq.~\eqref{eq:Lagrange2}.

An arbitrary vector field $\spvec{V}(\rv)$ can be decomposed into its
longitudinal $\spvec{V}_{\parallel}$ and transverse $\spvec{V}_\perp$
components
\beq
\spvec{V}=\spvec{V}_{\parallel} +
\spvec{V}_{\perp}\,,
\eeq
defined such that $\nabla \times \spvec{V}_\parallel \equiv 0$ and  $\nabla \cdot \spvec{V}_\perp \equiv 0$.
In the Coulomb gauge the EM vector potential is set purely transverse
by requiring that $\nabla\cdot \spvec{A}(\spvec{r})=0$.
It follows from Maxwell's equations that $\spvec{B}(\spvec{r})$ is
purely transverse and that the longitudinal component of the electric
field $\spvec{E}_\parallel$ is not a true dynamical variable, but is
given by an algebraic relation by the charge density.\cite{CohenT} In
particular, we may write
\begin{equation}
  \label{eq:E_long}
  \spvec{E}_{\parallel} = -\nabla U\,,
\end{equation}
where
\begin{equation}
  U(\rv) = \frac{1}{4\pi\epsilon_0} \int d^3r'\, \frac{\rho(\rv')}{\left| \rv -\rv'\right|}
\end{equation}
is the scalar potential.
The Coulomb energy $V_{\mathrm{Coul}}$ is given in terms of the
meta-atom charge densities in Eq.~\eqref{eq:V_Coul}, and can be
expressed directly in terms of $\spvec{E}_\parallel$ as
\begin{equation}
  \label{eq:V_Coul_E_par}
  V_{\mathrm{Coul}}  = \frac{\epsilon_0}{2} \int d^3r\, |\spvec{E}_\parallel|^2\,.
\end{equation}
The transverse component of
the electric field is given in terms of
the vector potential $\spvec{A}$ as
\begin{equation}
  \spvec{E}_\perp =  -\dot{\spvec{A}}\,.
\end{equation}

The standard Lagrangian in the Coulomb gauge may be written as
\begin{equation}
  \label{eq:StandardLagrange}
  \mathcal{L}_{\mathrm{C}} = \mathcal{K} - V_{\mathrm{Coul}}+
  \mathcal{L}_{\mathrm{EM}} + \mathcal{L}_{\mathrm{I}} \, \text{,}
\end{equation}
where
\begin{equation}
  \mathcal{L}_{\mathrm{I}} = \int d^3r\, \spvec{j}(\spvec{r},t) \cdot
  \spvec{A}(\spvec{r},t)
  \label{eq:Lint}
\end{equation}
accounts for the interaction between the matter and the free
EM field, and
$\spvec{j} \equiv \sum_j \spvec{j}_j$ is the
total current density with the contribution from meta-atom $j$.
The meta-atom current densities $\spvec{j}_j$ are given in terms of the generalized velocities $I_j =
\dot{Q}_j$ by Eq.~\eqref{eq:chargeAndCurrentDensities}.
The vector potential $\spvec{A}(\spvec{r},t)$ provides the continuum
of dynamic variables describing the evolution of the EM field.
The EM field dynamics in the absence of charge and
current sources is governed by the Lagrangian, $\mathcal{L}_{\mathrm{EM}}$ [Eq.~\eqref{eq:L_EM}].
The charge carriers that give rise to the charge and current
densities have an inertia, and hence the current in a meta-atom,
resulting from the motion of these carriers, must have an associated
kinetic energy.
This kinetic energy $\mathcal{K}$ is given in terms of
phenomenological inertial inductances in Eq.~\eqref{eq:KinEnergy}.

The canonical momentum for the fields in the Coulomb gauge is given in
terms of the time derivative of the vector potential and is
proportional to the transverse component of the electric field
\begin{equation}
  \Pi^{(\mathrm{C})}(\spvec{r}) \equiv
    \frac{\partial \mathcal{L}_{\mathrm{C}}}{\partial \dot{\spvec{A}}}
    = \epsilon_0 \dot {\rm A} ( \spvec{r})= - \epsilon_0 \spvec{E}_\perp (\spvec{r)
}\,.
\end{equation}
Similarly, the canonical momentum corresponding to the charges $Q_j$
is given by
\begin{equation}
  \phi_j^{(\mathrm{C})} \equiv \frac{\partial \mathcal{L}_C}{\partial I_j}
  = l_j I_j + \chi_j(t)  \text{ ,}
\end{equation}
where
\begin{equation}
  \chi_j \equiv \int d^3r\, \spvec{A}(\rv,t) \cdot
    \left[\spvec{p}_j(\rv) + \nabla \times \spvec{w}_j(\rv)\right] \text{.}
\end{equation}
The factor $\chi_j$ originates from the interaction Lagrangian
$\mathcal{L}_{\mathrm{I}}$ [Eq.~\eqref{eq:Lint}]; its specific form
arises from how the current density $\spvec{j}_j$ within each
meta-atom $j$ depends on that meta-atom's generalized velocity $I_j$
[see Eq.~\eqref{eq:currentDensity}].
This factor represents an averaged projection of the vector potential
onto the current oscillation's mode functions $\spvec{p}_j$ and
$\spvec{w}_j$.
The Hamiltonian in the Coulomb gauge may then be derived from the Lagrangian [Eq.~\eqref{eq:StandardLagrange}]
\beq
 \mathcal{H}^{(C)} = {1\over 2 l_j} [\phi_j - \chi_j]^2 +
  \mathcal{H}_{\mathrm{EM}}^{(C)} + V_{\mathrm{Coul}}
\label{eq:minimalcoupling}
\eeq
where the energy of the transverse EM field, or the radiation field, is responsible for the
excitations of the meta-atoms
\beq
{\cal H}_{\rm EM}^{(\mathrm{C})}= \frac{\epsilon_0}{ 2} \int d^3 r\, \left[\spvec{E}^2_\perp (\spvec{r})+ c^2 \spvec{B}^2(\spvec{r})\right]\,.
\eeq

The quantity $\chi_j$
originates
from the assumption
that a
mode of current oscillation
depends on
a single dynamic variable
with units of charge.
The amplitude of the charge distribution may change in time, but its
spatial distribution will not.
By contrast, in the more familiar scenario where one
describes the motion of particles with \emph{fixed} charge $q_j$ at a
\emph{time varying} position $\rv_j(t)$, the conjugate momentum for
the position coordinates is given by the vector $\dot{\spvec{r}}_j +
q_j \spvec{A}(\rv_j(t))$.
The scalar quantity $\chi_j$ arising from our model plays the same
role as the quantity $q_j\spvec{A}(\rv_j(t))$ appearing in the
familiar minimal coupling Hamiltonian for moving charged particles.

Although Eq.~\eqref{eq:minimalcoupling} is analogous to the standard
minimal coupling Hamiltonian description of charged particles in an EM
field, it does not turn out to be the most suitable representation to
study the interaction of discrete scatterers with the EM field.
We find it convenient to express the dynamics in terms of polarization
and magnetization densities rather than charge and current densities.
In this way, when the circuit elements are much smaller than a
wavelength of EM field with which they interact, we may more easily treat
the dynamics in terms of interacting electric and magnetic
multi-poles.
To that end, we employ the Power-Zienau-Woolley transformation.
\cite{PZW}
For any globally neutral charge distribution with respective charge
and current density $\rho$ and $\spvec{j}$, there exists a
corresponding polarization $\spvec{P}$ and magnetization
density $\spvec{M}$ such that
\begin{subequations}
  \begin{align}
    \rho(\rv,t) &= - \nabla \cdot \spvec{P}(\rv,t) \label{eq:rhoasP}
    \\
    \spvec{j}(\rv,t) & = \dot{\spvec{P}}(\rv,t) + \nabla \times
    \spvec{M}(\rv,t)  \label{eq:jasPandM}
  \end{align}
  \label{eq:PandMDefs}
\end{subequations}
Here,
the polarization density is a function of the dynamic
variables $Q_j$  and the magnetization density is a function of their
rates of change $I_j$
[Eq.~\eqref{eq:PandMDensDefs}].
One can modify the Lagrangian by adding the total time derivative
$dF/dt$ of a function to the original Lagrangian.
Here, we take
\begin{equation}
  \label{eq:Fdef}
  F = - \int d^3 r \,\spvec{P}(\rv,t) \cdot \spvec{A}(\rv,t) \,\text{,}
\end{equation}
and the equivalent Lagrangian in the {\em length} gauge is thus
\begin{equation}
  \mathcal{L} = \mathcal{L}_{\mathrm{C}} +
  \frac{dF}{dt} \label{eq:PZWTransform}  \text{.}
\end{equation}
Because $F$ is only a function of the dynamic variables $Q_j$ and
$\spvec{A}$, the Lagrange equations of motion are invariant under this
transformation.
Explicitly, adding $dF/dt$ to the interaction term $\mathcal{L}_I$  yields
\begin{equation}
  \label{eq:dFdt}
  \mathcal{L}'_{\mathrm{I}} \equiv \mathcal{L}_I + \frac{dF}{dt} =
  \int d^3r\, \left(\spvec{j} - \dot{\spvec{P}} \right)\cdot
  \spvec{A} - \int d^3r\, \spvec{P} \cdot \dot{\spvec{A}}  \,\text{.}
\end{equation}
From Eq.~\eqref{eq:jasPandM}, the first integral in
Eq.~\eqref{eq:dFdt} can be expressed as
\begin{equation}
  \label{eq:intermediate1}
  \int d^3r\, \left(\spvec{j} - \dot{\spvec{P}} \right)\cdot
  \spvec{A} = \int d^3r \spvec{A}\cdot \left(\nabla \times
    \spvec{M}\right) \text{.}
\end{equation}
Integrating this by parts, we obtain the interaction Lagrangian
\begin{equation}
  \label{eq:intLagrange}
  \mathcal{L}_{\mathrm{I}}' = - \int d^3r \spvec{B} \cdot
  \spvec{M} - \int d^3r \dot{\spvec{A}} \cdot \spvec{P} \,\text{.}
\end{equation}
To evaluate the second integral, we recognize that
$-\dot{\spvec{A}} =
\spvec{E} + \nabla U$, where $U(\rv,t)$ is the electric scalar
potential.
The last integral appearing in Eq.~\eqref{eq:intLagrange} thus becomes
\begin{equation}
  \label{eq:intermediate2}
  -\int d^3r  \dot{\spvec{A}} \cdot \spvec{P} = \int d^3r
  \spvec{E}\cdot \spvec{P} + \int d^3r \spvec{P}\cdot \nabla U \text{.}
\end{equation}
We integrate the last term of Eq.~\eqref{eq:intermediate2} by parts,
and because $U$ is the Coulomb gauge scalar potential, we obtain
\begin{align}
  \int d^3r \spvec{P}\cdot \nabla U &= -\int d^3r\,
  (\nabla\cdot\spvec{P}) U = \int d^3r\, \rho U \nonumber\\
  & = 2 V_{\mathrm{Coul}} \,\text{.}
  \label{eq:toVCoul}
\end{align}
Therefore, the Lagrangian in the Power-Zienau-Woolley picture can be
expressed in terms of the total electric and magnetic fields as
\begin{equation}
  \mathcal{L} = \mathcal{K} + V_{\mathrm{Coul}} +
  \mathcal{L}_{\mathrm{EM}} + \int d^3 r \left[\spvec{B}\cdot
    \spvec{M} + \spvec{E} \cdot \spvec{P}\right] \, \text{.}
  \label{eq:LquasiFin}
\end{equation}
Although we derived the Lagrangian in Eq.~\eqref{eq:LquasiFin} for a
system composed of ensembles of circuit elements, its form is valid
for any system of charges where the charge density is described by any
generalized dynamic variables and the current density is a function of
their generalized velocities.
In our system, the total polarization $\spvec{P} = \sum_j \spvec{P}_j$
and magnetization $\spvec{M} = \sum_j \spvec{M}_j$, with the
corresponding densities $\spvec{P}_j$ and $\spvec{M}_j$ expressed in
terms of the dynamic variable $Q_j$ and velocity $I_j$ for meta-atom
$j$ given by Eqs.~\eqref{eq:PandMDensDefs}.
Thus, in an ensemble of meta-atoms, the system Lagrangian is given by
Eq.~\eqref{eq:Lagrange2}.

\section{Elimination of instantaneous, non-local interactions in the Power-Zienau-Woolley picture}
\label{sec:elim-inst-non}

In this Appendix, we provide details of the derivation of the
Hamiltonian in the \emph{length} gauge obtained by the
Power-Zienau-Woolley transformation.
The derivation is analogous to the one discussed in
Ref.~\onlinecite{CohenT} in determining the Power-Zienau-Woolley
Hamiltonian for systems of charged particles. We begin by examining
the portion of the Hamiltonian, $\mathcal{H}_{\mathrm{E}}$ [Eq.~\eqref{eq:H_EDef}],
\begin{equation}
  \label{eq:H_EreDef}
  \mathcal{H}_{\mathrm{E}} = \int d^3r\,  \spvec{E}_{\perp} \cdot
  \spvec{D}
  - \int d^3r\, \spvec{E}\cdot \spvec{P} - V_{\mathrm{Coul}}
  -\mathcal{L}_{\mathrm{EM}}\,.
\end{equation}
We will show how the Coulomb potential is absorbed by the
Power-Zienau-Woolley Hamiltonian.
We will express each term in $\mathcal{H}_{\mathrm{E}}$ in terms of
the displacement field $\spvec{D}$, and the
polarization density $\spvec{P}$.
The various components then combine to yield the Hamiltonian for the
free EM field, $\mathcal{H}_{\mathrm{EM}}$ [Eq.~\eqref{eq:HEM}],
the local polarization contact interaction, and an interaction between the polarization density and the displacement field.

We noted in Sec.~\ref{sec:lagr-hamilt-meta} that an advantage of
working with the Hamiltonian formalism in the Power-Zienau-Woolley
picture is that long-range, instantaneous interactions between
meta-atoms do not appear in the Hamiltonian.
In any treatment of electrodynamics, the instantaneous non-causal
nature of the Coulomb interaction is cancelled by other non-causal
contributions to dynamics.
The form of this cancellation, however, is often rather subtle.
In the Power-Zienau-Woolley Hamiltonian, the Coulomb potential is
absorbed into a local polarization self-energy.
Interactions between meta-atoms are then mediated entirely by the
variables describing the scattered EM fields.

In carrying out the simplification, it is useful to note the following
properties of the longitudinal and transverse components of any two
vector fields $\spvec{V}_1$ and $\spvec{V}_{2}$.
The first is that
\begin{equation}
  \label{eq:Vparperp1}
  \int d^3r\, \spvec{V}_{1,\parallel}(\rv) \cdot
  \spvec{V}_{2,\perp}(\rv) = 0 \text{\,,}
\end{equation}
and as a consequence
\begin{eqnarray}
  \lefteqn{ \int d^3r\, \spvec{V}_1 \cdot \spvec{V}_2 } \nonumber\\
  & & = \int d^3r\,
  \left(\spvec{V}_{1,\parallel} \cdot \spvec{V}_{2,\parallel} +
  \spvec{V}_{1,\perp} \cdot \spvec{V}_{2,\perp}\right)\text{\,.}
  \label{eq:Vparperp2}
\end{eqnarray}
We also note, that because the charge density in our ensemble of
meta-atoms is accounted for entirely by the polarization
[Eq.~\eqref{eq:rhoasP}], the displacement field $\spvec{D}$ is
transverse, i.e., $\spvec{D} = \spvec{D}_\perp$.
We may therefore write the transverse and longitudinal electric fields as
\begin{equation}
  \label{eq:E_perpinDandP}
  \spvec{E}_{\perp} = \frac{1}{\epsilon_0} \left(\spvec{D} -
    \spvec{P}_\perp \right),\quad \spvec{E}_{\parallel} = -\frac{1}{\epsilon_0}
    \spvec{P}_\parallel\,.
\end{equation}
The Coulomb interaction energy [Eq.~\eqref{eq:V_Coul_E_par}] and the
Lagrangian for the free electromagnetic field [Eq.~\eqref{eq:L_EM}] then becomes
\begin{eqnarray}
 \label{eq:V_Coul_P_par}
  V_{\mathrm{Coul}} &=& \frac{1}{2 \epsilon_0} \int d^3r\, |\spvec{P}_\parallel|^2,\\
  \label{eq:L_EM_inDandP}
  \mathcal{L}_{\mathrm{EM}} &=& \frac{1}{2\epsilon_0}
  \int d^3r\, \left(\left|\spvec{D}\right|^2 -
    c^2\left|\spvec{B}\right|^2 \right) \nonumber \\
  && + \frac{1}{2\epsilon_0} \int d^3r\, \left| \spvec{P}_\perp
  \right|^2 - \frac{1}{\epsilon_0} \int d^3r\, \spvec{D} \cdot
  \spvec{P} \text{\,.}
\end{eqnarray}
Similarly, the other two integrals appearing in
Eq.~\eqref{eq:H_EreDef} can be expressed as
\begin{eqnarray}
  \int d^3r\, \spvec{E}_\perp \cdot \spvec{D} &=&
  \frac{1}{\epsilon_0} \int d^3r\, \left( \left|\spvec{D}\right|^2 -
    \spvec{D}\cdot\spvec{P} \right) \\
  \int d^3r\, \spvec{E} \cdot \spvec{P} &=&
  \frac{1}{\epsilon_0} \int d^3r\, \left(\spvec{D}\cdot\spvec{P}
    - \left|\spvec{P}\right|^2 \right)
\end{eqnarray}
By the property of Eq.~\eqref{eq:Vparperp2}, we may write the portion
of the Hamiltonian, $\mathcal{H}_{\mathrm{E}}$ as
\begin{equation}
  \label{eq:H_E_final}
  \spvec{H}_{\mathrm{E}} = \mathcal{H}_{\mathrm{EM}} +
  \frac{1}{2\epsilon_0} \int d^3r\, \left| \spvec{P}\right|^2 -
  \frac{1}{\epsilon_0} \int d^3r\, \spvec{D} \cdot \spvec{P} \text{\,,}
\end{equation}
where $\mathcal{H}_{\mathrm{EM}}$, given in Eq.~\eqref{eq:HEM}, is the
Hamiltonian for the electromagnetic field.
The second term in Eq.~\eqref{eq:H_E_final}
has absorbed
the Coulomb interaction and results only in a local meta-atom self-interaction as discussed in Sec.~\ref{sec:lagr-hamilt-meta}.
The final term of Eq.~\eqref{eq:H_E_final} accounts for interaction
between the distribution of electric dipoles in the polarization
density and the displacement field.
The total Hamiltonian for the system is then given in
Eq.~\eqref{eq:Ham}.

\section{Collective interactions of strongly interacting meta-atoms
outside the rotating wave approximation}
\label{sec:ensemble-meta-atoms}

In Sec.~\ref{sec:analysis-model}, we saw how the EM field scattered from the
metamaterial elements produces interactions between meta-atom current
oscillations in the RWA.
For this approximation to be strictly valid, the meta-atoms must
weakly interact with the field, radiatively decaying at rates much
slower than the oscillator frequencies.
In many metamaterial systems, however, such assumptions can be violated,
and the RWA may not be employed.
In this Appendix, we develop a more general framework for the
dynamics that allows us to account for very strong radiative
coupling between the oscillator variables.
We begin by reframing the equations of motion for the dynamic
variables $Q_j$ and their conjugate momenta in terms of column
vectors of scaled quantities
[Eqs.~\eqref{eq:QsvDef} and \eqref{eq:phiSVDef}]
\begin{align}
  \label{eq:DVarColVecDefs}
  \colvec{\tilde{Q}} \equiv & (\tilde{Q}_1, \tilde{Q}_2, \ldots,
  \tilde{Q}_N)^T \text{,}\\
  \colvec{\tilde{\phi}} \equiv & (\tilde{\phi}_1, \ldots, \tilde{\phi}_N)^T
\end{align}
in the frequency domain.
For each meta-atom $j$, the scaled charge $\tilde{Q}_j$ and
its scaled conjugate momentum $\tilde{\phi}_j$ are slowly varying,
with bandwidths comparable to that of the incident field's positive
frequency component.
In the time domain, they are related to the physical quantities
$Q_j$ and $\phi_j$ by
\begin{align}
    \label{eq:QtildePhitildeRevisited}
    Q_j(t) =& \sqrt{\omega_jC_j}
    \left[e^{-i\Omega_0 t}
      \tilde{Q}_j(t) + e^{i\Omega_0 t}
      \tilde{Q}_j^\ast(t)\right] \\
    \phi_j(t) =& \sqrt{\omega_jL_j} \left[
      e^{-i\Omega_0 t}
      \tilde{\phi}_j(t) + e^{i\Omega_0 t}
      \tilde{\phi}_j^\ast(t)
    \right] \,\textrm{.}
  \end{align}
In deriving the coupling between the elements, we will find that
$\colvec{\tilde{\phi}}$ is related to the scaled currents
[Eq.~\eqref{eq:scaledCurrent}]
\begin{equation}
  \colvec{\tilde{I}} \equiv (\tilde{I}_1, \ldots, \tilde{I}_N)^T
\label{eq:7}
\end{equation}
through a dimensionless mutual inductance matrix $\mathcal{M}$, and
similarly, that the vector of scaled EMFs
[Eq.~\eqref{eq:svEMFDef}]
\begin{equation}
  \colvec{\tilde{\mathcal{E}}} \equiv (\tilde{\mathcal{E}}_1, \ldots,
  \tilde{\mathcal{E}}_N)^T
  \label{eq:4}
\end{equation}
is related to $\colvec{\tilde{Q}}$ through a
matrix resembling a mutual capacitance.
Since the meta-atoms are separated by significant fractions of a
wavelength and interactions between them are mediated by the radiated
field, $\colvec{\tilde{\phi}}$ contains an additional contribution from
$\colvec{\tilde{Q}}$. In addition, $\colvec{\tilde{\mathcal{E}}}$ is linearly
coupled to $\colvec{\tilde{\phi}}$.
This is because oscillating dipoles,
whether electric or magnetic,
produce both electric and magnetic
fields which drive $\colvec{\tilde{\mathcal{E}}}$ and
$\colvec{\tilde{\phi}}$, respectively.
We find that, in general, this produces an additional non-trivial
coupling between resonators.

We first examine the behavior of the Fourier components of
$\colvec{\tilde{Q}}$ and $\colvec{\tilde{\phi}}$ for a frequency
$\Omega>0$, detuned from the central frequency of the incident field
by $\delta = \Omega - \Omega_0$.
Since, by construction, $\colvec{\tilde{Q}}$, $\colvec{\tilde{\phi}}$,
$\colvec{\tilde{\mathcal{E}}}$ and $\colvec{\tilde{\Phi}}$ are related
only to the positive frequency components of $Q_j$, $\phi_j$,
$\mathcal{E}_j$ and $\Phi_j$, the scaled variables have no Fourier
components for $\delta < -\Omega_0$.
From the equations of motion for the unscaled variables
[Eqs.~\eqref{eq:HamEqs1}] and the definitions of the scaled
variables, we arrive at the relations for $\delta >
-\Omega_0$
\begin{subequations}
  \label{eq:eqsOfMColvec}
  \begin{eqnarray}
    \label{eq:QColVecEqmInI}
    -i(\delta+\Omega_0)\colvec{\tilde{Q}}(\delta)
    & = & \omega \colvec{\tilde{I}}(\delta) \label{eq:QeqmColvec} \\
    -i(\delta + \Omega_0)\colvec{\tilde{\phi}}(\delta) & = &
    \colvec{\tilde{\mathcal{E}}}(\delta)
  \end{eqnarray}
\end{subequations}
where $\omega$ represents a diagonal matrix whose elements
$[\omega]_{j,j} \equiv \omega_j$ are the resonance frequencies of the
individual meta-atoms.
Equations~\eqref{eq:eqsOfMColvec} represent the coupling of the meta-atom
dynamic variables to the EM fields, including the
incident field, the fields emitted by all other meta-atoms, and the
field generated from the meta-atom itself.
The self-interactions were derived in Sec.~\ref{sec:single-meta-atom} [Eqs.~\eqref{eq:emfSingleMA} and \eqref{eq:fluxSingleMA}],
while we obtained the contributions from the scattered fields in
Sec.~\ref{sec:coll-inter-rotat} [Eqs.~\eqref{EMF_jjprime_rwa_final}
and \eqref{eq:Phi_jjprime_rwa_final} with $\delta+\Omega_0$
substituted for $\Omega_0$].

As presently written, Eq.~\eqref{eq:QColVecEqmInI}, states in terms of scaled variables in frequency space that the
rate of change of the meta-atom charge is equal to its current.
Here we are interested in how these rates of change are
related to the states of the meta-atom dynamic variables and their
conjugate momenta.
To express these currents $\colvec{\tilde{I}}$ in terms of charges and
conjugate momenta, we recognize that the conjugate momentum is the sum
of the magnetic flux and the current multiplied by the kinetic
inductance $l_j$
[Eq.~\eqref{eq:conjMomofQ_j}],
\begin{equation}
  \tilde{\phi}_j = \frac{l_j}{L_j}\tilde{I}_j +\tilde{\Phi}_j \text{.}
  \label{eq:conjMomScaled}
\end{equation}
The scaled fluxes, $\tilde{\Phi}_j$
[Eq.~\eqref{eq:svFluxDev}], contain
 contributions from
the meta-atoms' self-generated fields
  [Eq.~\eqref{eq:fluxSingleMA}], which result in magnetic
  self-inductances, as well as to the magnetic fields generated
  externally.
  The kinetic and magnetic self-inductances combine to provide the total
  self-inductance [Eq.~\eqref{eq:LjDef}].
  Equation~\eqref{eq:relationBetweendimlessIandPhi} thus
  relates the conjugate momentum, $\tilde{\phi}_j$, for meta-atom $j$
  to its current $\tilde{I}_j$ and the externally generated flux.
  The contributions to the external flux from
the incident magnetic field
and
fields scattered from other meta-atoms in the system
[Eq.~\eqref{eq:Phi_jjprime_rwa_final} with $\Omega$ substituted for
$\Omega_0$]
combine to provide the external magnetic driving of individual
  meta-atoms.
We synthesize these contributions to obtain the
relationship between the column  vectors of scaled
conjugate momenta $\colvec{\tilde{\phi}}$, currents
$\colvec{\tilde{I}}$, charges $\colvec{\tilde{Q}}$, and
fluxes induced by the incident field $\colvec{\tilde{\Phi}}_{\mathrm{in}}$
expressed as
\begin{equation}
  \label{eq:colvecPhiExp}
  \colvec{\tilde\phi} (\delta) = \mathcal{M}(\Omega) \colvec{\tilde{I}}(\delta)
  - \omega^{-1} \Upsilon_{\mathrm{M}}^{\frac{1}{2}} \mathcal{G}_\times^T(\Omega)
  \Upsilon_{\mathrm{E}}^{\frac{1}{2}} \colvec{\tilde{Q}}(\delta) +
  \colvec{\tilde{\Phi}}_{\mathrm{in}}(\delta) \text{,}
\end{equation}
where we have
redefined
the diagonal matrices $\Upsilon_{\mathrm{E}}$
and $\Upsilon_{\mathrm{M}}$
containing the meta-atom electric and magnetic
dipole emission rates so that they
reflect the frequency dependence of meta-atom scattering rates
  outside the RWA.
These matrices have the diagonal matrix elements
\begin{eqnarray}
  [\Upsilon_{\mathrm{E}}]_{j,j} &\equiv& \left(\frac{\Omega}{\omega_j}\right)^3
  \Gamma_{\mathrm{E},j} \text{ ,}\\
  \left[\Upsilon_{\mathrm{M}}\right]_{j,j} & \equiv &
  \left(\frac{\Omega}{\omega_j}\right)^3
  \Gamma_{\mathrm{E},j} \text{ .}
\end{eqnarray}
The scaled mutual inductance $\mathcal{M}$, is given by
\begin{equation}
  \mathcal{M}(\Omega)  =
  \left( 1 + i \omega^{-1}\Upsilon_{\mathrm{M}}
    +\omega^{-1}\Upsilon_{\mathrm{M}}^{\frac{1}{2}}\mathcal{G}_{\mathrm{M}}
    \Upsilon_{\mathrm{M}}^{\frac{1}{2}}
  \right) \text{,}
\end{equation}
and the matrices $\mathcal{G}_{\mathrm{M}}$ and
  $\mathcal{G}_{\times}$ are given in Eq.~\eqref{eq:GCouplingMats}.
The diagonal portion of $ \mathcal{M}$ has both real and imaginary
components: the real part is the self-inductances' contribution to
this scaled mutual inductance matrix, while the imaginary part
arises from emission of magnetic dipole radiation from the meta-atoms
current oscillations.
Solving Eq.~\eqref{eq:colvecPhiExp} for $\colvec{\tilde{I}}$, yields
\begin{align}
  \colvec{\tilde{I}}(\delta) = &\mathcal{M}^{-1}(\Omega) \Big(
    \colvec{\tilde{\phi}}(\delta) -
    \colvec{\tilde{\Phi}}_{\mathrm{in}}(\delta) \nonumber\\
 &  +
    \omega^{-1} \Upsilon_{\mathrm{M}}^{\frac{1}{2}}
  \mathcal{G}_\times^T(\Omega) \Upsilon_{\mathrm{E}}^{\frac{1}{2}}
  \colvec{\tilde{Q}}(\delta) \Big)  \text{.}
  \label{eq:colVecCurrents}
\end{align}
The current $\tilde{I}_j$ on an element $j$ is not just related to the
conjugate momentum $\tilde{\phi}_j$, but to the conjugate momenta and
charges of all other meta-atoms in the system, as well as the flux
from the incident field.
In the absence of electric dipole radiation $\Upsilon_{\mathrm{E}}=0$,
we recover the relationship between currents and magnetic field fluxes
found in systems of interacting, radiating, inductive circuits.
The additional coupling that results from oscillating electric dipoles
when $\Upsilon_{\mathrm{E}}\ne 0$ adds some richness to the dynamics of metamaterial systems outside the RWA.

As with the magnetic fluxes, the EMFs $\colvec{\tilde{\mathcal{E}}}$
contain contributions from the self-generated electric fields of the
meta-atom [Eq.~\eqref{eq:emfSingleMA}] and from electric fields
generated by all other meta-atoms in the system
[Eq.~\eqref{EMF_jjprime_rwa_final} with $\Omega$ substituted for
$\Omega_0$].
From the previous results, we can express the column vector of EMFs as
\begin{align}
  \colvec{\tilde{\mathcal{E}}}(\delta) = & -\left(\omega -i
    \Upsilon_{\mathrm{E}} - \Upsilon_{\mathrm{E}}^{\frac{1}{2}}
    \mathcal{G}_{\mathrm{E}} (\Omega)
    \Upsilon_{\mathrm{E}}^{\frac{1}{2}} \right)
  \colvec{\tilde{Q}}(\delta) \nonumber \\
  & + \Upsilon_{\mathrm{E}}^{\frac{1}{2}}
  \mathcal{G}_{\times}(\Omega) \Upsilon_{\mathrm{M}}^{\frac{1}{2}}
  \colvec{\tilde{I}}(\delta) +
  \colvec{\tilde{\mathcal{E}}}_{\mathrm{in}}(\delta) \text{.}
  \label{eq:emfcolvec}
\end{align}
The diagonal matrix, $\omega - i \Upsilon_{\mathrm{E}}$, results from
the coupling of each element with its self-generated field where
$\Upsilon_{\mathrm{E}}$ accounts for decay due to electric dipole
radiation.
The matrix $\Upsilon_{\mathrm{E}}^{\frac{1}{2}}
\mathcal{G}_{\mathrm{E}} (\Omega)
\Upsilon_{\mathrm{E}}^{\frac{1}{2}}$, where
  $\mathcal{G}_{\mathrm{E}}$ is given in Eq.~\eqref{eq:GCouplingMats}, provides dipole-dipole coupling between
electric polarization densities of distinct
meta-atoms, while $\Upsilon_{\mathrm{E}}^{\frac{1}{2}}
\mathcal{G}_{\times}(\Omega) \Upsilon_{\mathrm{M}}^{\frac{1}{2}}$
provides radiated contributions of oscillating magnetic dipoles to the
EMFs.
Since we wish to express the EMFs exclusively in terms of the charges
and their conjugate momenta, we eliminate $\colvec{\tilde{I}}$ by
substituting Eq.~\eqref{eq:colVecCurrents} into
Eq.~\eqref{eq:emfcolvec} to obtain
\begin{align}
  \colvec{\tilde{\mathcal{E}}}(\delta) = & -\omega\Xi^{-1}(\Omega)
  \colvec{\tilde{Q}}(\delta)  + \Upsilon_{\mathrm{E}}^{\frac{1}{2}}
  \mathcal{G}_{\times}(\Omega) \Upsilon_{\mathrm{M}}^{\frac{1}{2}}
  \mathcal{M}^{-1}(\Omega)\colvec{\tilde{\phi}}(\delta) \nonumber\\
  & +
  \colvec{\tilde{\mathcal{E}}}_{\mathrm{in}}(\delta)
  - \Upsilon_{\mathrm{E}}^{\frac{1}{2}}
  \mathcal{G}_{\times}(\Omega) \Upsilon_{\mathrm{M}}^{\frac{1}{2}}
  \mathcal{M}^{-1}(\Omega) \colvec{\tilde{\Phi}}_{\mathrm{in}}(\delta)\text{,}
  \label{eq:emfColVec2}
\end{align}
where $\Xi(\Omega)$ is an effective dimensionless mutual capacitance
matrix defined such that
\begin{align}
  &\Xi^{-1}(\Omega) \equiv  1 - i\omega^{-1}\Upsilon_{\mathrm{E}}
  -
  \omega^{-1} \Upsilon_{\mathrm{E}}^{\frac{1}{2}}
  \mathcal{G}_{\mathrm{E}}(\Omega) \Upsilon_{\mathrm{E}}^{\frac{1}{2}}   \nonumber\\
  & - \omega^{-1} \Upsilon_{\mathrm{E}}^{\frac{1}{2}}
  \mathcal{G}_{\times}(\Omega) \Upsilon_{\mathrm{M}}^{\frac{1}{2}}
  \mathcal{M}^{-1}(\Omega) \omega^{-1}
  \Upsilon_{\mathrm{M}}^{\frac{1}{2}} \mathcal{G}_{\times}^T(\Omega)
  \Upsilon_{\mathrm{E}}^{\frac{1}{2}}
  \label{eq:XiDef}
\end{align}
where the matrix expression on the final line arises from the
expression for current in terms of conjugate momenta and charges.
This matrix expression is reminiscent of a
scattering process in which oscillating charges couple to oscillating
conjugate momenta in other meta-atoms via
$\Upsilon_{\mathrm{M}}^{\frac{1}{2}} \mathcal{G}_{\times}^T(\Omega)
\Upsilon_{\mathrm{E}}^{\frac{1}{2}}$, these conjugate momenta are transformed
into currents by $\mathcal{M}^{-1}(\Omega)$, and these currents
produce electric fields in neighboring  meta-atoms through
$\Upsilon_{\mathrm{E}}^{\frac{1}{2}} \mathcal{G}_{\times}(\Omega)
\Upsilon_{\mathrm{M}}^{\frac{1}{2}}$.

Having expressed the currents [Eq.~\eqref{eq:colVecCurrents}]
and EMFs [Eq.~\eqref{eq:emfColVec2}] exclusively in terms of
charges, conjugate momenta and driving due to the incident field, we
may finally write the equations of motion in frequency space for the
slowly varying charges $\colvec{\tilde{Q}}(\delta)$ and conjugate
momenta $\colvec{\tilde{\phi}}(\delta)$.
\begin{widetext}
  \begin{equation}
    -i(\delta+\Omega_0)\left(
      \begin{array}{c}
        \colvec{\tilde{Q}}(\delta) \\
        \colvec{\tilde{\phi}}(\delta)
      \end{array}
    \right)
     =
     \left(
       \begin{array}{cc}
         \omega \mathcal{M}^{-1} \omega^{-1}
         \Upsilon_{\mathrm{M}}^{\frac{1}{2}}
         \mathcal{G}_\times^T
         \Upsilon_{\mathrm{E}}^{\frac{1}{2}} &
         \omega \mathcal{M}^{-1} \\
         - \omega\Xi^{-1} & \Upsilon_{\mathrm{E}}^{\frac{1}{2}}
         \mathcal{G}_\times \Upsilon_{M}^{\frac{1}{2}}
         \mathcal{M}^{-1}
       \end{array}
     \right)
     \left(
       \begin{array}{c}
         \colvec{\tilde{Q}}(\delta) \\
         \colvec{\tilde{\phi}}(\delta)
       \end{array}
    \right)
    + \left(
      \begin{array}{c}
        - \omega \mathcal{M}^{-1}
        \colvec{\tilde{\Phi}}_{\mathrm{in}}(\delta) \\
        \colvec{\tilde{\mathcal{E}}}_{\mathrm{in}}(\delta) -
        \Upsilon_{\mathrm{E}}^{\frac{1}{2}}
        \mathcal{G}_{\times}
        \Upsilon_{\mathrm{M}}^{\frac{1}{2}} \mathcal{M}^{-1}
        \colvec{\tilde{\Phi}}_{\mathrm{in}}(\delta)
      \end{array}
    \right)
    \label{eq:EqsOfMFullMat}
  \end{equation}
  The physical content of Eq.~\eqref{eq:EqsOfMFullMat} becomes more
  evident in the limits where the meta-atoms are spaced far enough
  apart that their interactions can be seen as interactions between
  point sources [i.e., where the approximation in
  Eq.~\eqref{eq:GCouplingMatsDip} holds] and when the magnetic
  interaction is weak ($\Gamma_{\mathrm{M},j} \ll \Omega_0$).
  Then, writing Eq.~\eqref{eq:EqsOfMFullMat} to lowest order in
  $\Gamma_{\mathrm{M},j}/\omega_j$, we obtain dynamic equations where
  $\tilde{Q}_j$ and $\tilde{\phi}_j$ are driven by fields scattered
  from the meta-atom electric
  dipoles $\spvec{d}^{(+)}_{j'}(\Omega) \equiv h_{j'}
    \sqrt{\omega_{j'}C_{j'}} \tilde{Q}_{j'}(\delta)$ and magnetic dipoles
  $\spvec{m}_{j'}^{(+)}(\Omega) \equiv A_{j'}
    \sqrt{\omega_{j'}/L_{j'}}
  \tilde{I}_{j'}(\delta)$
  [see Eq.~\eqref{eq:dipoles} and
  Eqs.~\eqref{eq:QsvDef},\eqref{eq:phiSVDef},\eqref{eq:scaledCurrent}
  and \eqref{eq:GCouplingMatsDip}]
  \begin{subequations}
    \label{eq:DipoleEqsMNonRWA}
    \begin{align}
      -i\Omega \tilde{Q}_j(\delta) &+ \left[\omega_j -
      i\left(\frac{\Omega}{\omega_j}\right)^3 \Gamma_{\mathrm{M},j}
      \right] \tilde{\phi}_j(\delta) =
      -\omega_j\tilde{\Phi}_{\mathrm{in},j}(\delta) \nonumber\\
    & - \alpha_j \unitvec{m}_j \cdot \frac{\mu_0k^3}{4\pi}\sum_{j'\ne j} \left(
      \sptensor{G}(\spvec{r}_j - \spvec{r}_{j'}, \Omega) \cdot
      \spvec{m}_{j'}^{(+)}(\Omega) - c \sptensor{G}_{\times}(\spvec{r}_j -
      \spvec{r}_{j'},\Omega) \cdot \spvec{d}_{j'}^{(+)}(\Omega)\right) \\
    -i\Omega \tilde{\phi}_j(\delta) &+\left[\omega_j -
    i\left(\frac{\Omega}{\omega_j}\right)^3 \Gamma_{\mathrm{E},j}
    \right] \tilde{Q}_j(\delta) = \tilde{\mathcal{E}}_{{\rm in},j}
    \nonumber\\
    & +\chi_j \unitvec{d}_j \cdot \frac{k^3}{4\pi\epsilon_0}\sum_{j' \neq j}
    \left(\sptensor{G}(\spvec{r}_j - \spvec{r}_{j'},\Omega) \cdot
      \spvec{d}_{j'}^{(+)}(\Omega) +\frac{1}{c}
      \sptensor{G}_{\times}(\spvec{r}_j - \spvec{r}_{j'},\Omega) \cdot
      \spvec{m}_{j'}^{(+)}(\Omega) \right) \text{ ,}
  \end{align}
 \end{subequations}
 where $\alpha_j \equiv (2 A_j/3)\sqrt{\omega_j/L_j} / 3$ and
 $\chi_j \equiv 2h_j\sqrt{\omega_jC_j}/3$.
 The incident field drives the meta-atom electric and magnetic
 dipoles, producing the terms $-\omega_j\tilde{\Phi}_{\mathrm{in},j}$
 and  $\tilde{\mathcal{E}}_{\mathrm{in},j}$.
 The effects of the meta-atoms' self-generated fields are included on the left
 hand side of Eqs.~\eqref{eq:DipoleEqsMNonRWA}.
 The magnetic fields produced by all other meta-atomic dipoles in the
 system $j'\ne j$ drive the dynamics of $\tilde{Q}_j$, while the
 electric fields scattered from these meta-atoms drive the
 dynamics of the conjugate momentum $\tilde{\phi}_j$.
\end{widetext}

In principle, one could solve Eq.~\eqref{eq:EqsOfMFullMat} in
the narrow bandwidth approximation in which the driving field
envelopes $\tilde{\spvec{E}}_{\mathrm{in}}$ and
$\tilde{\spvec{B}}_{\mathrm{in}}$ vary on time scales much larger than
$1/\Omega_0$.
One would accomplish this by substituting $\Omega_0$ for $\Omega$ in
the coupling matrices $\mathcal{G}_{\mathrm{E/M/}\times}$ and
$\Upsilon_{\mathrm{E/M}}$, then inverse Fourier transforming
Eq.~\eqref{eq:EqsOfMFullMat}.
This procedure, however, may not be particularly illuminating.
We find it useful to explore the dynamics in terms of the oscillator
normal variables $\colvec{b}$.
But first, we will revisit the basic characteristics of these normal
variables to understand how they behave outside the RWA.

\subsection{The normal meta-atom variables revisited}
\label{sec:normal-meta-atom-revisit}

In Sec.~\ref{sec:analysis-model},
we have assumed the meta-atom normal variables
$\colvec{b}=(b_1,b_2,\ldots,b_N)^T$ are slowly varying, and are
proportional to the slowly varying envelopes of the charges
$\colvec{\tilde{Q}}$ and conjugate momenta $\colvec{\tilde{\phi}}$.
This was a good approximation when we assumed each meta-atom coupled
to its self-generated fields much more strongly than it couples to the
external fields, i.e., when
$\Gamma_{\mathrm{E},j},\Gamma_{\mathrm{M},j} \ll \omega_j,\Omega_0$.
In those limits, the external field interactions act as a perturbation
that slowly alters the dominant behavior of the meta-atoms
oscillating as effective LC circuits.
When the coupling to the external field is stronger, however, we will
see that this is no longer the case.

While the normal variables are slowly varying in the RWA, in general
they contain fast oscillating components even when
$\colvec{\tilde{Q}}$ and $\colvec{\tilde{\phi}}$ have narrow bandwidths.
To see this, we rewrite the normal variables [defined in
Eq.~\eqref{eq:bSingAtDef}] in terms of the slowly varying dynamic
variables.
Recall that in terms of the slowly varying quantities, the original
charges and conjugate momenta are expressed as
\begin{subequations}
  \label{eq:13}
  \begin{align}
    \frac{Q_j(t)}{\sqrt{\omega_jC_j}} &= e^{-i\Omega_0 t}
    \tilde{Q}_j(t) + e^{i\Omega_0 t} \tilde{Q}_j^\ast(t) \\
    \frac{\phi_j(t)}{\sqrt{\omega_j L_j}} & = e^{-i\Omega_0 t}
    \tilde{\phi}_j(t) + e^{i\Omega_0 t} \tilde{\phi}_j^\ast(t) \text{.}
  \end{align}
\end{subequations}
We therefore express the column vector of normal variables as
\begin{equation}
  \colvec{b}(t) = \colvec{\tilde{b}}^{(+)}(t) +
  e^{2i\Omega_0}[\colvec{\tilde{b}}^{(-)}(t)]^\ast \,,
  \label{eq:normalVarsInSVVars}
\end{equation}
where we have defined the slowly varying normal variable components
\begin{equation}
  \colvec{\tilde{b}}^{(\pm)} (t) \equiv \frac{1}{\sqrt{2}}(\colvec{\tilde{Q}}(t) \pm i
  \colvec{\tilde{\phi}}(t)) \,. \label{eq:svNormVarsDef}
\end{equation}
The normal variables $\colvec{b}$ therefore have a bimodal spectrum
with a slowly varying component peaked at zero frequency and fast
oscillating component peaked at frequency $-2\Omega_0$.
Note that despite the apparent similarity between the
  definition of $b_j$ [Eq.~\eqref{eq:bSingAtDef}] and
  $\colvec{\tilde{b}}^{(\pm)}$ [Eq.~\eqref{eq:svNormVarsDef}],
  $\colvec{\tilde{b}}^{(-)}$ is not equal to $\left[\colvec{\tilde{b}}^{(+)}\right]^\ast$.
In the RWA, for example, we ignore the rapidly rotating term [$ \exp
(2i\Omega_0)[\colvec{\tilde{b}}^{(-)}(t)]^\ast$ in
Eq.~\ref{eq:normalVarsInSVVars}] and make the approximation
$\colvec{\tilde{b}}^{(-)} \approx 0$.
Physically, the RWA implies that the
normalized conjugate momentum
$\tilde{\phi}_j(t)$ has the same amplitude as $\tilde{Q}_j(t)$ but its
oscillation lags by a definite phase $\pi/2$ in accordance
  with Eq.~\eqref{eq:svVarsAndbs}.
However,
outside the
RWA,
the contributions of $\colvec{\tilde{b}}^{(-)}$ cannot be
neglected.

\subsection{Normal variable dynamics outside the RWA}
\label{sec:norm-fari-dynam-nonRWA}

We can obtain the normal variable dynamics from those for the slowly
varying charges and conjugate momenta from
Eq.~\eqref{eq:EqsOfMFullMat}.
In terms of the normal variables the vectors $\colvec{\tilde{Q}}$ and
$\colvec{\tilde{\phi}}$ are given by
\begin{equation}
  \left(
    \begin{array}{c}
      \colvec{\tilde{Q}} \\
      \colvec{\tilde{\phi}}
    \end{array}
  \right) = \frac{1}{\sqrt{2}}
  \left(
    \begin{array}{cc}
      1 & 1 \\
      -i & i
    \end{array}
  \right)
  \left(
    \begin{array}{c}
      \colvec{\tilde{b}^{(+)}}\\
      \colvec{\tilde{b}^{(-)}}
    \end{array}
  \right) \text{.}
  \label{eq:NormalVarTransform}
\end{equation}
The vectors of normal variables $\colvec{\tilde{b}}^{(\pm)}$ therefore
evolve according to
\begin{widetext}
  \begin{equation}
    \label{eq:EqsOfM}
    -i\delta
    \left(
      \begin{array}{c}
        \colvec{\tilde{b}}^{(+)}(\delta) \\
        \colvec{\tilde{b}}^{(-)}(\delta)
      \end{array}
    \right)
    =
    \left(
      \begin{array}{cc}
        \mathcal{C}^{(+,+)}(\Omega) & \mathcal{C}^{(+,-)}(\Omega) \\
        \mathcal{C}^{(-,+)}(\Omega)  & \mathcal{C}^{(-,-)}(\Omega)
      \end{array}
    \right)
    \left(
      \begin{array}{c}
        \colvec{\tilde{b}}^{(+)}(\delta) \\
        \colvec{\tilde{b}}^{(-)}(\delta)
      \end{array}
    \right)
    +
    \left(
      \begin{array}{c}
        \colvec{f}^{(+)}_{\mathrm{in}}(\delta) \\
        \colvec{f}^{(-)}_{\mathrm{in}}(\delta)
      \end{array}
    \right) \text{,}
  \end{equation}
  where $\mathcal{C}^{(s_1,s_2)}$ ($s_1,s_2=\pm$) are $N\times N$
  matrices that provide a linear coupling between the column vectors
  $\tilde{\colvec{b}}^{(s_2)}$ and $\tilde{\colvec{b}}^{(s_1)}$.
  These coupling matrices are given by
  \begin{align}
    \mathcal{C}^{(s_1,s_2)} &\equiv
    i\left(\Omega_0 - s_1\omega\right) \delta_{s_1,s_2} - \frac{s_1
      \Upsilon_{\mathrm{E}} + s_2\Upsilon_{\mathrm{M}}
      \mathcal{M}^{-1}}{2} \nonumber\\
     &+ i \frac{s_1
      \Upsilon_{\mathrm{E}}^{\frac{1}{2}} \mathcal{G}_{\mathrm{E}}
      \Upsilon_{\mathrm{E}}^{\frac{1}{2}}  + s_2
      \Upsilon_{\mathrm{M}}^{\frac{1}{2}} \mathcal{G}_{\mathrm{M}}
      \Upsilon_{\mathrm{M}}^{\frac{1}{2}} \mathcal{M}^{-1}} {2} + \frac{1}{2}
    \left(
      \omega + is_1 \Upsilon_{\mathrm{E}}^{\frac{1}{2}}
      \mathcal{G}_\times \Upsilon_{\mathrm{M}}^{\frac{1}{2}}
    \right)
    \mathcal{M}^{-1}\omega^{-1} \Upsilon_{\mathrm{M}}^{\frac{1}{2}}
    \mathcal{G}_\times^T \Upsilon_{\mathrm{E}}^{\frac{1}{2}} +
    \frac{s_1s_2 \Upsilon_{\mathrm{E}}^{\frac{1}{2}}
      \mathcal{G}_{\times} \Upsilon_{\mathrm{M}}^{\frac{1}{2}}
      \mathcal{M}^{-1}} {2} \textrm{.}
    \label{eq:CNRWA_Def}
\end{align}
The top line of Eq.~\eqref{eq:CNRWA_Def} contains the diagonal
elements of the matrices $\mathcal{C}^{(s_1,s_2)}$ that arise from the
interaction of the meta-atoms with their self-generated fields.
For example the diagonal elements of $\mathcal{C}^{(+,+)}$ contain the
detunings $\mathrm{\Delta}$
[Eq.~\eqref{eq:DeltaMatDef}]
and the total radiative decay rate
$\Upsilon_{\mathrm{E}} + \Upsilon_{\mathrm{M}}$.
The interaction matrices
$\mathcal{C}^{(s_1,s_2)}$
contain the effects of all scattering
processes, including those resulting from scattered electric fields
emitted from electric and magnetic dipoles and those resulting from
magnetic fields emitted from magnetic and electric dipoles.
Interaction with the incident field produces the driving represented
by the column vectors
\begin{equation}
  \label{eq:fpmDef}
  \colvec{f}_{\mathrm{in}}^{(\pm)}(\delta) \equiv \frac{1}{\sqrt{2}}
  \left[ \pm i \tilde{\colvec{\mathcal{E}}} - \left(\omega \pm i
      \Upsilon_{\mathrm{E}}^{\frac{1}{2}} \mathcal{G}_\times
      \Upsilon_{\mathrm{M}}^{\frac{1}{2}}\right) \mathcal{M}^{-1}
    \tilde{\colvec{\Phi}}_{\mathrm{in}}\right]
\end{equation}

The EM mediated
interactions simplify greatly if we assume the self-inductance
of a meta-atom is much greater than the mutual inductance between any
two meta-atoms.
A necessary condition for this is that $\Gamma_{\mathrm{M},j} \ll
\omega_j$ for all $j$ since $\mathcal{M} = 1 +
O(\Gamma_{\mathrm{M}}\omega^{-1})$.
In this limit, we neglect all contributions of order
$\Gamma_{\mathrm{M},j}/\omega_j$ to the mutual inductance, allowing us
to make the substitution $\mathcal{M}^{-1} \approx 1$.
This yields
\begin{equation}
  \label{eq:CNRWA_approx}
  \mathcal{C}^{(s_1,s_2)} \approx i\left(\Omega_0 - s_1\omega\right)
  \delta_{s_1,s_2}  - \frac{s_1
      \Upsilon_{\mathrm{E}} + s_2\Upsilon_{\mathrm{M}}}{2}
    + i \frac{s_1
    \Upsilon_{\mathrm{E}}^{\frac{1}{2}} \mathcal{G}_{\mathrm{E}}
    \Upsilon_{\mathrm{E}}^{\frac{1}{2}}  + s_2
    \Upsilon_{\mathrm{M}}^{\frac{1}{2}} \mathcal{G}_{\mathrm{M}}
    \Upsilon_{\mathrm{M}}^{\frac{1}{2}}} {2} +
    \frac{ \Upsilon_{\mathrm{M}}^{\frac{1}{2}}
    \mathcal{G}_\times^T \Upsilon_{\mathrm{E}}^{\frac{1}{2}} + s_1s_2
    \Upsilon_{\mathrm{E}}^{\frac{1}{2}}
      \mathcal{G}_{\times} \Upsilon_{\mathrm{M}}^{\frac{1}{2}}} {2} \,\textrm{.}
  \end{equation}
  Under the additional assumption that all meta-atom resonance frequencies
  lie in a narrow bandwidth around $\Omega_0$,
  the matrix providing the dynamic coupling
  between the various $\tilde{b}^{(+)}_{j}$ and $\tilde{b}^{(+)}_{j'}$ is
  identical to the coupling matrix between normal variables in the
  RWA [Eq.~\eqref{eq:C_rwa}], i.e. $\mathcal{C}^{(+,+)}
  \approx \mathcal{C}$.

  \subsection{Temporal dynamics and collective modes outside the RWA}
  \label{sec:temp-dynam-coll}

  When the incident field possesses a narrow bandwidth around
  $\Omega_0$ and varies much more slowly than the time it takes for
  light to propagate across the metamaterial sample, we can obtain a
  simple expression for the collective temporal evolution of the
  metamaterial.
  With this slowly varying incident field, we can approximate the
  dynamics by replacing the frequencies $\Omega \equiv \Omega_0 +
  \delta$ appearing in the interaction matrices with $\Omega_0$.
  We then inverse Fourier transform Eq.~\eqref{eq:EqsOfM} to obtain
  \begin{equation}
    \label{eq:EqsOfM_timeDomain}
    \frac{d}{dt}
    \left(
      \begin{array}{c}
        \colvec{\tilde{b}}^{(+)}(t) \\
        \colvec{\tilde{b}}^{(-)}(t)
      \end{array}
    \right)
    =
    \left(
      \begin{array}{cc}
        \mathcal{C}^{(+,+)}(\Omega_0) & \mathcal{C}^{(+,-)}(\Omega_0) \\
        \mathcal{C}^{(-,+)}(\Omega_0)  & \mathcal{C}^{(-,-)}(\Omega_0)
      \end{array}
    \right)
    \left(
      \begin{array}{c}
        \colvec{\tilde{b}}^{(+)}(t) \\
        \colvec{\tilde{b}}^{(-)}(t)
      \end{array}
    \right)
    +
    \left(
      \begin{array}{c}
        \colvec{f}^{(+)}_{\mathrm{in}}(t) \\
        \colvec{f}^{(-)}_{\mathrm{in}}(t)
      \end{array}
    \right) \text{,}
  \end{equation}
  These equations describe the collective response of a
  metamaterial to a narrow bandwidth incident field where the
  inter-resonator interactions and emission rates can be arbitrarily
  large.  Unlike the simplified collective dynamics derived in
  Sec.~\ref{sec:coll-inter-rotat}, Eq.~\eqref{eq:EqsOfM_timeDomain} is not
  subject to the constraints of the RWA.
\end{widetext}

\subsection{Recovering the dynamics of the RWA}
\label{sec:recov-dynam-rwa}

The dynamics in the RWA that we explored
earlier in Sec.~\ref{sec:analysis-model} amounted to neglecting the fast oscillating
components of the normal variables $\colvec{b}$.
Here, this equates to assuming $\colvec{\tilde{b}}^{(-)} = 0$, and therefore
$\colvec{b} = \tilde{\colvec{b}}^{(+)}$.
We argued earlier that this approximation is valid in the limits of
weak interaction -- i.e., $\Gamma_{\mathrm{E},j},\, \Gamma_{\mathrm{M},j} \ll
\Omega_0$ -- and in which all single meta-atom resonance frequencies
lie within a narrow bandwidth about the driving frequency -- i.e.,
$|\Delta_j| \ll \Omega_0$.
Indeed, when the interactions are sufficiently weak, the diagonal
elements of $\mathcal{C}^{(-,-)}$
[Eq.~\eqref{eq:EqsOfM}] [$i(\Omega_0 + \omega) + (\Gamma_\mathrm{E} +
\Gamma_{\mathrm{M}}) /2$] dominate over every other element of the
coupling.
As a result, in the response of the metamaterial to the incident field,
the elements of $\tilde{\colvec{b}}^{(-)}$ would be negligible in
comparison to $\tilde{\colvec{b}}^{(+)}$.
Note, however, that although $\Gamma_{\mathrm{E/M}} \ll \Omega_0$ is a
necessary condition for the validity of the RWA, it is not sufficient
in and of itself.
This is because the interaction between elements can still become very
strong if the separation between them is much less than a
wavelength.
Here, however, we will assume the inter-element separation is sufficiently large that the limits on $\Gamma_{\mathrm{E/M}}$ are sufficient.
In the RWA, we may therefore expand $\mathcal{C}^{(+,+)}$ to lowest
order in $\Gamma_{\mathrm{E}}$ and $\Gamma_{\mathrm{M}}$ and make the
approximation  $(\Omega/\omega_j)^3\approx 1$.
Upon doing this, we recover precisely the dynamics given in
Eq.~\eqref{eq:rwa_b_eqm} of subsection \ref{sec:coll-inter-rotat}.

%\bibliography{metamaterials}

\end{document}